\pdfoutput=1
\documentclass[11pt,twoside,a4paper,cmspaper,final,collab]{cms-tdr}
\def\svnVersion{321c892-D}\def\svnDate{2020/11/25}\def\cmsCernNoTag{CERN-EP-2020-180}\def\cmsCernDate{\today}\def\cmsMessage{"Published in the Journal of Instrumentation as \href{http://dx.doi.org/10.1088/1748-0221/16/02/P02010}{\doi{10.1088/1748-0221/16/02/P02010}.}"}
\begin{document}\cmsNoteHeader{PRF-18-002}

\providecommand{\cmsLeft}{left\xspace}
\providecommand{\cmsRight}{right\xspace}

\newcommand{\sqrts}{\ensuremath{{\sqrt{s}}}\xspace}
\newcommand{\ECR}{\ensuremath{{E_{\mathrm{CR}}}}\xspace}
\newcommand{\Epp}{\ensuremath{E_{{\Pp}{\Pp}}}\xspace}
\newcommand{\Ejet}{\ensuremath{{E_{\text{jet}}}}\xspace}
\newcommand{\Egen}{\ensuremath{{E_{\text{}{gen}}}}\xspace}
\newcommand{\Edet}{\ensuremath{{E_{\text{det}}}}\xspace}
\newcommand{\Erec}{\ensuremath{{E_{\text{rec}}}}\xspace}
\newcommand{\Ecal}{\ensuremath{{E_{\text{cal}}}}\xspace}
\newcommand{\SLED}{\ensuremath{{S_{\text{LED}}}}\xspace}
\newcommand{\Sdet}{\ensuremath{{S_{\text{det}}}}\xspace}
\newcommand{\Spp}{\ensuremath{{S_{{\Pp}{\Pp}}}}\xspace}
\newcommand{\Cpp}{\ensuremath{{C_{{\Pp}{\Pp}}}}\xspace}
\newcommand{\pdqz}{\ensuremath{{p_{\mathrm{dqz}}}}\xspace}
\newcommand{\Npe}{\ensuremath{{N_{\text{p.e.}}}}\xspace}
\newcommand{\gpmt}{\ensuremath{{g_{\text{PMT}}}}\xspace}
\newcommand{\felec}{\ensuremath{{f_{\text{electrons}}}}\xspace}
\newcommand{\QE}{\ensuremath{{\epsilon_{\text{quant}}}}\xspace}
\newcommand{\NgamCh}{\ensuremath{{N^\gamma_{\text{Ch}}}}\xspace}
\newcommand{\Nch}{\ensuremath{{N_{\text{channel}}}}\xspace}
\newcommand{\nmumin}{\ensuremath{{n_{\mu,\text{min}}}}\xspace}
\newcommand{\nextramax}{\ensuremath{{n_{\text{extra,max}}}}\xspace}
\newcommand{\gHV}{\ensuremath{{g_{\text{HV}}}}\xspace}

\cmsNoteHeader{PRF-18-002}
\title{The very forward CASTOR calorimeter\\of the CMS experiment}

\date{\today}

\abstract{
The physics motivation, detector design, triggers, calibration,
alignment, simulation, and overall performance of the very forward
CASTOR calorimeter of the CMS experiment are reviewed. The CASTOR
Cherenkov sampling calorimeter is located very close to the 
LHC beam line, at a radial distance of about
1\unit{cm} from the beam pipe, and at 14.4\unit{m} from the CMS
interaction point, covering the pseudorapidity range of
$-6.6<\eta<-5.2$.  It was designed to withstand high ambient
radiation and strong magnetic fields.  The performance of the detector
in measurements of forward energy density, jets, and processes
characterized by rapidity gaps, is reviewed using data collected in
proton and nuclear collisions at the LHC.}

\hypersetup{
  pdfauthor={CMS Collaboration},
  pdftitle={The very forward CASTOR calorimeter of the CMS experiment},
  pdfsubject={CASTOR very-forward calorimeter},
  pdfkeywords={CASTOR, CMS, forward detector, Cherenkov calorimeter}}

\maketitle 
\tableofcontents

\section{Introduction}
\label{sec:intro}

The CASTOR (Centauro And STrange Object Research) calorimeter was
proposed~\cite{Angelis:1997vx,Angelis:2001mr},
built~\cite{Aslanoglou:2007fx,Aslanoglou:2007wv,Basegmez:2008zza,Gottlicher:2010zz,Andreev:2010zzb},
and installed in the CMS experiment~\cite{Chatrchyan:2008aa} with the
purpose of studying very forward particle production in heavy ion (HI) and
proton-proton ({\Pp}{\Pp}) collisions at the CERN LHC. The
calorimeter extends the CMS
acceptance to the very forward pseudorapidity range, $-6.6<\eta<-5.2$.

The location and design of CASTOR are optimized for the study of the
longitudinal development of electromagnetic and hadronic showers
produced by particles emitted at very small polar angles (between 0.16$^\circ$ and 0.64$^\circ$) with respect
to the beam direction. This prime detector motivation---focused on
searches for deeply penetrating particles, such as
strangelets~\cite{Angelis:1997vx,Angelis:2001mr}, connected to exotic
events observed in high-energy cosmic ray (CR)
interactions~\cite{Bjorken:1979xv,Arisawa:1994eq,Kempa:2012zz}---was
 extended also to measurements of generic
properties of particle production at forward rapidities in inelastic
proton and nuclear collisions, as well as to identify rapidity gaps
(regions in the detector devoid of any particle production) in
diffractive and exclusive processes~\cite{Albrow:2006xt,Deile:2010mv}.
The physics program and reach of the multipurpose CASTOR calorimeter
are broad and unique, because no comparable instrumentation exists in this
$\eta$ range at any other interaction point (IP) of the LHC.

The installation of a calorimeter at the forward rapidities covered by
CASTOR involves significant challenges. The beam pipe is just
$\approx$1\unit{cm} away from the detector and must be carefully
protected. The calorimeter is immersed in the forward shielding of
CMS, which channels the strong magnetic field from the central
region of the apparatus. Small gaps in the shielding produce strong local
variations in the magnetic field structure.  The field has a
magnitude of about $B=0.2\unit{T}$ in the vicinity of the
detector and has an impact on the shower development inside the
calorimeter and also, more importantly, on the gain of the
photomultiplier tubes (PMTs) used to record the signals. Because of the
enormous Lorentz boost of the particles produced at rapidities
approaching the beam direction, the energy absorbed by the
calorimeter reaches hundreds of \TeV per collision event~\cite{D'Enterria:2007xr}, and
thus the radioactive activation of the calorimeter itself, as well as
of the surrounding shielding, is one of the largest at the LHC per
unit of integrated luminosity. Because of the lack of precise
vertex-pointing capabilities of the calorimeter, it is not possible to
distinguish particles produced in different proton or nucleus
collisions occurring simultaneously at the IP (pileup), within the default
25\unit{ns} readout integration time of the detector. Thus, operation
of the CASTOR calorimeter is most useful for LHC luminosities
corresponding to a maximum average number of collisions around
one per bunch crossing. Lower luminosities are often
required for physics analyses that are very sensitive to pileup
backgrounds.

\begin{figure}[tbp!]
 \centering
 \includegraphics[width=.75\textwidth]{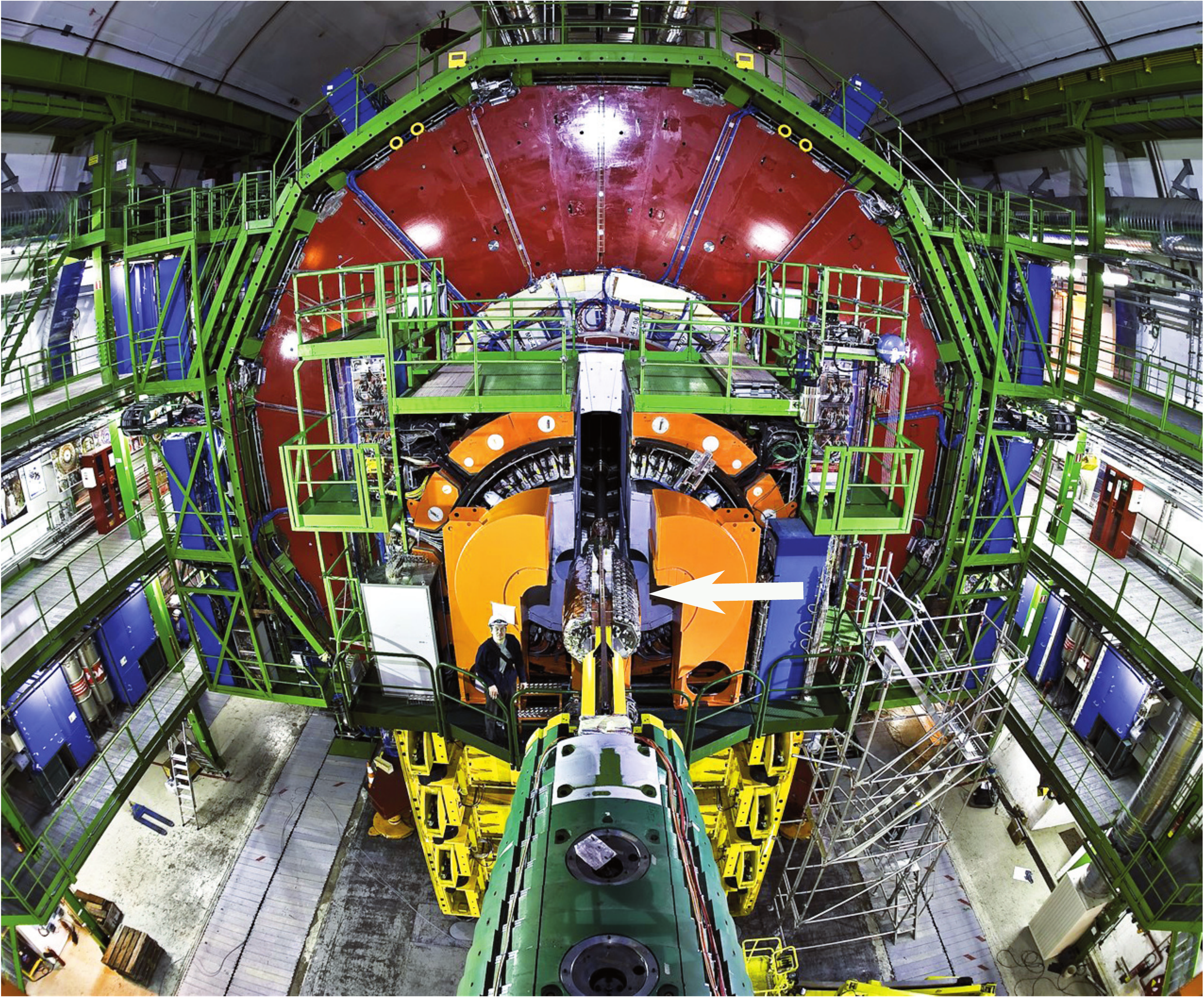}
 \caption{Picture of the forward region of the CMS experiment with the
  two half-cylinders of the CASTOR calorimeter (indicated by an arrow) visible in the zone
  where the collar shielding (orange) structures are open. The green
  cylindrical structure in the foreground belongs to the target
  absorber region. The massive shielding around the calorimeter is in
  the open position. The two half-cylinders of CASTOR are not yet
  closed around the LHC beam pipe that is visible emerging from
  the target absorber region in the foreground and disappearing
  towards the IP inside the CMS detector.}
 \label{fig:photo1}
\end{figure}

\begin{figure}[tbp!]
 \centering
 \includegraphics[width=0.45\textwidth]{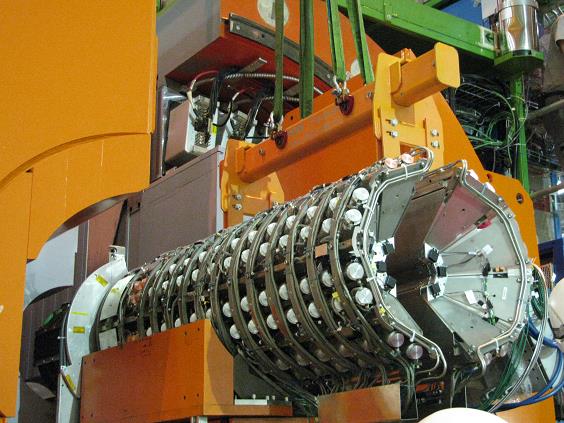}\hfill
  \includegraphics[width=0.45\textwidth]{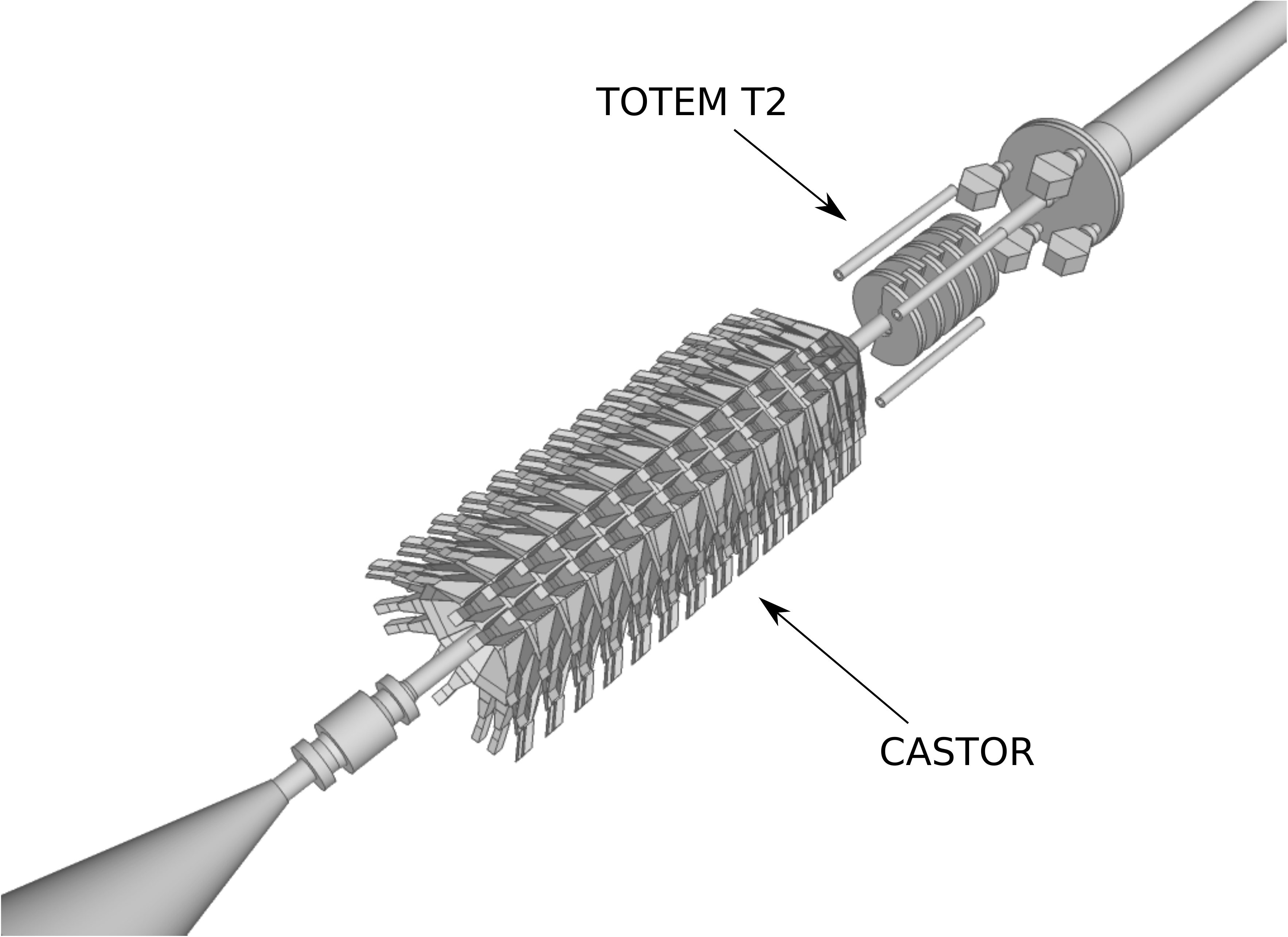}
 \caption{Left: Closeup image of the CASTOR calorimeter surrounded by the 
  CMS collar shielding in the open position during an integration test,
  about 5\unit{m} below the beam pipe. Right: Visualization
  of the CASTOR detector integrated around the beam pipe. The CMS
  interaction point is at 14.4\unit{m} upstream (towards the upper
  right). The TOTEM T2 tracking station, and several pieces of LHC/CMS
  infrastructure, are visible along this
  direction. }
 \label{fig:photo2}
\end{figure}

Figure~\ref{fig:photo1} shows the CASTOR calorimeter installed around
the beam pipe in front of the central part of the CMS detector, and
Fig.~\ref{fig:photo2}~(\cmsLeft) displays a closeup view of the calorimeter surrounded by the open collar shielding.
A visualization of the calorimeter
integration around the beam pipe is shown in
Fig.~\ref{fig:photo2}~(\cmsRight).

Many of the results discussed in this work
are based on the various CASTOR measurements carried out so far in
{\Pp}{\Pp}~\cite{Chatrchyan:2011wm,Chatrchyan:2013gfi,Khachatryan:2015gka,Sirunyan:2017nsj,Sirunyan:2018jwn,Sirunyan:2018nqx,Sirunyan:2019dfx,Sirunyan:2019rqy},
proton-lead
({\Pp}Pb)~\cite{Sirunyan:2018ffo,Sirunyan:2018nqr,Sirunyan:2019nog}, and
lead-lead (PbPb)~\cite{dEdEtaPbPb} collisions at the LHC.
The calorimeter took data during LHC Runs 1 (2009--13) and 2 (2015--2018).
It was decommissioned in 2019--20, since a redesign of the beam pipe renders
the detector mechanically incompatible with LHC Run 3 operation (expected to start in 2022). 
The CASTOR data are being made publicly available via the CERN open
data initiative~\cite{opendata,Mccauley:2019uis}. The data are
accompanied with all tools, calibration,
and running conditions as described and summarized here.

The paper is organized as follows. Section~\ref{sec:phys} reviews the
main physics motivations of the CASTOR calorimeter project. The
detector design is summarized in Section~\ref{sec:design}, whereas the
CASTOR operation and triggers are described in
Section~\ref{sec:operation}.  The event reconstruction, including the
calorimeter calibration, and the detector geometry and alignment are
discussed in Sections~\ref{sec:reco} and~\ref{sec:align},
respectively. The CASTOR simulation and its validation are presented
in Section~\ref{sec:sim}.  The summary of the paper 
is provided in Section~\ref{sec:summ}.

\section{Physics motivation}
\label{sec:phys}

\begin{figure}[tbp!]
\centering
 \includegraphics[width=.61\textwidth]{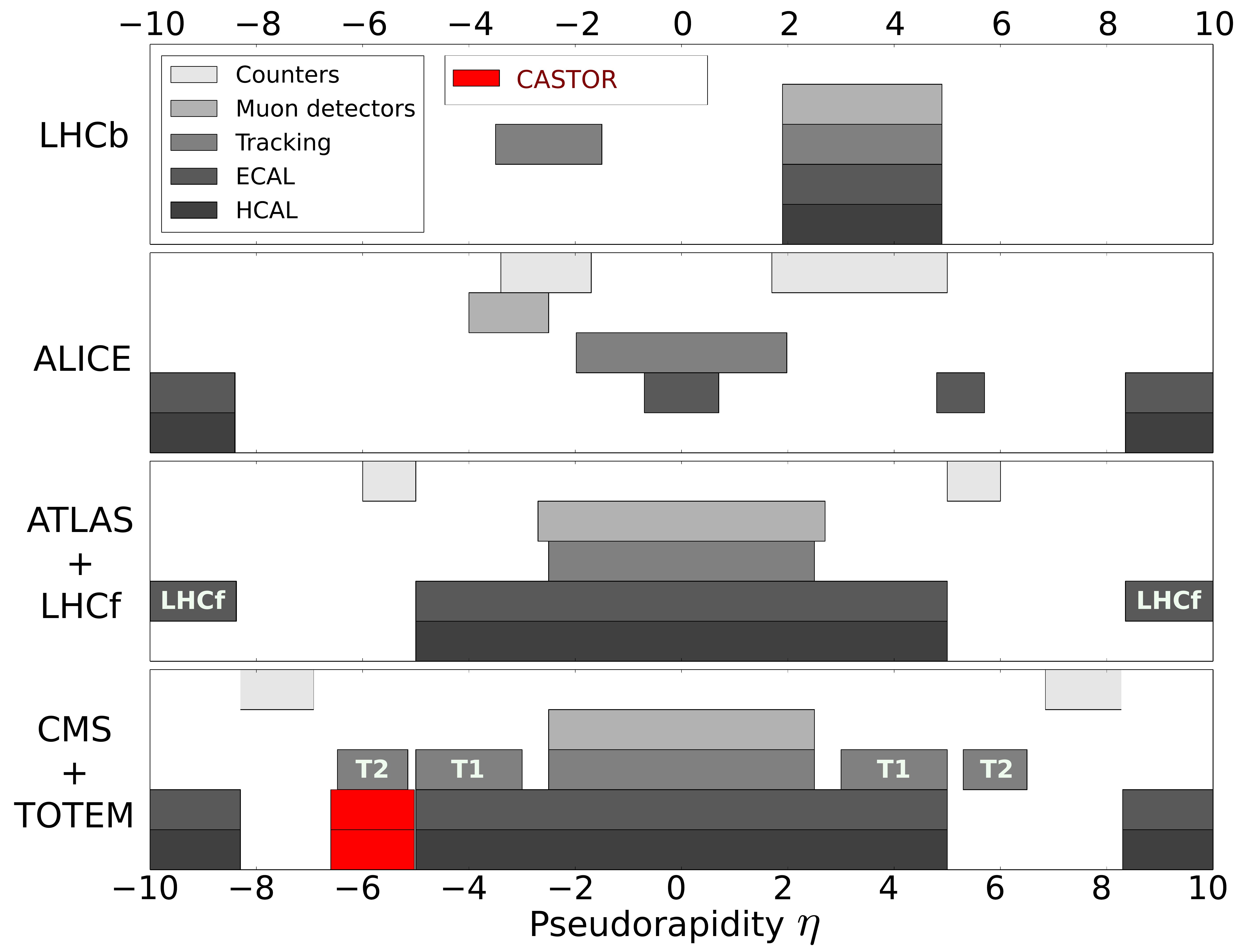}
 \includegraphics[width=.38\textwidth]{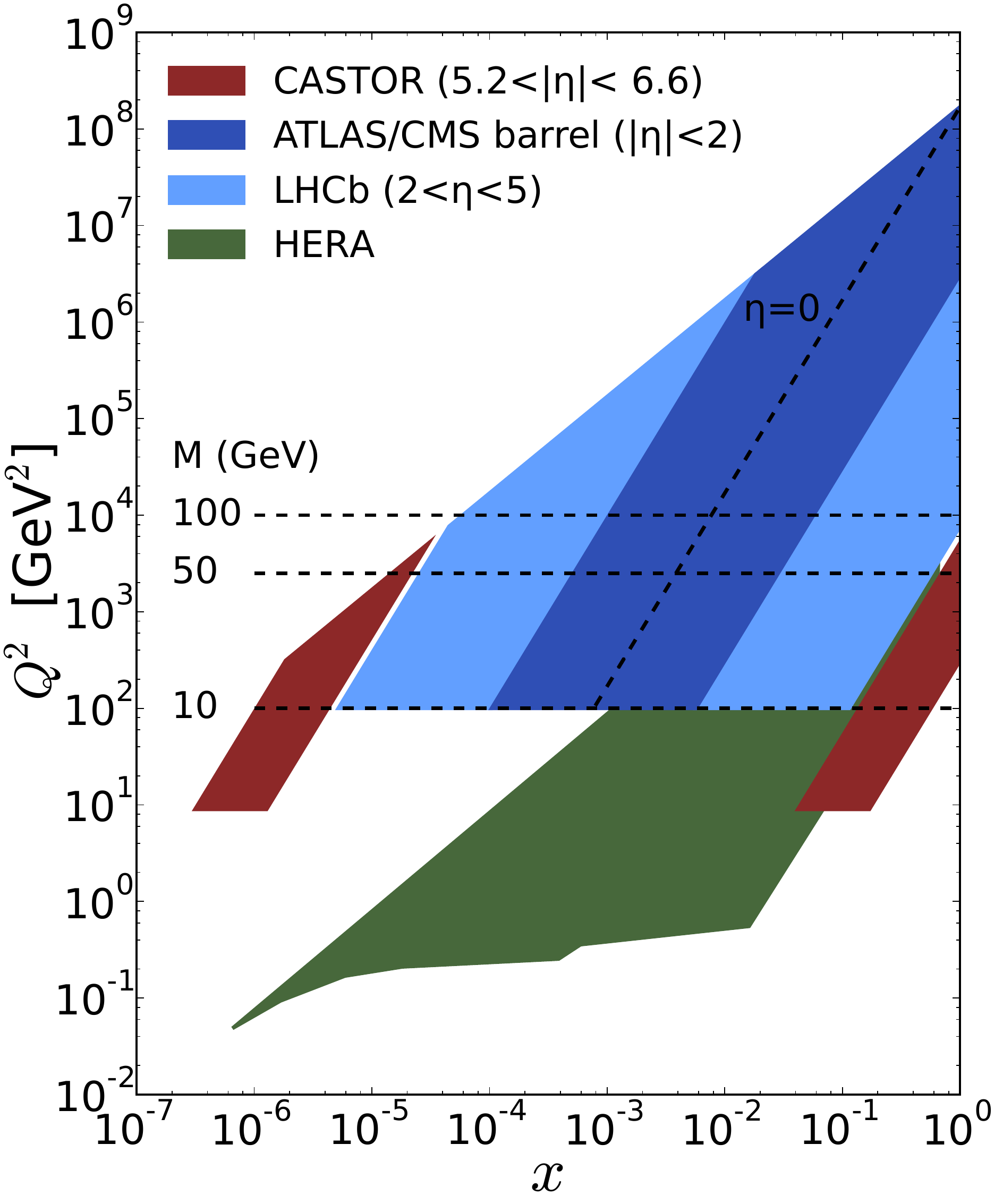}
  \caption{Left:
  Comparison of the pseudorapidity acceptances of all the LHC
  experiments~\cite{Aad:2008zzm,Aamodt:2008zz,Adriani:2008zz,Alves:2008zz,Anelli:2008zza}. 
  The ATLAS ALFA and CMS TOTEM proton spectrometers (``Roman pots'')
  installed inside the beam tunnel at $\approx$200\unit{m} around IPs 1
  and 5 are not plotted. The TOTEM forward tracking T1 and T2
  telescopes~\cite{Anelli:2008zza} are individually identified.
  Right: Kinematic coverage in the parton fractional momentum $x$ 
  and momentum transfer $Q^2$ plane corresponding to the
  CASTOR detector, the central
  CMS/ATLAS~\cite{Aad:2008zzm,Aamodt:2008zz}, the
  LHCb~\cite{Alves:2008zz}, and the DESY HERA~\cite{zeus,h1} experiments.
\label{fig:acceptance}} 
\end{figure}

The CASTOR detector closes to a large extent the gap in the
calorimetric coverage between the central detector
($\abs{\eta}\lesssim5.2$) and the zero-degree calorimeter ($\abs{\eta}>8.4$
for neutral particles)~\cite{Grachov:2008qg} on one side of the CMS
experiment.  The addition of the CASTOR calorimeter provides the CMS
detector with the most complete geometric coverage a the LHC
(Fig.~\ref{fig:acceptance}, left), thereby opening up unique
possibilities for physics measurements in {\Pp}{\Pp}, {\Pp}Pb, and PbPb
collisions. This section summarizes the main research topics
accessible with the CASTOR detector.

\subsection{Forward physics in proton-proton collisions}

The physics program of the CASTOR calorimeter includes typical
``forward physics''
studies~\cite{Albrow:2006xt,Deile:2010mv,N.Cartiglia:2015gve}, such as
those connected with low-$x$ parton dynamics, underlying event (UE),
multiparton interactions (MPI), as well as with diffractive,
photon-induced, and central exclusive processes.  Key measurements in
the investigation of these different topics are very forward energy
densities, single-particle (muon, electron) spectra, very forward
jets, as well as soft- and hard-diffractive and exclusive processes
with a large rapidity gap. 
Measurements of diffractive, photon-induced, and central exclusive
production benefit from the possibility to tag events with large gaps
devoid of particle production that are typical of processes with
color-singlet (pomeron and/or photon) exchanges, over a broad range of
rapidities in the forward acceptance.

By measuring the very forward energy density deposited in the CASTOR
acceptance, the characteristics of the UE can be studied in a rapidity
region far away from the central hard scattering processes.  The
impact of CASTOR has been particularly important for underlying event
studies~\cite{Chatrchyan:2013gfi,Sirunyan:2018jwn,Sirunyan:2019rqy},
and has helped to constrain and tune models used to describe MPI, 
which are responsible of a large fraction of the UE activity,
in Monte Carlo event generators~\cite{Sirunyan:2019dfx}.

For the study of processes characterized by rapidity gaps, CASTOR
further extends the accessible range in terms of $\xi$, the fractional
momentum loss of the proton, from about $\xi>10^{-6}$ for the
central part of the CMS experiment alone, down to
$\xi>10^{-7}$~\cite{Sirunyan:2018nqx}.  The low noise level of the
CASTOR calorimeter, equivalent to a few hundred \MeV of energy,
allows the reduction in the rate of misidentified rapidity
gaps, and improves the rapidity gap tagging efficiency. The use of the
CASTOR detector not only extends the kinematic range in which
diffraction can be observed, but also helps to disentangle single- and
double-diffractive dissociation processes~\cite{Khachatryan:2015gka}.

Measurements of very forward jets in the CASTOR
acceptance~\cite{SunarCerci:2017lwc,Sirunyan:2018ffo} have opened up
the possibility to study parton dynamics in a region of very small
parton fractional momenta, $x\approx\pt\exp{(\pm\eta)}/\sqrts\approx
10^{-6}$ (Fig.~\ref{fig:acceptance}, right), which has never been
accessible before.  In this low-$x$ regime, where the
standard Dokshitzer--Gribov--Lipatov--Altarelli--Parisi (DGLAP) parton
evolution equations~\cite{Gribov:1972ri,Altarelli:1977zs,Dokshitzer:1977sg}
are expected to fail, alternative evolutions  described by the
Balitsky--Fadeev--Kuraev--Lipatov (BFKL)~\cite{Kuraev:1977fs,Kuraev:1976ge,Balitsky:1978ic,Lipatov:1985uk}
or gluon saturation~\cite{Iancu:2002xk} 
dynamics should become important. Forward (di)jets, as
proxies of the underlying low-$x$ parton
scatterings~\cite{Cerci:2008xv,dEnterria:2009vln}, have long been
identified as useful probes of beyond-DGLAP phenomena.  When one jet
is measured in CASTOR and the other in the central CMS region, unique
dijet rapidity separations of up to $\Delta\eta\approx10$ can be
reached. Such Mueller--Navelet dijet topologies~\cite{Mueller:1986ey}
are sensitive probes of the BFKL parton
dynamics~\cite{Khachatryan:2016udy}.

\subsection{Ultrahigh-energy cosmic ray air showers} 

Improvements in our understanding of particle production in collisions
of ultrahigh-energy cosmic rays (mostly protons, with energies in the
range $\ECR\approx10^{17}$--$10^{20}\unit{eV}$) with air nuclei in the
upper atmosphere, are among the top motivations of the CASTOR research
program.  Such CRs lead to collisions at nucleon-nucleon
center-of-mass energies of $\sqrtsNN\approx14$--450\TeV, which are
around or well above those reachable at the LHC. The subsequent shower
of secondary particles generated in the atmosphere, called extensive
air shower (EAS), needs to be well-reproduced by the hadronic Monte
Carlo (MC) models to determine the nature and energy of the incoming
cosmic particle. The description of semihard and diffractive
processes, which dominate multiparticle production in hadronic
collisions within the EAS, is based on phenomenological approaches
whose parameters are tuned to particle accelerator
data~\cite{Ulrich:2010rg,dEnterria:2011twh}.  Since in hadronic
collisions most of the primary energy is transported along the forward
direction, collider data at the highest accessible energies, and in
particular in the poorly known forward region, are needed for tuning
and refinement of the MC models used in CR
physics~\cite{Ulrich:2010rg,dEnterria:2011twh}.  The importance of
CASTOR to cosmic ray research is driven by its very forward location
that allows detailed studies of particle production in a region not
covered so far by collider experiments. The highest nucleon-nucleon
center-of-mass energy of 13\TeV reached at the LHC corresponds to that
of a fixed-target collision of a primary cosmic ray proton of
$\ECR=10^{16.9}\unit{eV}$ lab energy against a nucleus at rest in the
upper atmosphere.

The highest energy particles of an EAS drive the peak of the shower
activity deeper into the atmosphere and determine the most important
features of EAS, such as the position of its shower maximum, used to identify
the CR energy and identity. In this context, the CASTOR measurement of
the forward-directed energy flow up to 13\TeV center-of-mass
energies~\cite{Chatrchyan:2011wm,Haevermaet:2016ytn} provides a
powerful benchmark for air-shower modeling. In addition, the muon
component in air showers remains poorly described by the current MC
models~\cite{Dembinski:2019uta}, clearly pointing to an insufficient
understanding of the underlying hadronic processes in the core of the
shower.  The fraction of collision energy going into neutral pions
does not contribute to the production of muons in the downstream air
shower processes and, thus, by independently measuring hadronic
(mostly charged pions) and electromagnetic (mostly neutral pions)
energy densities in the CASTOR detector, the production mechanisms
that influence muon production in air showers can be directly
studied~\cite{Sirunyan:2017nsj}.  The understanding of the development
of extensive air showers also depends on nuclear effects, since EAS
develop via hadronic collisions with nitrogen and oxygen nuclei in the
atmosphere. Thus, measurements in proton-nucleus collisions are also
of high interest (in particular with lighter nuclei in the near
future~\cite{Citron:2018lsq}), as discussed next.

\subsection{Proton-nucleus and nucleus-nucleus collisions} 

{\tolerance=800
The menu of heavy ion physics studies in the very forward region
is very rich and mostly theoretically and experimentally
unexplored~\cite{D'Enterria:2007xr}.  The original motivation for the
CASTOR calorimeter construction was the search for new phenomena in HI
collisions~\cite{GladyszDziadus:2006yu,Angelis:1997vx}, such as
strangelets~\cite{Farhi:1984qu} and disoriented chiral condensates
(DCCs)~\cite{Anselm:1989pk}. The exotic Centauro events observed in
CR collisions in the upper
atmosphere~\cite{Arisawa:1994eq,Kempa:2012zz} have been interpreted in
terms of deconfined quark matter in the forward fragmentation
region~\cite{Bjorken:1979xv}, where the net baryon number
(baryochemical potential) is very high due to nuclear
transparency. Strangelets require a similar environment for
strangeness distillation. The DCCs are theoretical states of low
\pt and can therefore be best found in the forward
region. The CASTOR calorimeter occupies the peak of the net
baryon distribution in HI collisions and is thus well suited for all these
searches. To identify irregular longitudinal shower developments in
the calorimeter, CASTOR is equipped with a 14-fold longitudinal
segmentation.\par}

Besides its potential for the discovery of exotic phenomena, CASTOR is
particularly well suited to distinguish hadronic, photon-nucleus,
and purely electromagnetic (photon-photon) processes in HI
collisions~\cite{Baltz:2007kq}. Because of the
large charge of the colliding nuclei, photons are directly involved as
an exchange particle in a significant fraction of the ion-ion
interactions.  The absence of activity in CASTOR leading to large
rapidity gaps, spanning over an increasingly larger fraction of the
central detectors, can be combined with signals from the zero-degree
calorimeters, to separate hadronic and electromagnetic
interactions. This is a powerful tool in particular
to study photon-pomeron scattering, as for
example, in exclusive vector meson production that is sensitive to the
nuclear parton distribution functions~\cite{Sirunyan:2019nog}.
As aforementioned, the distribution of energy emitted in the forward
direction in {\Pp}Pb and PbPb collisions provides constraints on CR
physics models~\cite{dEdEtaPbPb,Sirunyan:2018nqr}.  The energy
transport in the forward phase space~\cite{Wohrmann:2013nta}, as well
as forward jets in {\Pp}Pb collisions~\cite{Sirunyan:2018ffo}, can be used
to study saturation effects in the nuclear parton densities. Last but not least, the
fine segmentation of the calorimeter allows the measurement of elliptic
flow in HI collisions, a signal of final-state parton
collective effects, far away in rapidity from the hard
scattering~\cite{D'Enterria:2007xr}.

\section{Detector design}
\label{sec:design}

\begin{figure}[tbp!]
 \centering
 \includegraphics[width=.7\textwidth]{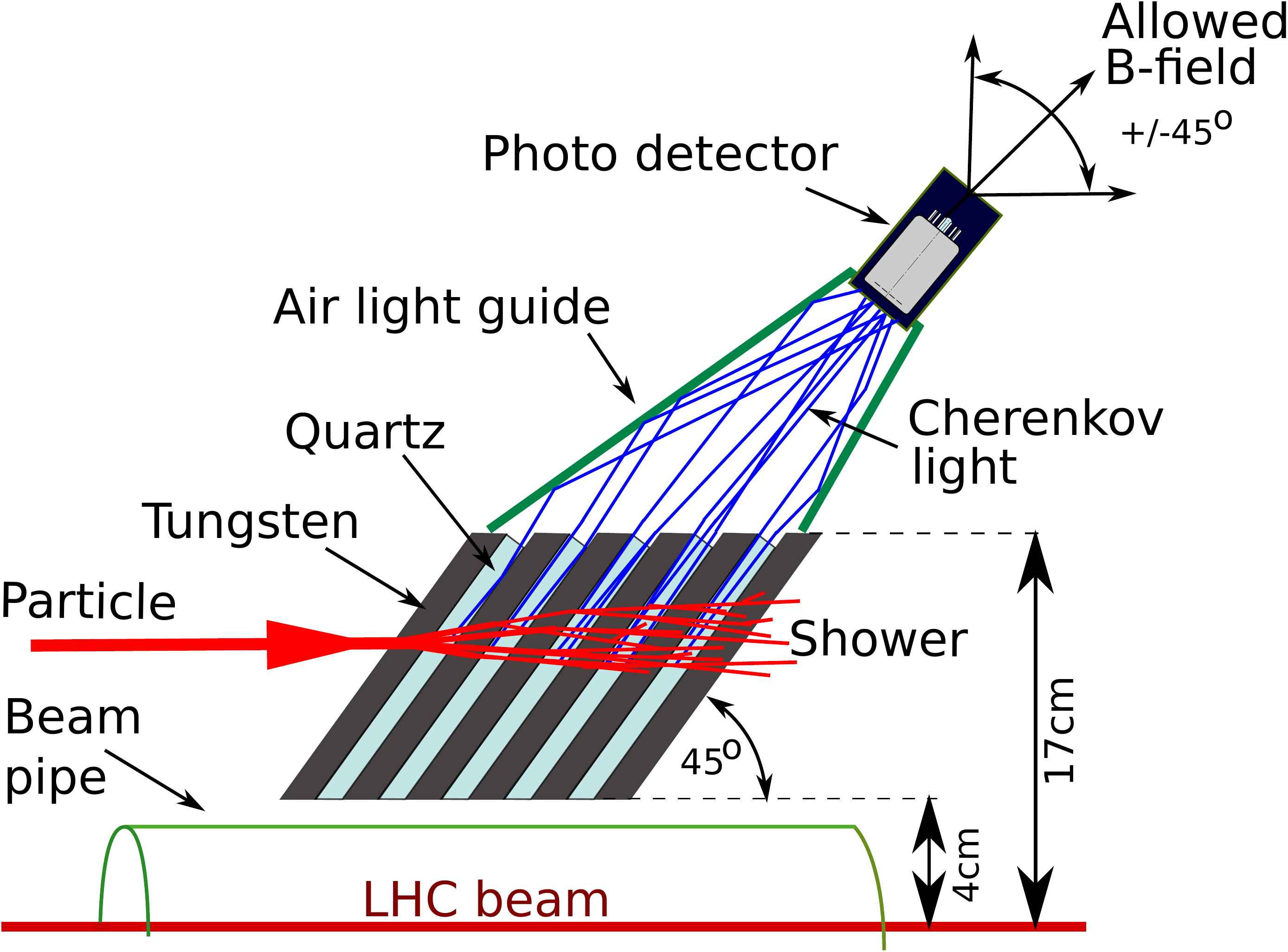}
 \caption{Schematics of one CASTOR readout module, illustrating
  the signal production mechanism. 
  Each module consists of five quartz and five tungsten absorber plates, an
  air-core light guide, and a PMT photosensor. Incident particles
  develop a shower in the tungsten absorber, and Cherenkov light is
  produced in the interleaved quartz plates. Total internal reflection
  transports very efficiently the Cherenkov photons through the quartz
  and the air light-guide to the PMT. One tower consists of 14 such
  modules. Note that six tungsten plates are visible here since the air
  light-guides are attached 
  on top of the tungsten plates between neighboring modules. \label{fig:castor_func}}
\end{figure}

{\tolerance=800
The CASTOR detector is a Cherenkov sampling calorimeter constructed from two
half-cylinders with an outer radius of 40\unit{cm} and a length of
160\unit{cm}, placed around the beam pipe at a distance of
14.4\unit{m} in the negative $z$ direction from the CMS interaction
point.  The acceptance of the calorimeter is $-6.6<\eta<-5.2$, and
$2\pi$ in azimuth $\phi$.  The whole calorimeter is made of nonmagnetic
materials, where the bulk of the mass is tungsten used as absorber,
and the active material is quartz in which Cherenkov photons are
produced.  The integration in CMS is described in
Ref.~\cite{Chatrchyan:2008aa}.  \par}

The CASTOR calorimeter has an electromagnetic section of 20 radiation
lengths ($20\mathrm{X}_0$) and a total depth of 10 interaction lengths
($10\lambda_\mathrm{I}$). The segmentation is 16-fold in $\phi$ and
14-fold in depth, resulting in 224 individual channels. The sum of
all channels in one $\phi$-segment form one calorimeter tower that is,
thus, constructed from 14 layers in depth, which are called modules.
The first two modules of a tower correspond to the electromagnetic
section, where the tungsten (quartz) plates have a thickness of 5\mm (2\mm), 
while the remaining 12 modules form the 
hadronic section, where the thickness of tungsten (quartz) plates is
10\mm (4\mm).  Thus, the material depth of a
hadronic module is twice that of the electromagnetic one.

\begin{figure}[tbp!]
  \centering
  \includegraphics[width=.6\textwidth]{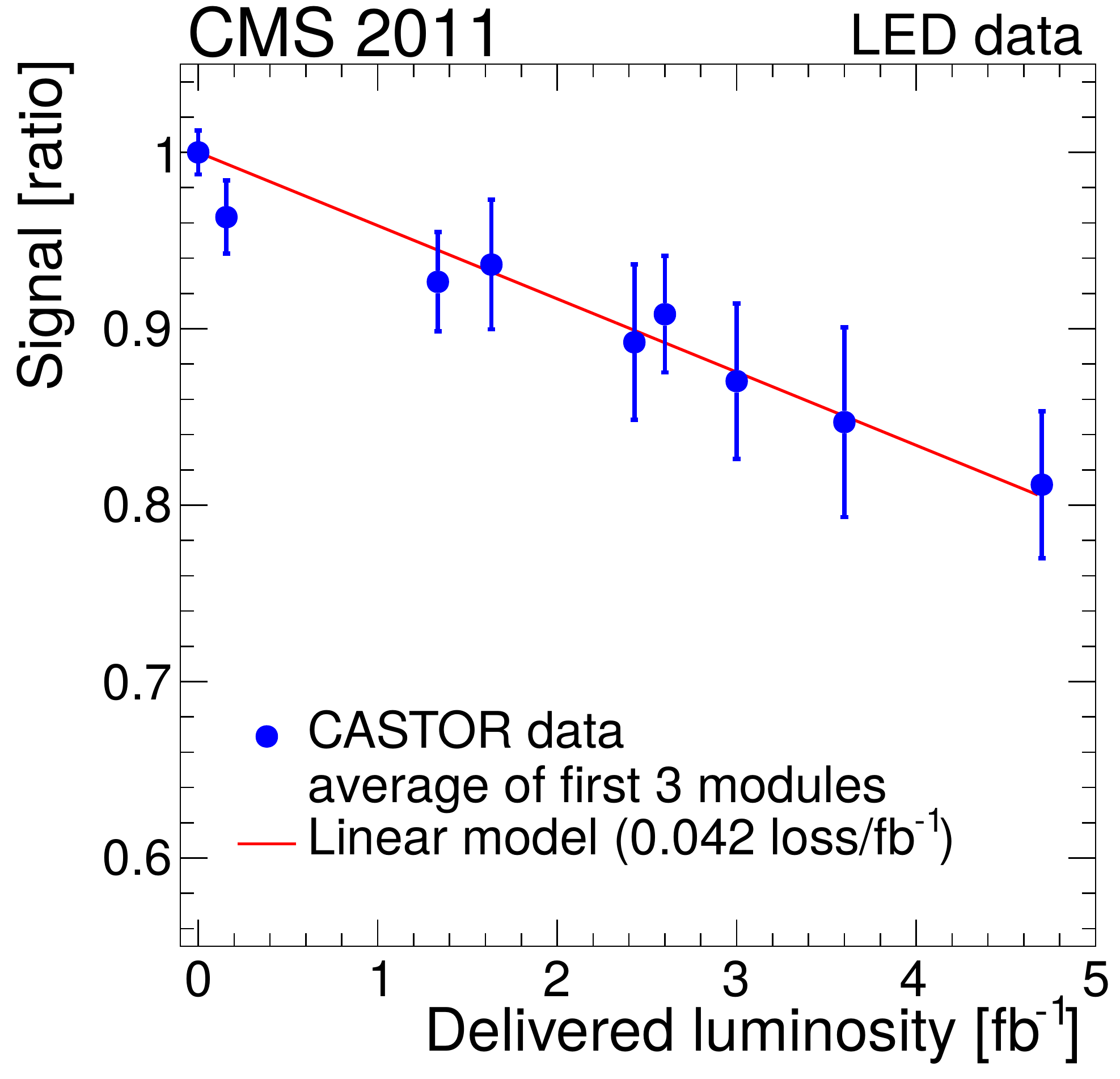}
  \caption{Observed 
  loss of sensitivity of the CASTOR photocathodes, attributed to the
  aging of the borosilicate PMT windows, during the 2011
  high-intensity proton-proton LHC operation. The relative loss of the
  signal (normalized to its original strength) is plotted as a
  function of integrated luminosity.
  The uncertainties indicate the RMS of the respective 48 PMTs involved in
  the measurement (3 modules times 16 towers). \label{fig:aging}}
\end{figure}

The schematics
of one readout module is depicted in Fig.~\ref{fig:castor_func}.
The signal is generated in radiation-hard quartz plates in which
Cherenkov photons are produced.  Interleaved with the quartz plates
are tungsten plates as absorber material to generate and contain the
electromagnetic and hadronic particle showers. The quartz plates are
tilted at 45$^\circ$ to coincide with the Cherenkov light emission
angle in quartz, and to obtain the best photon transport efficiency.  The
photosensors are mechanically coupled to the body of the calorimeter,
and photons are transported from the quartz to the photocathodes via
air-core light guides. Five tungsten and five quartz plates are grouped
together and read out by a single PMT to form one readout module.  The
inner surface of the light guides is covered with Dupont polyester film
reflector coated with AlO and a reflection enhancing
SiO$_{2}$+TiO$_{2}$ dielectric layer stack, which is both very hard
against exposure to radiation and an extremely good specular reflector
of ultraviolet light~\cite{Bayatian:2006jz}.  The reflectivity of the
foil has a low-wavelength cutoff of 400\unit{nm} that suppresses the
wavelength region where the Cherenkov yield from quartz becomes
significantly dependent on radiation exposure.  

Because of severe space restrictions, as well as stringent
demands on radiation hardness and magnetic
field resistance, the photosensors must meet special
requirements. Photomultiplier tubes with fine-mesh dynodes were
selected after extensive test-beam
studies~\cite{Aslanoglou:2007fx,Aslanoglou:2007wv,Andreev:2010zzb}.
The original set of Hamamatsu R5505~\cite{Appuhn:1997sv} photosensors
was provided from the decommissioned H1
calorimeter~SpaCal~\cite{Appuhn:1996na} at the DESY HERA collider. However,
their photocathode window, made of borosilicate glass, was observed to
degrade noticeably under exposure to intense radiation during the higher luminosity LHC
data-taking period of 2011, as shown in Fig.~\ref{fig:aging}. To
prepare for the LHC Run 2 and data taking at 13\TeV, these original
PMTs were replaced by R7494 PMTs from Hamamatsu during 2012.  These new
PMTs have a fused silica entrance window, which is very
radiation tolerant. The PMTs are driven by a passive base. The
last dynode with the largest current is powered by a dedicated low-voltage
power supply of typically 100\unit{V}.  The rest of the
dynodes and the cathode are powered by a high-voltage (HV) supply in the
range 800 to 1800\unit{V}, depending on the physics measurement. The
separation of the last dynode from the HV supply allows the
safe powering of four PMTs by a single HV channel. The
response of all new and exchanged PMTs has been characterized in a
dedicated external calibration setup as a function of the HV
for several light-emitting diode (LED) wavelengths.

\begin{figure}[tpb!]
 \centering
 \includegraphics[width=0.49\textwidth]{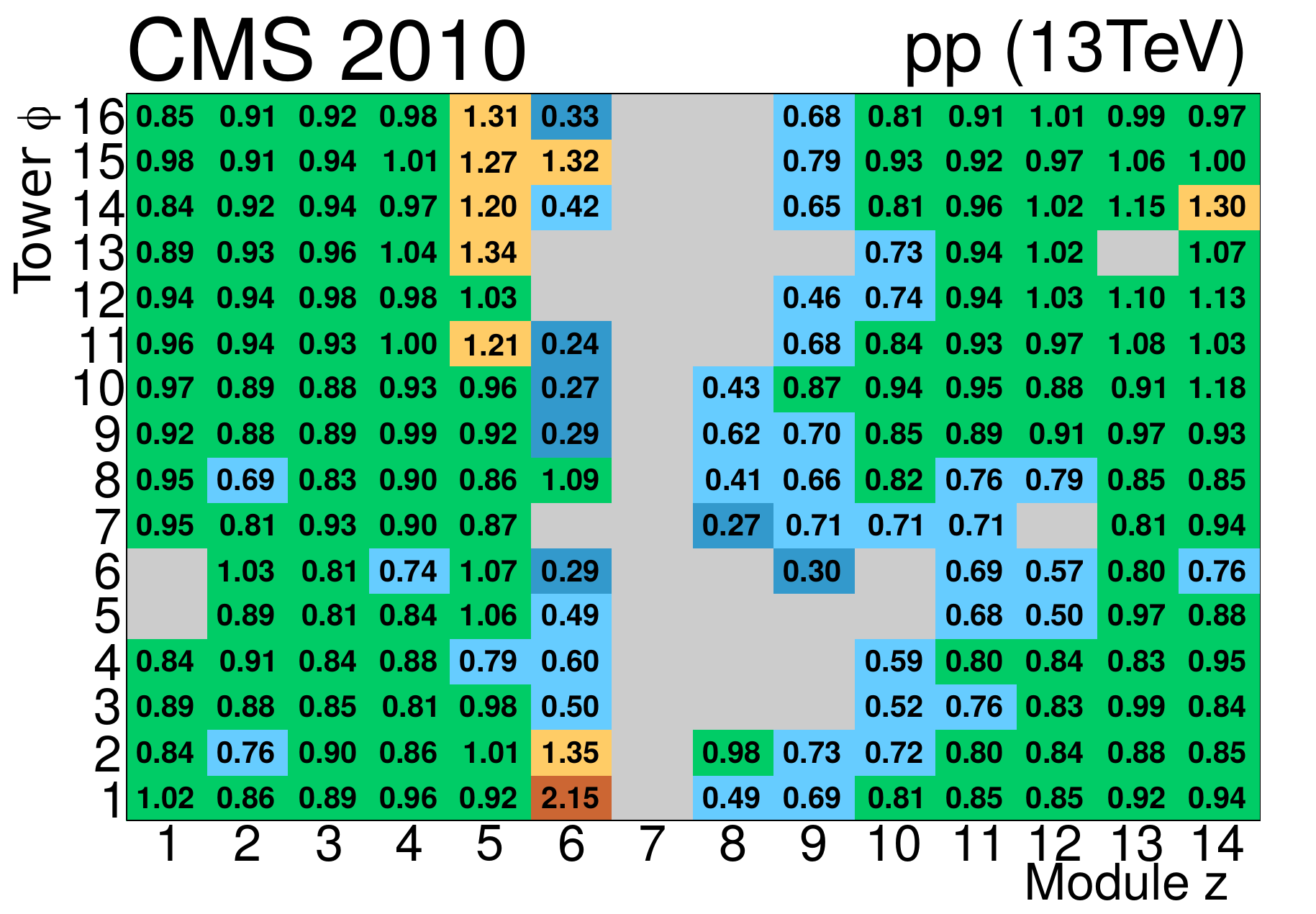}~
 \includegraphics[width=0.49\textwidth]{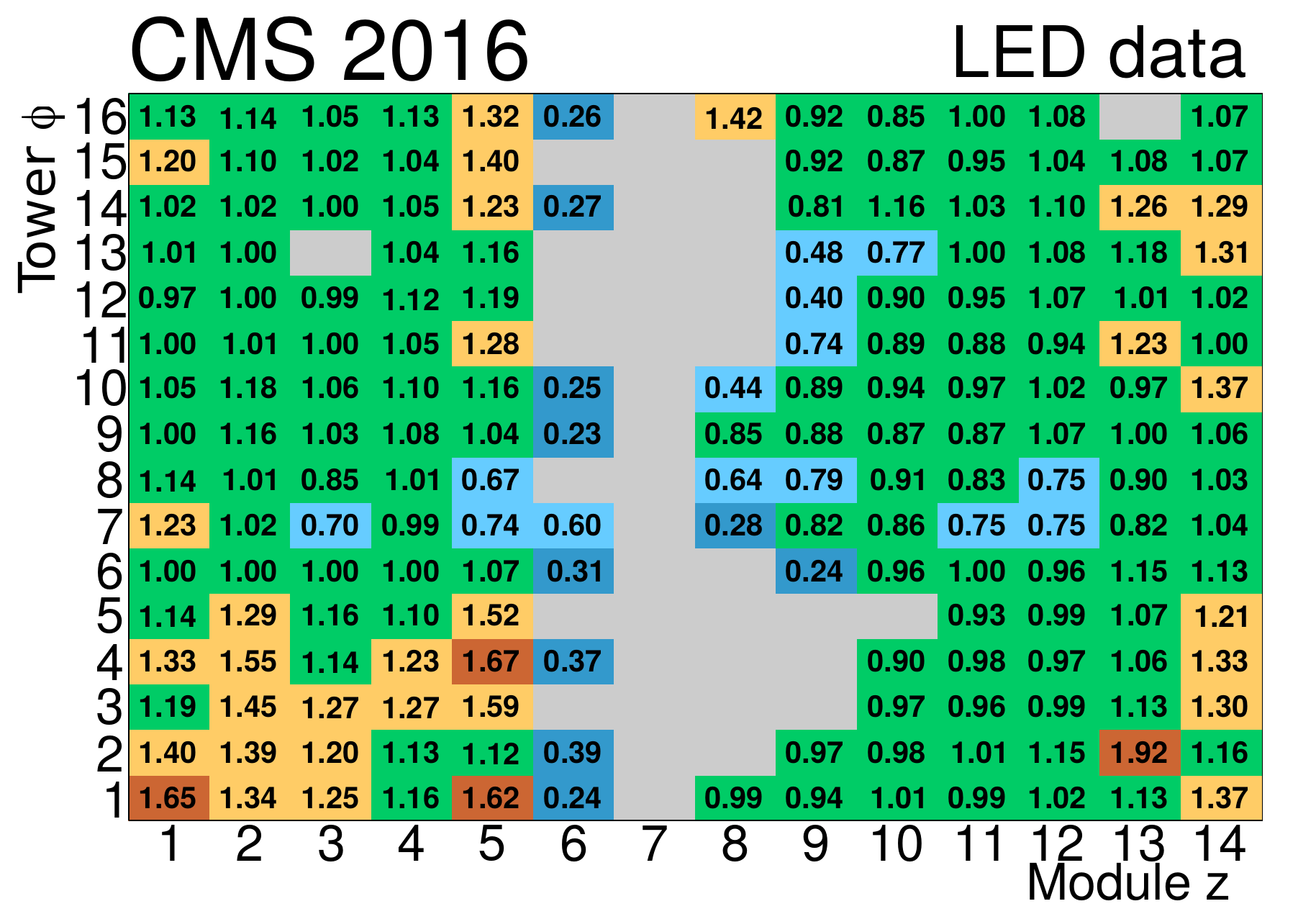}
 \caption{Impact of the CMS magnet on the signals observed in the
  CASTOR calorimeter in the $(z,\phi)$ plane for 2010 {\Pp}{\Pp} minimum-bias
  (\cmsLeft) and 2016 LED pulser (\cmsRight) data. The numerical values at
  each bin correspond to the ratio of signals measured at 3.8\unit{T}
  over those at 0\unit{T}.  The channels in gray have normalized
  signals below 0.2, or are excluded for technical reasons.
 \label{fig:magnet}}
\end{figure}

An in-situ monitoring system is available for the evaluation of the
performance of all channels of the detector at all times during data
taking. This is done using an LED pulser system that can illuminate
the entrance window of the PMTs when CASTOR is installed inside
CMS. The light pulse of wavelength 470\unit{nm} and pulse duration
20\unit{ns} is generated by a dedicated pulser in the electronics
crate and is transported through an optical fiber to the calorimeter,
where it is distributed via optical splitters to all 224 channels.
The fibers are positioned so that they 
illuminate roughly the entire photocathode area. The LED signals are used to
study the response of the electronics while there are no collisions at
the IP. In this way, dead channels are identified, and the PMT gain is
measured. Gain correction factors are obtained for various high-voltage
operation modes, as well as magnetic field environment
situations. Since the LED signals are neither absolutely calibrated, nor
are identical during different installation periods, they are mainly
useful for characterizing the time-dependence and stability of single
channels. However, assuming small PMT gain variances, the gain $g$ can
be estimated approximately from the signal $S$ and variance
$\sigma^2_{S}$, according to $g\approx\sigma^2_{S}/S$
(Section~\ref{section:gaincorrectionfactors}).

The PMTs are read out by fast charge integrating circuits
(QIEs)~\cite{qie} in time samples of 25\unit{ns} duration. Before
digitization, the analog signals are stored in a cyclic buffer
consisting of four capacitors.  This is also relevant for the
calibration and event reconstruction, since each capacitor has
slightly different properties. One recorded event consists of ten
(before 2012) or six (after 2012) time samples. The pulse shape of
CASTOR is distributed over more than one time slice, which is
essentially a consequence of the dispersion in the long cables from
the PMTs to the digitizers. For this reason, CASTOR is not optimized
to operate at LHC bunch spacings smaller than 50\unit{ns} (two time
slices). CASTOR only recorded data with bunch spacings of 50\unit{ns},
with typically separations being much larger. Typical running conditions, 
\eg, during HI runs, have much larger bunch spacings. However, the finite
pulse width is exploited during the reconstruction to retrieve the
signal size for events where the signal in one of the time slices is
saturated, as explained below.

While the Hamamatsu fine-mesh PMTs are very tolerant to ambient
magnetic fields of the magnitude observed in the vicinity of CASTOR,
there is a well known dependence of the PMT performance on the
relative field direction with respect to the PMT axis. Unfortunately,
because of the massive shielding around the calorimeter, the direction of
the magnetic field varies considerably.  This has, for some channels, a
very strong and complex impact on the PMTs, as shown in
Fig.~\ref{fig:magnet}. The left panel illustrates the impact of the
field on the PMT response measured with {\Pp}{\Pp} data in 2010, 
and the right plot is the result from LED data taken in 2016.
Some general features, as the insensitivity in the range of the
modules 6--9,
are visible in both years, but there are also important differences
clearly asking for a careful time-dependent calibration.
There were
major modifications to CMS between 2010 and 2018, \eg, 
the installation of an additional 125-ton yoke endcap shielding
disc. Details of the shieldings directly around CASTOR (rotating
and collar shields) were also changed, and finally the exact positioning
and geometry of CASTOR is slightly different in each installation
period. For data taking with the nominal CMS
magnetic field, the recordings by most channels in modules 7 and 8
are significantly less useful, and modules 6, 9 and 10 are also
compromised. Furthermore, in 2016 there appears to be a general larger
impact on the first towers (numbers 1 to 5), which are located in the
top part of the calorimeter, where displacements of the PMTs caused by the
ramp-up of the CMS magnet are the largest.  These observations
clearly show the importance of properly calibrating the CASTOR
channels during physics data taking with a stable magnetic field.

The insensitivity in the central parts of CASTOR, induced by the
magnetic field channeling through the massive radiation shielding of
CMS, was discovered after the first data taking in 2009, and could not be
mitigated without a major redesign of the shielding and/or the
detector. However, the impact on almost all typical physics
applications remains limited: the front modules 1 to 5 are sufficient
to detect $\approx$100\% of the electromagnetic showers, as well as
$\approx$75\% of the hadronic showers in the calorimeter. Furthermore,
also the tails of hadronic showers (and through-going muons)
are seen in the back modules 10 to 14.

\begin{figure}[tpb!]
\centering
\includegraphics[width=.49\textwidth]{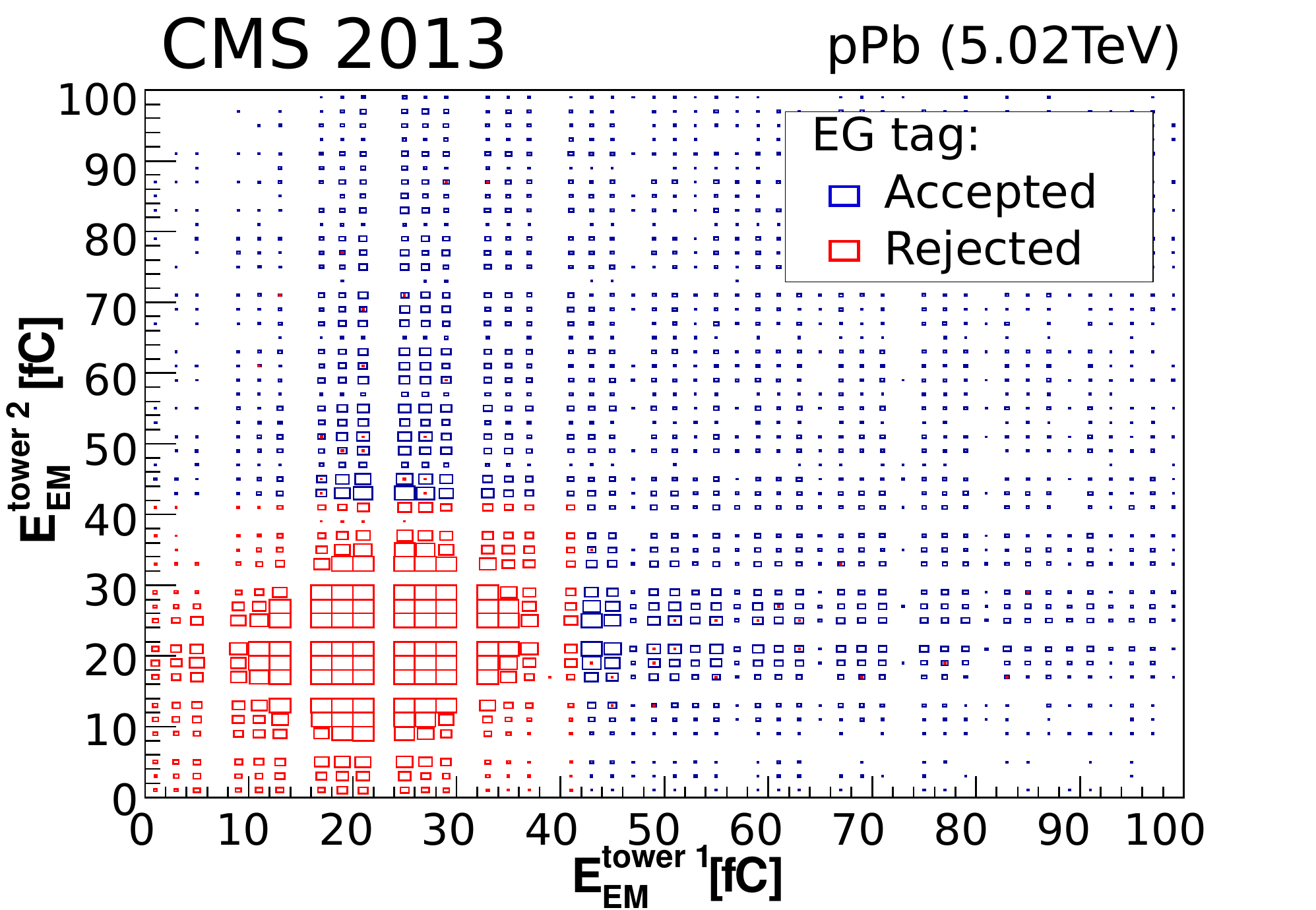}~
\includegraphics[width=.49\textwidth]{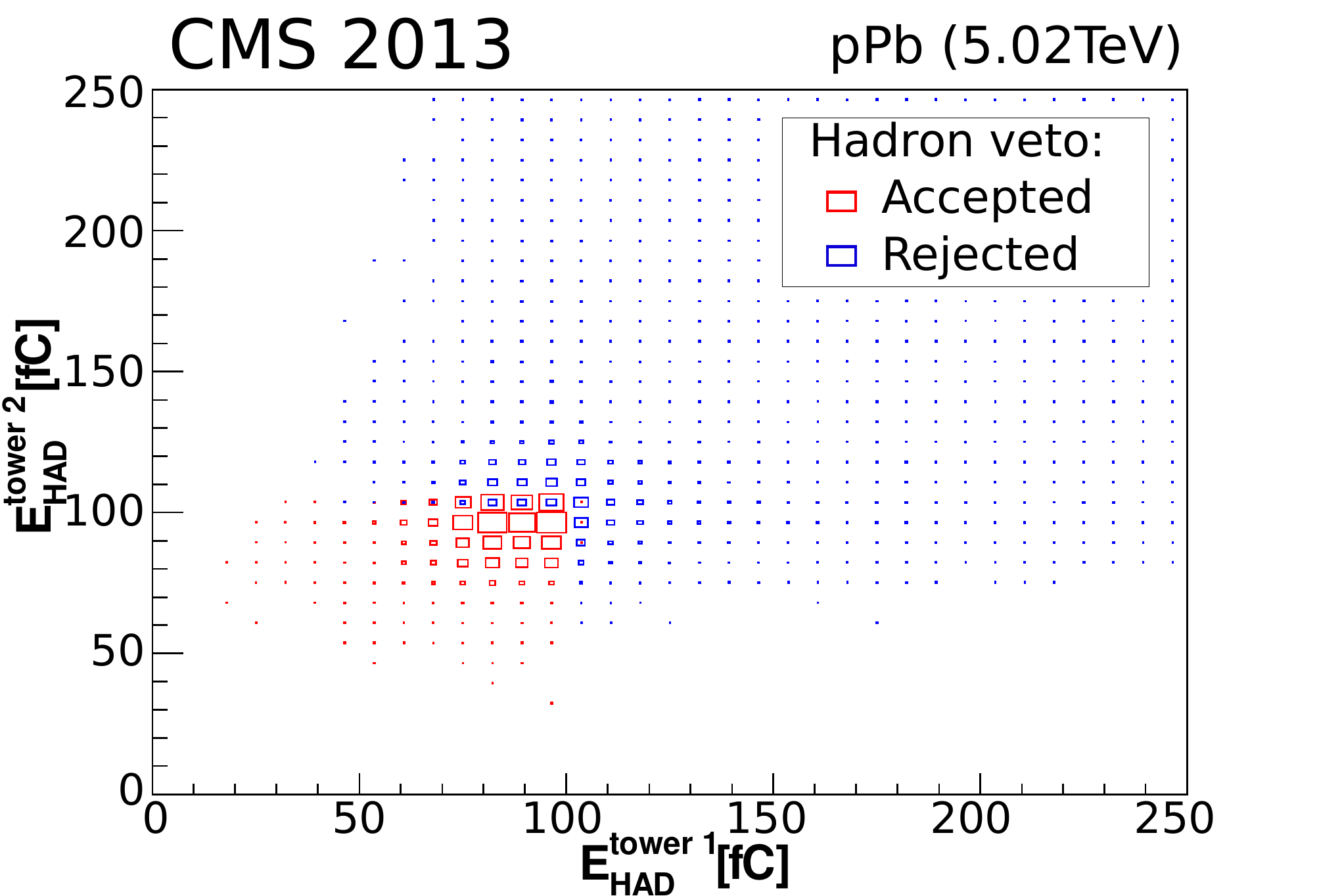}
\caption{Validation of the electron/gamma (EG) tag (\cmsLeft) and
  the hadron veto (\cmsRight) triggers 
  in {\Pp}Pb collisions at $\sqrtsNN=5.02\TeV$.  Shown are density
  distributions of the number of events with digitized signals (in \unit{fC})
  in pairs of towers in one octant.  A trigger is issued for each pair
  of towers when the logical condition ``$\text{EG
  tag}\wedge\text{Hadron veto}$'' is met.  The number of
  events that (do not) produce a trigger signal are in blue (red).
  At very small signals, one can see additional quantization
  effects from the nonlinear behavior of the QIE digitizers.
  \label{fig:trigger1}}
\end{figure}

\section{Triggers and operation}
\label{sec:operation}

\begin{table}[b!]
\centering
\topcaption{Overview of the running periods of the CASTOR detector, with 
indication of the year, colliding system, nucleon-nucleon center-of-mass energy, 
and the triggers provided to the CMS L1 global trigger system. }
\label{tab:operations}
\begin{tabular}{crll@{}m{0pt}@{}}
\hline
Year & $\sqrtsNN$ & Colliding system & CASTOR trigger(s) & \\[5pt]
\hline 
2009 & 0.9\TeV & proton-proton &  & \\
2010 & 0.9\TeV & proton-proton & halo muon & \\
	 & 2.76\TeV & proton-proton  & halo muon& \\
	 & 7\TeV & proton-proton  & halo muon& \\
	 & 2.76\TeV & lead-lead  & halo muon& \\
2011 & 7\TeV & proton-proton & halo muon & \\
	 & 2.76\TeV & lead-lead & halo muon& \\
2013 & 5.02\TeV & proton-lead  & halo muon \& e.m.\ cluster & \\
	 & 2.76\TeV & proton-proton  & halo muon \& e.m.\ cluster & \\
2015 & 13\TeV & proton-proton  & halo muon \& jet & \\
	 & 5.02\TeV & proton-proton  & halo muon \& jet& \\
	 & 5.02\TeV & lead-lead  & halo muon \& jet& \\
2016 &  5.02\TeV & proton-lead  & halo muon \& jet& \\
	& 8.16\TeV & proton-lead  & halo muon \& jet& \\
2018 & 5.02\TeV & lead-lead & halo muon \& jet & \\
\hline
\end{tabular}
\end{table}

Together with the data recorded during the various run periods, CASTOR
delivered signals to the CMS level-1 (L1) trigger system~\cite{Khachatryan:2016bia}. 
These triggers, listed in Table~\ref{tab:operations}, were able to
calibrate the detector (halo muon trigger), as well as to carry out
dedicated physics analyses (jet trigger, electromagnetic (e.m.) cluster
trigger).  Depending on the physics goals, the triggers
varied during the various running periods.  During all listed
years, CASTOR recorded a total luminosity of about 5\fbinv of data
with a large variety of colliding systems and center-of-mass
energies. These samples have been partly analysed for various physics
measurements, and further studies are still ongoing or planned in the
future with the collected data~\cite{opendata,Mccauley:2019uis}.

The CASTOR electronics chain provides four simultaneous hardware trigger
outputs using the standard CMS hadron calorimeter trigger readout (HTR)
boards. There are eight HTR boards; each analyzing the data of
one octant (two towers) of the calorimeter. Since one HTR board can
handle 24 front-end channels, the last two modules per tower are not
included in the trigger system. The output of all octants is further
processed in the ``trigger timing control'' (TTC) board to form
four trigger output signals for the whole CASTOR calorimeter. 
These are sent to the CMS L1 global trigger system. 

\begin{figure}[tpb!]
 \centering
 \includegraphics[width=.6\textwidth]{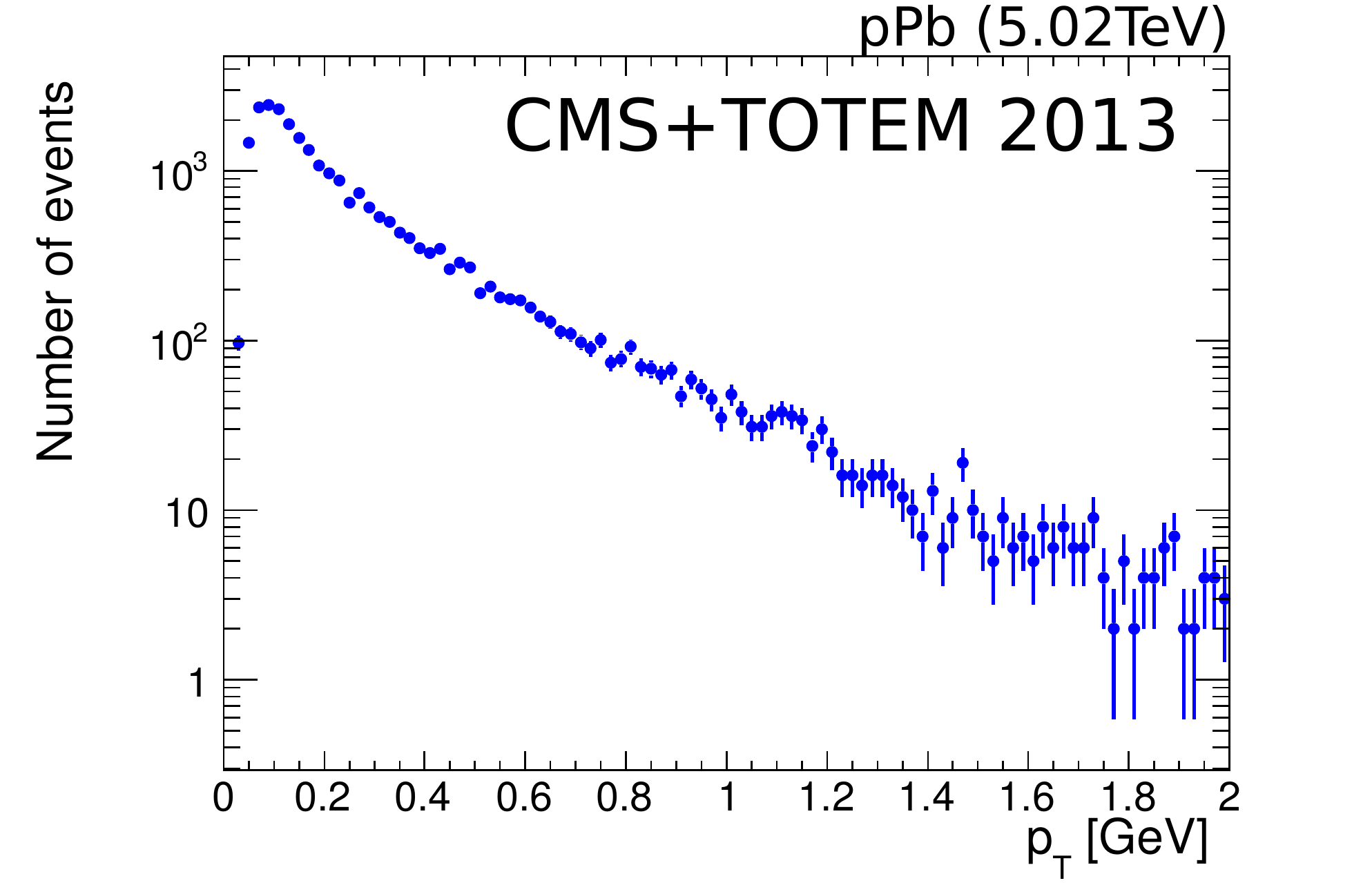}
 \caption{Detector-level \pt\ distribution of isolated electron
  candidates recorded in
  proton-lead collisions in 2013 with a dedicated trigger~\cite{CMS-DP-2014-014}. It is a
  unique feature that the \pt of these electrons can be
  precisely determined since the angles of their associated tracks are
  measured with the TOTEM T2 tracker. The uncertainties are purely statistical. \label{fig:electron_pt}}
\end{figure}

In terms of operation, the CASTOR detector was installed in CMS prior
to the first LHC beams and took data continuously during years
2009--2011. Over this period, {\Pp}{\Pp} collisions were recorded at 
$\sqrts=0.9$, 2.76, and 7\TeV.
The CASTOR calorimeter was removed at the end of 2011, before the high-intensity 
LHC operations in 2012, because data recorded at high luminosities are 
significantly less useful for physics analyses since
the calorimeter is unable to distinguish particles originating from different
pileup vertices because of its location, geometry, timing capabilities,
and granularity. The CASTOR detector was then reinstalled in 2013 for the {\Pp}Pb run
at $\sqrtsNN=5.02\TeV$, and for the {\Pp}{\Pp} run at
$\sqrts=2.76\TeV$. During LHC Run 2, CASTOR was never installed in
high-luminosity {\Pp}{\Pp} data taking to avoid needless radiation damage, but
it was installed in all HI runs and in a few
special low-luminosity {\Pp}{\Pp} runs. In 2015, after the first long shutdown of
the LHC, the detector was operated during the low-luminosity phase of
the first weeks of {\Pp}{\Pp} collisions at 13\TeV and during the PbPb run at
the end of the year. In 2016, the {\Pp}Pb runs at $\sqrtsNN=5.02$ 
and 8.16\TeV were recorded, as well as the PbPb runs at
$\sqrtsNN=5.02\TeV$ in 2018. A detailed summary of all running
periods is shown in Table~\ref{tab:operations}.

In 2013, the CASTOR calorimeter had an active trigger for isolated
electron/photon objects. The calorimeter trigger was combined during
part of the data taking with a TOTEM T2 tracker low-multiplicity
tag at the L1 trigger.  The resulting data set contains a very large
and clean sample of isolated electron candidates.  The validation of
the calorimeter trigger for each octant is shown in
Fig.~\ref{fig:trigger1}. Since the trigger is generated from the
combination of two adjacent towers of one octant, the trigger
validation corresponds to a two-dimensional analysis. The trigger
requires a single tower in one octant in which the first two
(electromagnetic) modules are above a charge threshold of 40\unit{fC}
(corresponding to $\approx$0.6\GeV energy), and applies an
additional veto on any hadronic activity in the modules at depths 4 to
6 in this octant. No more than 100\unit{fC} (corresponding to 
$\approx$1.6\GeV) of hadronic energy is allowed.

The data recorded with this trigger during the {\Pp}Pb run leads to the
\pt\ distribution of candidate electrons shown in Fig.~\ref{fig:electron_pt}. 
Events are only shown if they have an isolated electromagnetic energy
cluster in CASTOR that was associated to one isolated track in T2
during offline analysis.  Thus, the energy of the electron candidate
is measured with CASTOR and its trajectory with T2 (no acceptance
corrections are applied in the plot).  This combination of the two detectors
provides a unique case where individual charged particles can be fully
reconstructed in the very forward direction at the LHC.  It is
interesting to note that the lower threshold for the observation of
electron candidates, $\pt\approx100\MeV$, is similar to the
performance of the CMS pixel tracking detector at central
pseudorapidities~\cite{Chatrchyan:2014fea}.

\begin{figure}[tpb!]
 \centering
 \includegraphics[width=.57\textwidth]{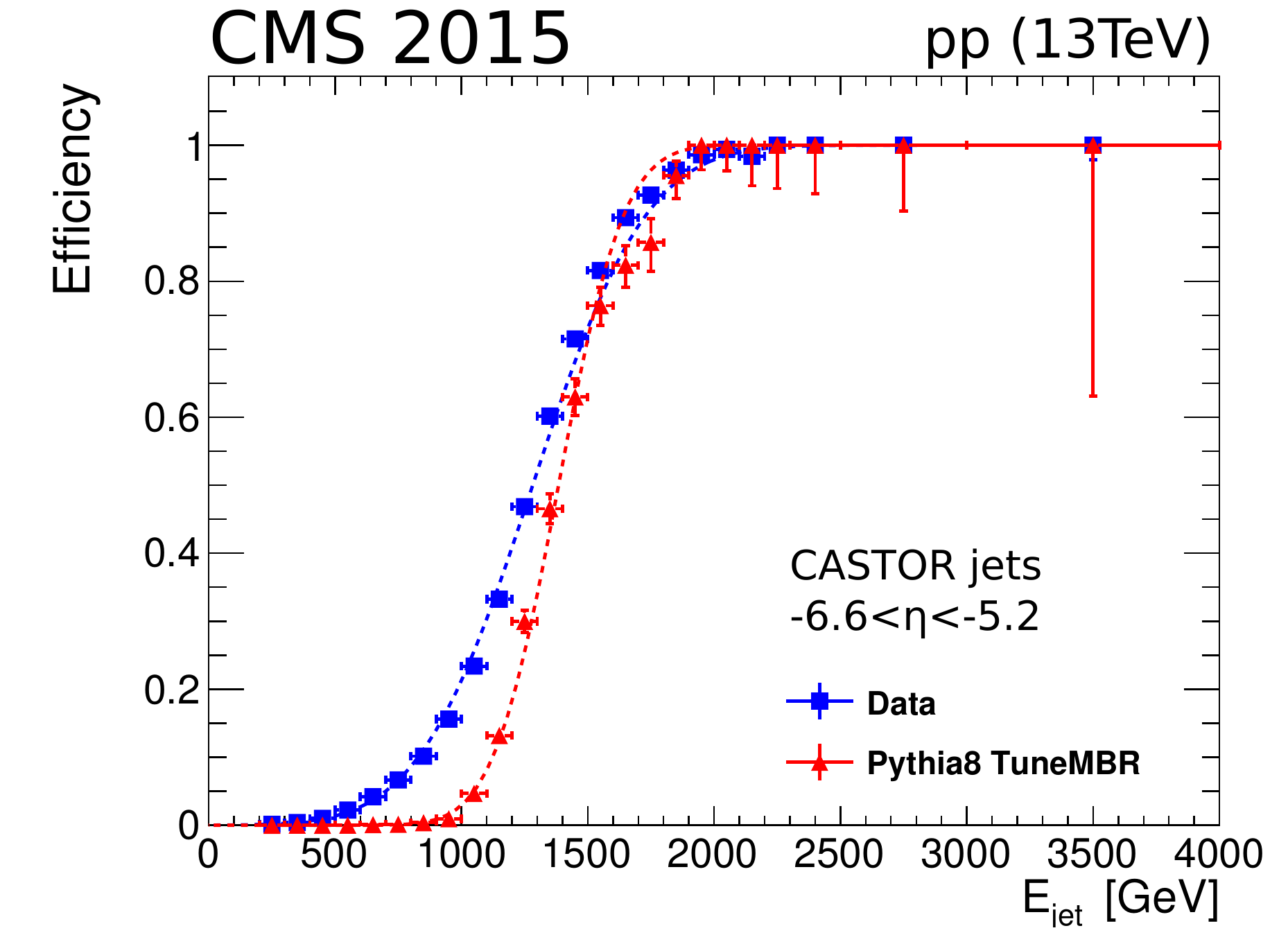}~
 \includegraphics[width=.42\textwidth]{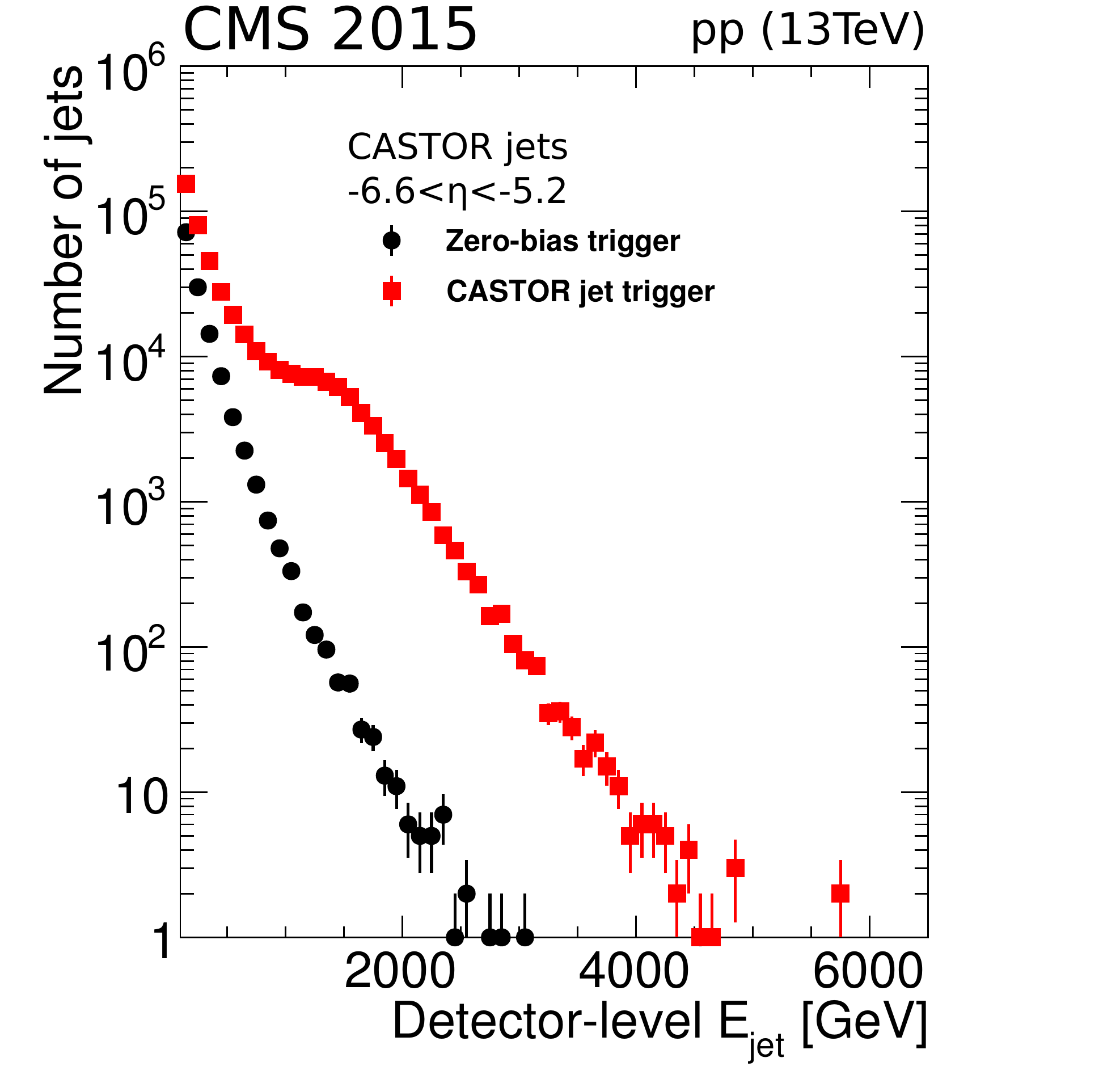}
 \caption{Performance of the CASTOR jet trigger in {\Pp}{\Pp} collisions
  during the LHC Run 2. 
  Left: Trigger efficiency turn-on curve in data and simulations
  generated with the \PYTHIA~8~\cite{Sjostrand:2007gs} with the MBR tune~\cite{Ciesielski:2012mc}.
  Right: Number of jets as a function of $\Ejet$
  collected with the CMS zero-bias and CASTOR jet triggers.
  In both figures the uncertainties are statistical only.
 \label{fig:jettrigger}}
\end{figure}

During the 2015, 2016, and 2018 runs, CASTOR provided a jet trigger
with the simple requirement of at least one tower with energy above a
given threshold.  The jet trigger has a well defined turn-on curve,
high efficiency and is sufficiently well reproduced by the detector
simulations as shown in Fig.~\ref{fig:jettrigger}~(\cmsLeft). 
The slightly slower rise of the turn-on seen below 2\TeV in the data with 
respect to the simulations is a consequence of the nominal MC reconstructed
jet resolution that is better than the actual one, since not all detector 
misalignment and miscalibration uncertainties are propagated into the simulation.
The trigger is used only in the range of $\Ejet$ above 2\TeV,
where the efficiency reaches a plateau in both data and simulations.
This assures that the efficiency corrections remain minimal and that the
predictions of the detector simulation in this region are well compatible
with the data. In Fig.~\ref{fig:jettrigger}~(\cmsRight), the impact on the
size of the event sample is illustrated by comparing the jet spectra
collected with the CMS zero-bias and CASTOR jet triggers.

\section{Event reconstruction and calibration}
\label{sec:reco}

\begin{figure}[tpb!]
 \centering
 \includegraphics[width=.42\textwidth]{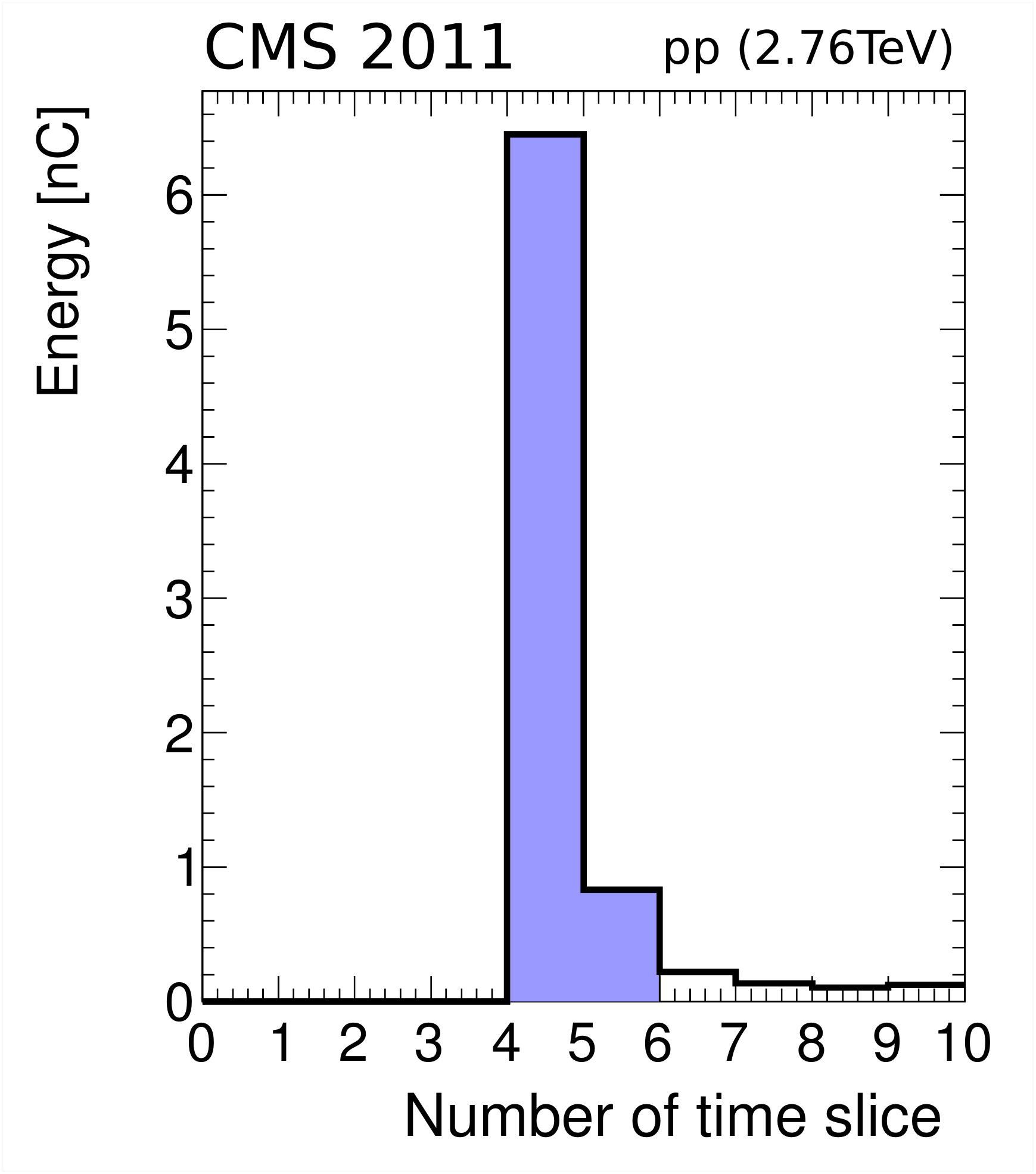}
 \includegraphics[width=.57\textwidth]{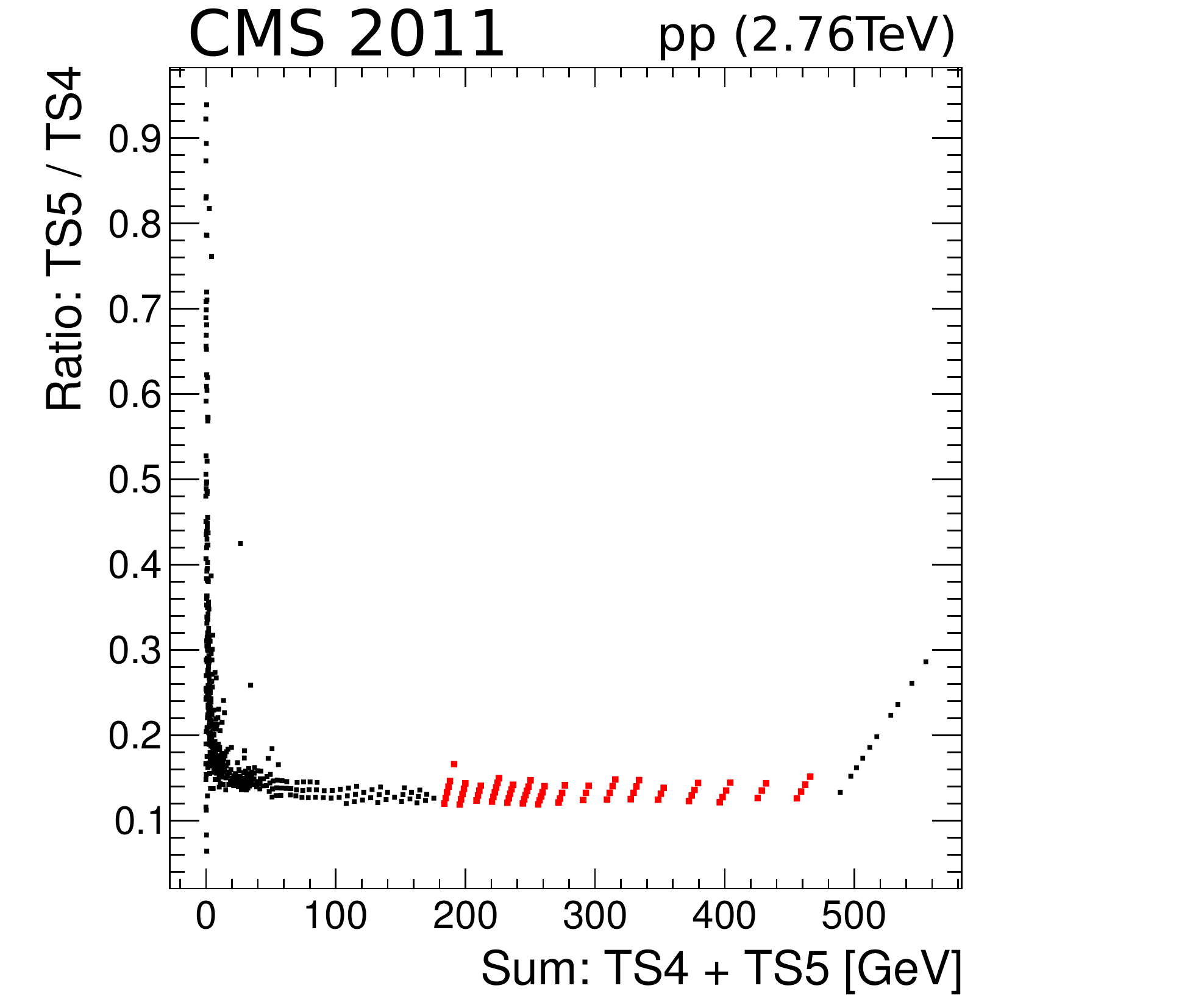}
 \caption{Left:
 Typical example of a pulse shape in CASTOR showing the energy
 (in \unit{nC}) distributed over several time slices, in data recorded
 in March 2011 using minimum-bias {\Pp}{\Pp} collisions at $\sqrts=2.76\TeV$ (the shaded
 area of the pulse is used to derive the signal energy in the event
 reconstruction).  Right: Ratio of the second-largest (TS5) to largest
 (TS4) signal time bin for one typical CASTOR channel. The small scale
 structure in this ratio of neighboring bins apparent for TS4+TS5
 larger than $\approx$100\GeV is a feature of the QIE
 digitizer. Saturation is visible at very high signals, but before
 saturation sets in, the signal ratio has a stable region (in red)
 that is used to estimate the channel signal at higher energies.
\label{fig:pulse}}
\end{figure}

The fundamental quantity measured by the CASTOR calorimeter is the
charge per channel.  Here, the procedure for obtaining a measurement
of the energy per channel from the signals digitized by the detector
is described. One signal pulse is distributed over several time slices
(TSs) of 25\unit{ns} duration, as shown in Fig.~\ref{fig:pulse}~(\cmsLeft).
The timing is adjusted so that the time sample with the
maximum energy sits at the TS number 4. The channel energy is
estimated from the sum of the energies in time samples 4 and
5. Although the fourth time sample saturates at around 400\GeV of
energy per channel, for the high-voltage settings applied during
{\Pp}{\Pp} collisions, particle energies of up to almost a factor
of 10 higher can be reconstructed from the energy in the TS number 5,
since the ratio between the energies in these time slices is constant
at high channel energies, as shown in Fig.~\ref{fig:pulse}~(\cmsRight).
Saturation is not a common problem in CASTOR, but can affect
particularly interesting events, such as those with high-energy
electrons, high-energy jets, or from central PbPb
collisions. Considering the normal HV settings during PbPb
collisions, where the PMT gain is reduced by a factor of $\approx$20
with respect to {\Pp}{\Pp}, and using the fifth time slice to
estimate the energy when
saturation occurs, signals can be measured up to a maximum of around
$400\GeV\times10\times20=80\TeV$ per single channel.

The fourteen channels belonging to one $\phi$-segment of CASTOR are
grouped into one tower during offline event reconstruction.  The
energy of a tower is the sum of its two electromagnetic and twelve
hadronic channels. Channels are nonactive only if they have been
flagged as \textit{bad channels} during detector commissioning.  The
signals from the towers are zero suppressed, to remove noise, using a
typical threshold of ($650\sqrt{\Nch})\MeV$,
where $\Nch$ is the number of active channels
in this tower. This threshold has shown to yield very consistent
results for data and simulations in many different data analysis
applications.

\begin{figure}[tpb!]
\centering
  \includegraphics[width=.6\textwidth]{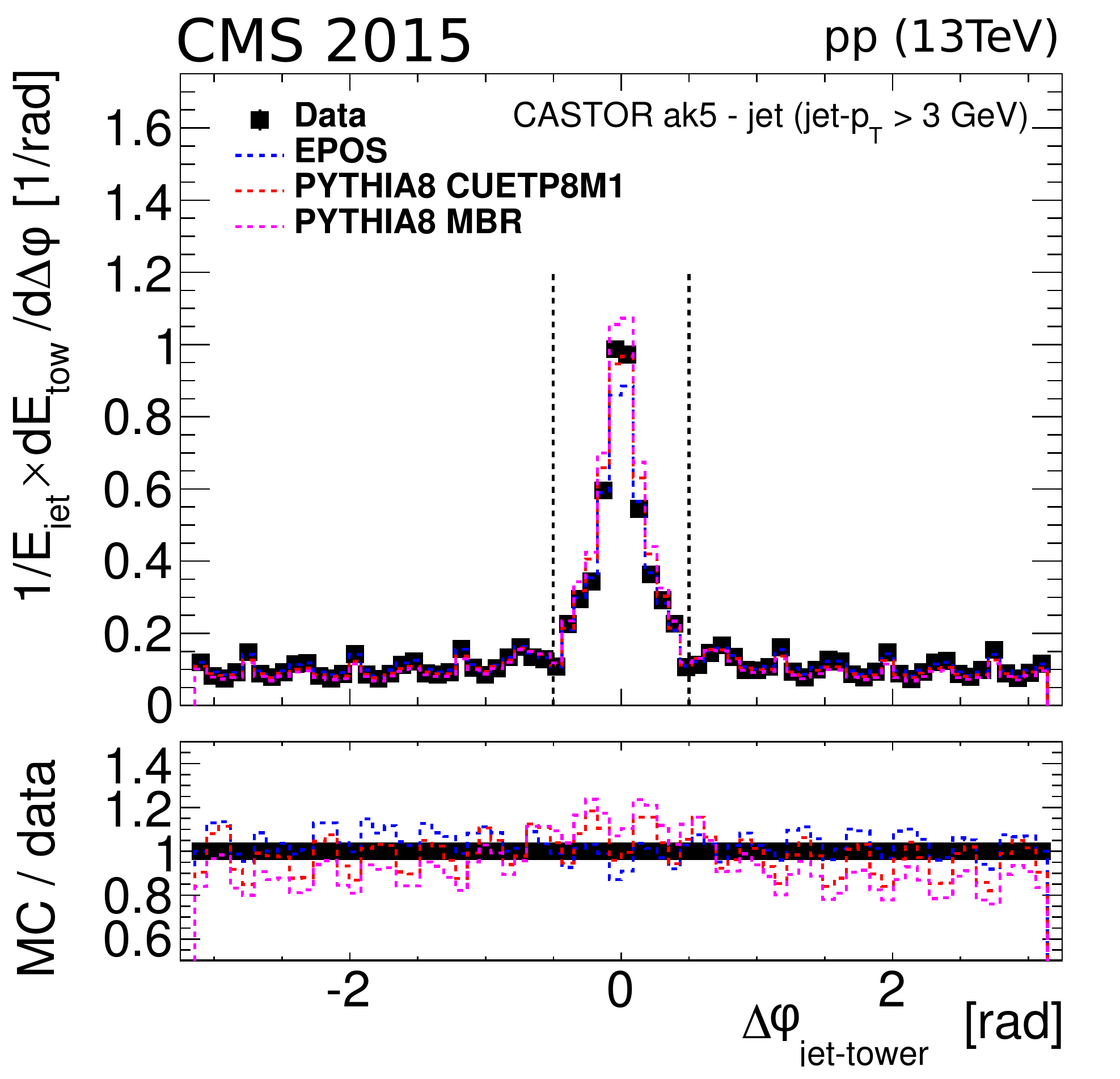}
  \caption{Average CASTOR tower energies, as a function of their azimuthal separation
  from the reconstructed jet axis (reconstructed with the anti-\kt
  algorithm with $R=0.5$) measured in data (squares) and in minimum-bias
  \PYTHIA~8 and \textsc{Epos-LHC} MC simulations (histograms) in {\Pp}{\Pp} collisions at
  $\sqrts=13\TeV$.  \label{fig:jetshape}}
\end{figure}

{\tolerance=800
Energy-momentum vectors are constructed from the towers using the
measured energies and the geometry of the detector.
Those four-vectors are then clustered into jet objects using the anti-\kt
clustering algorithm~\cite{Cacciari:2008gp} with a distance parameter
of $R=0.5$, a value optimized taking into account the given detector segmentation and matching
the typical size of jets at very forward
rapidities~\cite{Sirunyan:2018ffo}. The distribution of azimuthal
separations of the towers to the reconstructed jet axis for data and
MC simulations, generated with \PYTHIA~8 (MBR and CUETP8M1~\cite{Khachatryan:2015pea} tunes) 
and \textsc{Epos-LHC}~\cite{Werner:2005jf} models, are compared in Fig.~\ref{fig:jetshape}.\par}

\subsection{Noise and baseline}
\label{section:NoiseandBaseline}

\begin{figure}[tpb!]
\centering
\includegraphics[width=.6\textwidth]{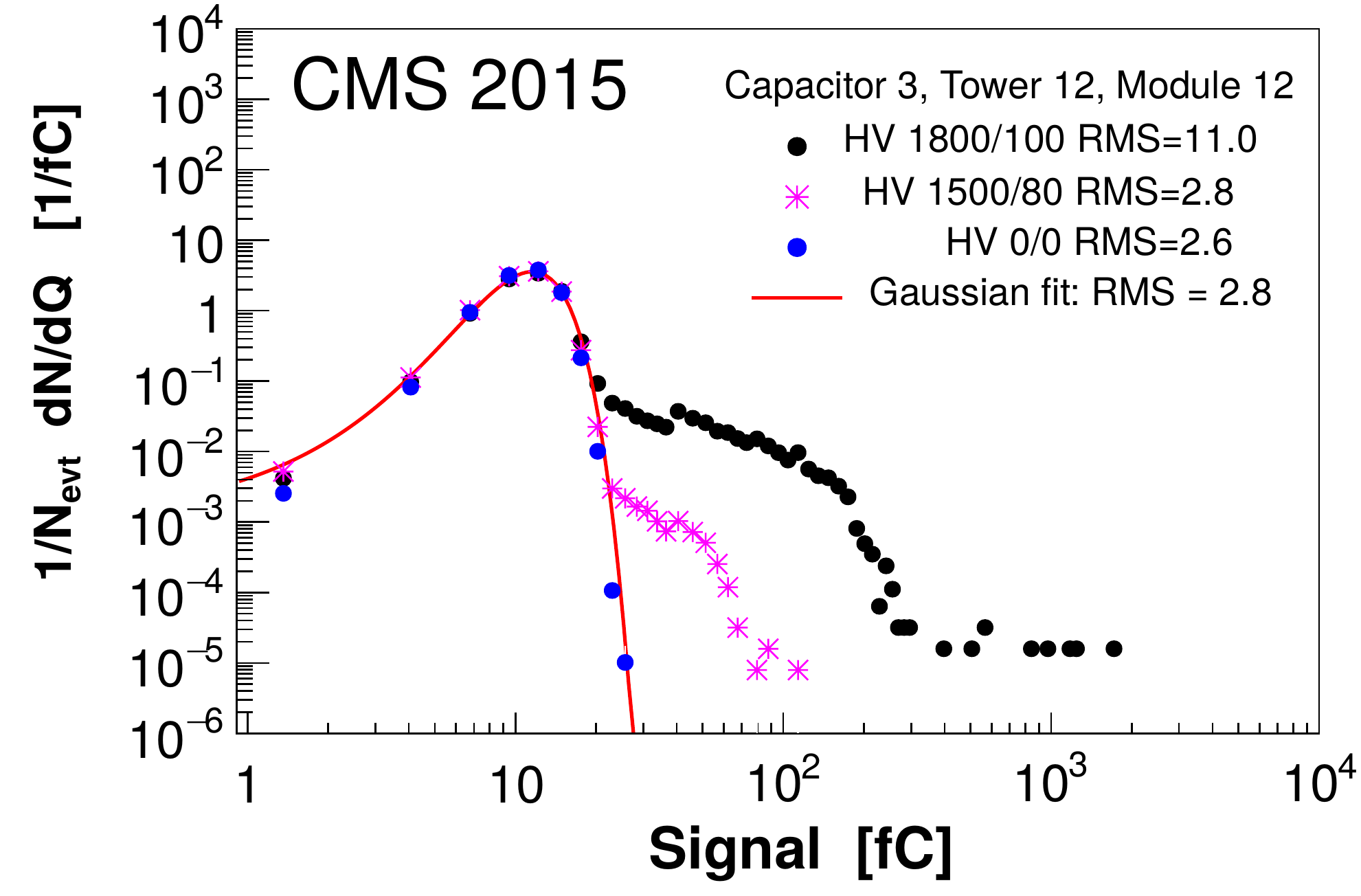}
\caption{\label{FIGPEDESTALANALYSIS}Charge spectrum for the
  noisiest capacitor of a typical CASTOR channel (tower~12, module~12)
  for various cathode/last dynode voltage settings (1800/100, 1500/80,
  and 0/0\unit{V}).  A Gaussian fit to the data recorded with
  0\unit{V} is also shown.}
\end{figure}

The baseline (pedestal) and noise levels are estimated for each
channel by analyzing a large set of events recorded in-situ in the
absence of beams in the LHC. These data samples typically comprise
around one million events, and take less than one hour to be
collected. The charge spectrum for one of the noisiest analog buffer
capacitors is shown in Fig.~\ref{FIGPEDESTALANALYSIS}. This noise
spectrum is depicted for HV off (blue), 1500\unit{V} (magenta), and
1800\unit{V} (black).  During data taking, the high voltage typically
corresponds to 1500\unit{V} where, even for this noisy capacitor, the
random noise probability per event is much below 1\% for signals in
the high tail of the charge distribution ($\approx$20\unit{fC} in this
example).

The measurement without high voltage represents the purely electronic
noise from the cables, the amplification, and the digitization
chain. After applying a voltage to the PMT, additional noise
components start to contribute. The first shoulder in Fig.~\ref{FIGPEDESTALANALYSIS} corresponds to the
thermal emission of single electrons from the PMT cathode.  For
CASTOR's fine-mesh PMTs, single photoelectrons produce a smooth
spectrum instead of a clear peak. Signals of ion feedback
(also known as afterpulses) start to become visible at very high voltages, but with a
tiny noise rate of $<10^{-4}$ per event. The maximum noise signal
observed here is $\approx$2000\unit{fC} corresponding to about 10 to
100\GeV of energy depending on the gain setting.

\begin{figure}[tb!]
\centering
\includegraphics[width=0.49\textwidth]{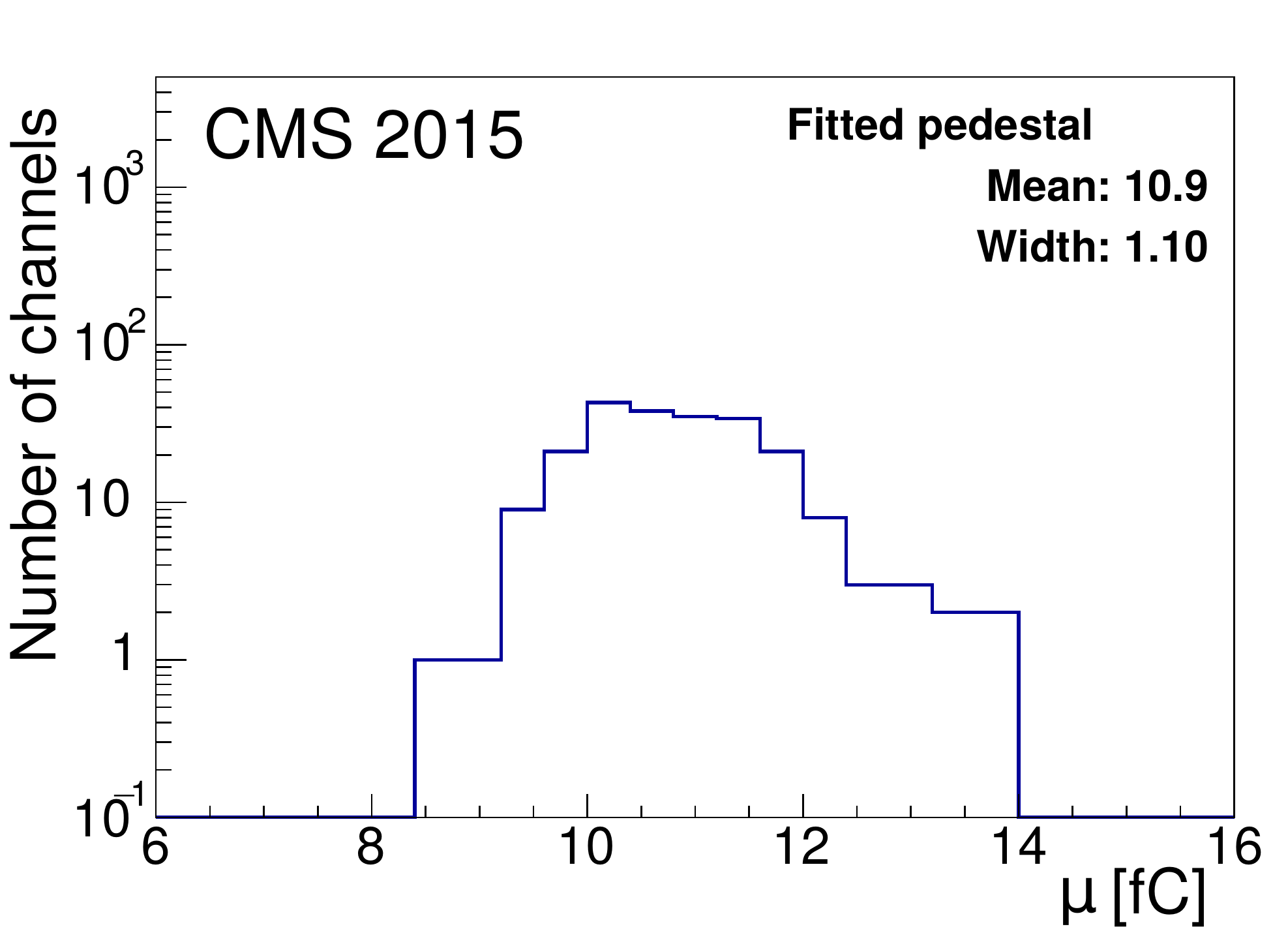}~
\includegraphics[width=0.49\textwidth]{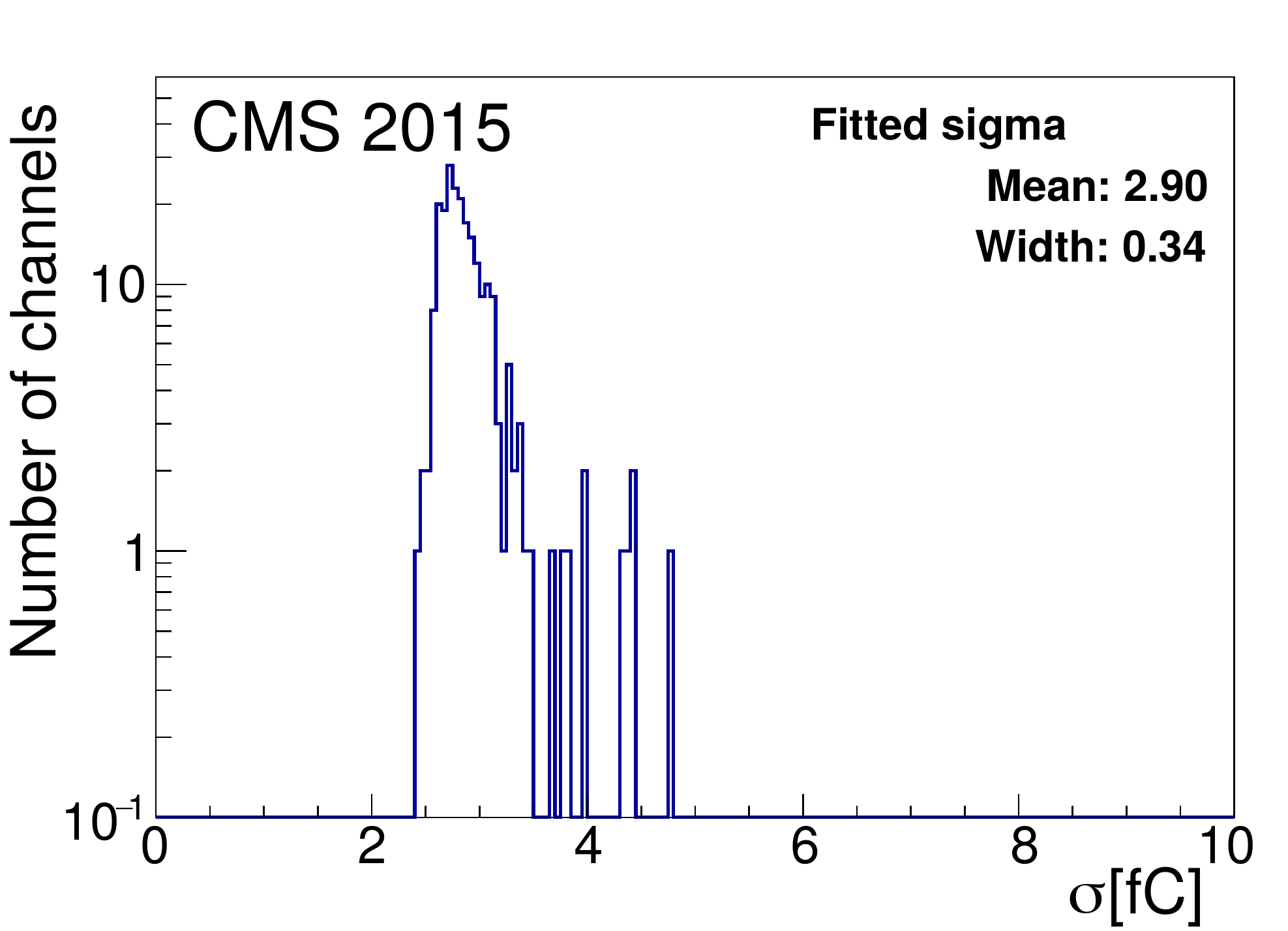}\\
\includegraphics[width=0.49\textwidth]{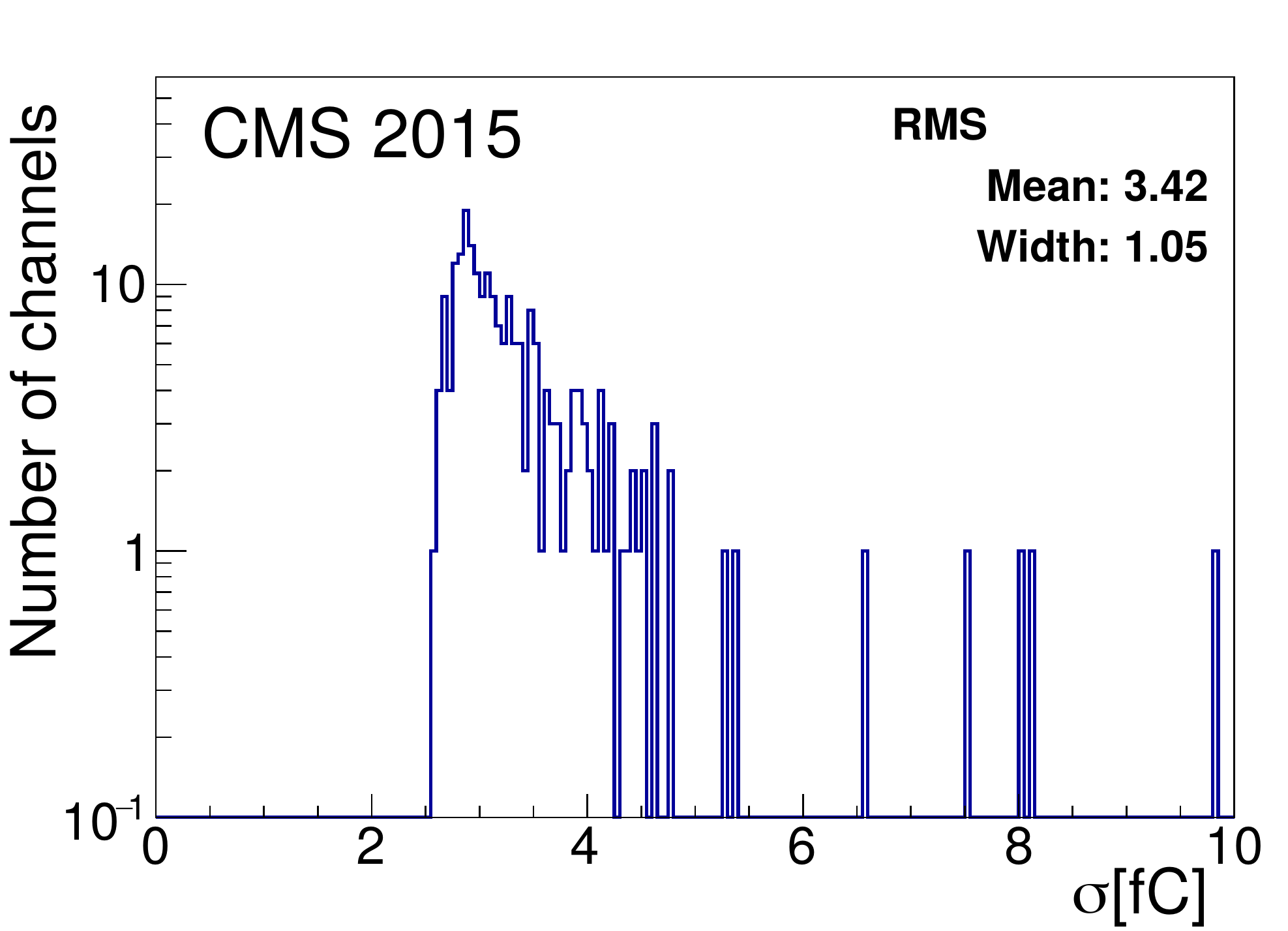}~
\includegraphics[width=0.49\textwidth]{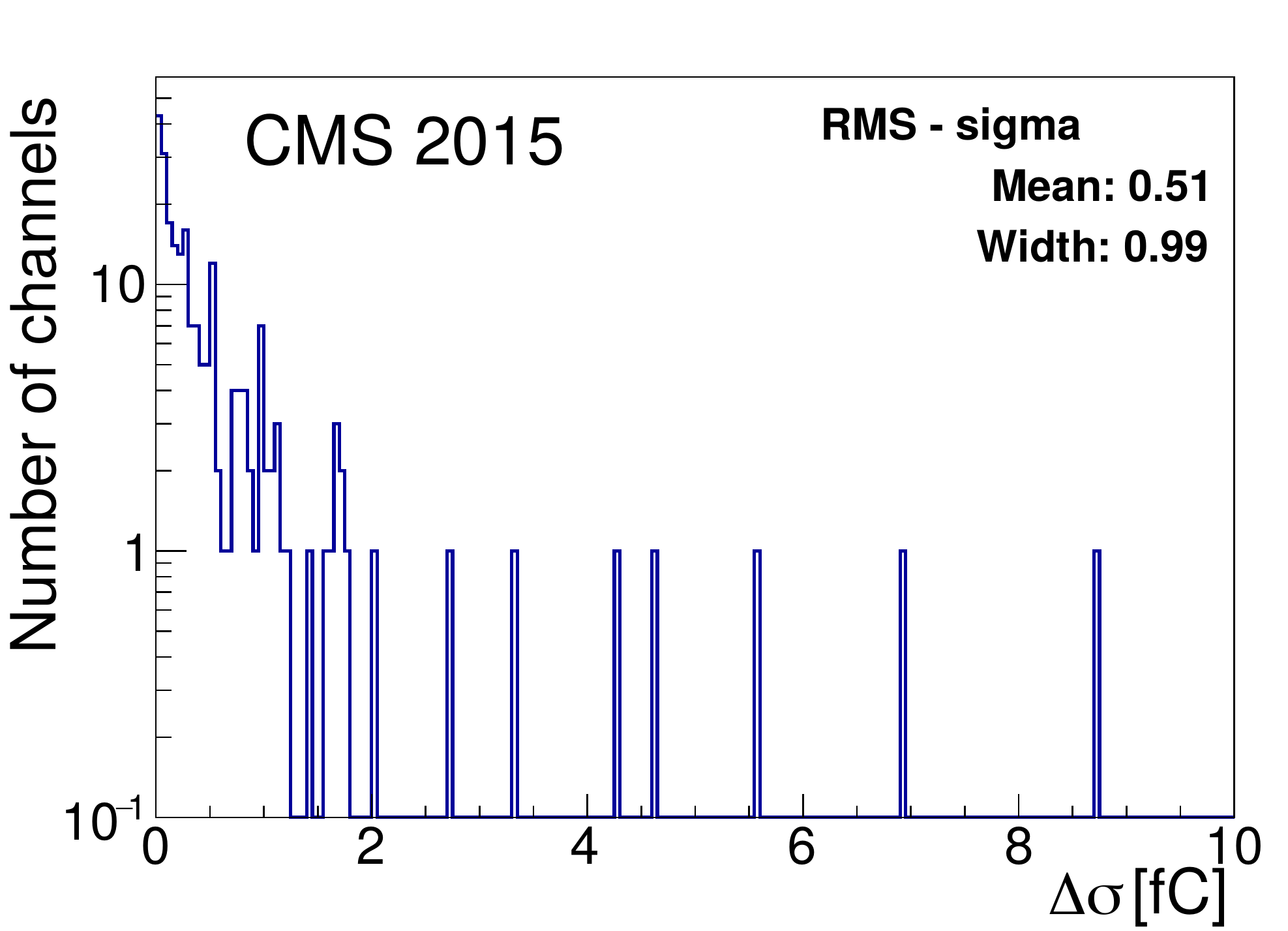}\\
\caption{Distributions of the baseline (upper left) and variances 
  (width $\sigma$, upper right; RMS, lower left;
  and $\Delta\sigma=$RMS-$\sigma$, lower right) of the CASTOR pedestals, with
  1500\unit{V} supplied to the PMTs, from data recorded in
  2015. \label{FIGStatisticalParametersNoise}}
\end{figure}

The mean of the Gaussian fit of the charge distribution for zero HV
is used as an estimate of the pedestal value, and subtracted
from all subsequent measurements. The width of the Gaussian fit is
used to monitor the response of a given channel.
In Fig.~\ref{FIGStatisticalParametersNoise}~(upper left), the
distribution of the fitted charge value of the pedestal is shown,
which is 11\unit{fC} on average. The fitted width of the Gaussian is
displayed in the upper right plot, and the root-mean-square (RMS)
value of the data in the lower left plot. On average, the RMS is
larger by 10\% than
the width obtained from the Gaussian fit, 
because of the aforementioned rare non-Gaussian tails. Comparing
the RMS to the Gaussian width has proven to be a useful method for
identifying noisy or malfunctioning channels. In the lower right plot,
the distribution of the difference between the RMS and the Gaussian
width per channel is displayed. A channel with a difference in excess
of 2\unit{fC} is considered noisy, and is subject
to additional inspection.  Channels that cannot be reliably reconstructed
are masked in the online trigger and excluded from physics
analyses. The number of such channels is kept to a minimum by
repairing PMTs, cables, and HV supplies, where
needed. This was the case, \eg, for eleven CASTOR channels in 2015, of
which none were located in 
the first three modules, where most of the energy is deposited.

\subsection{Gain correction factors}
\label{section:gaincorrectionfactors}

The CASTOR high-voltage settings are adapted to the varying LHC
operation configurations, depending on the instantaneous luminosity
and bunch spacing, and physics goals, with the heavy ion running being a
particular case. In central lead-lead collisions, the energy that is
deposited within the CASTOR acceptance can be enormous---the total
energy stored in a Pb nucleus at the LHC can be up to $5.5\TeV\times
208\approx1\PeV$ ($\approx$0.2\unit{mJ}).  Since the PMTs are always
operated at gains in the
range with an optimal signal-to-noise ratio, the
difference in gain settings between {\Pp}{\Pp} and PbPb collisions is about a
factor of twenty. Furthermore, in physics data taking, the PMTs close
to the shower maximum are configured with lower gains compared to the
ones measuring the shower tails in the back of the calorimeter.  On
the other hand, the signals from LHC beam halo muons are close to the
noise levels and very high amplification can be advantageous.

Correction factors are used to adjust the signals for these different
high-voltage settings. Three different methods are used, and they are
cross-checked against each other. In this section, we compare and
contrast the different methods and their results.  These methods have
different performances depending on the presence of magnetic fields.
Thus, we compare the results obtained with the
CMS magnet on and off.

\begin{figure}[tb!]
\centering
\includegraphics[width=.49\textwidth]{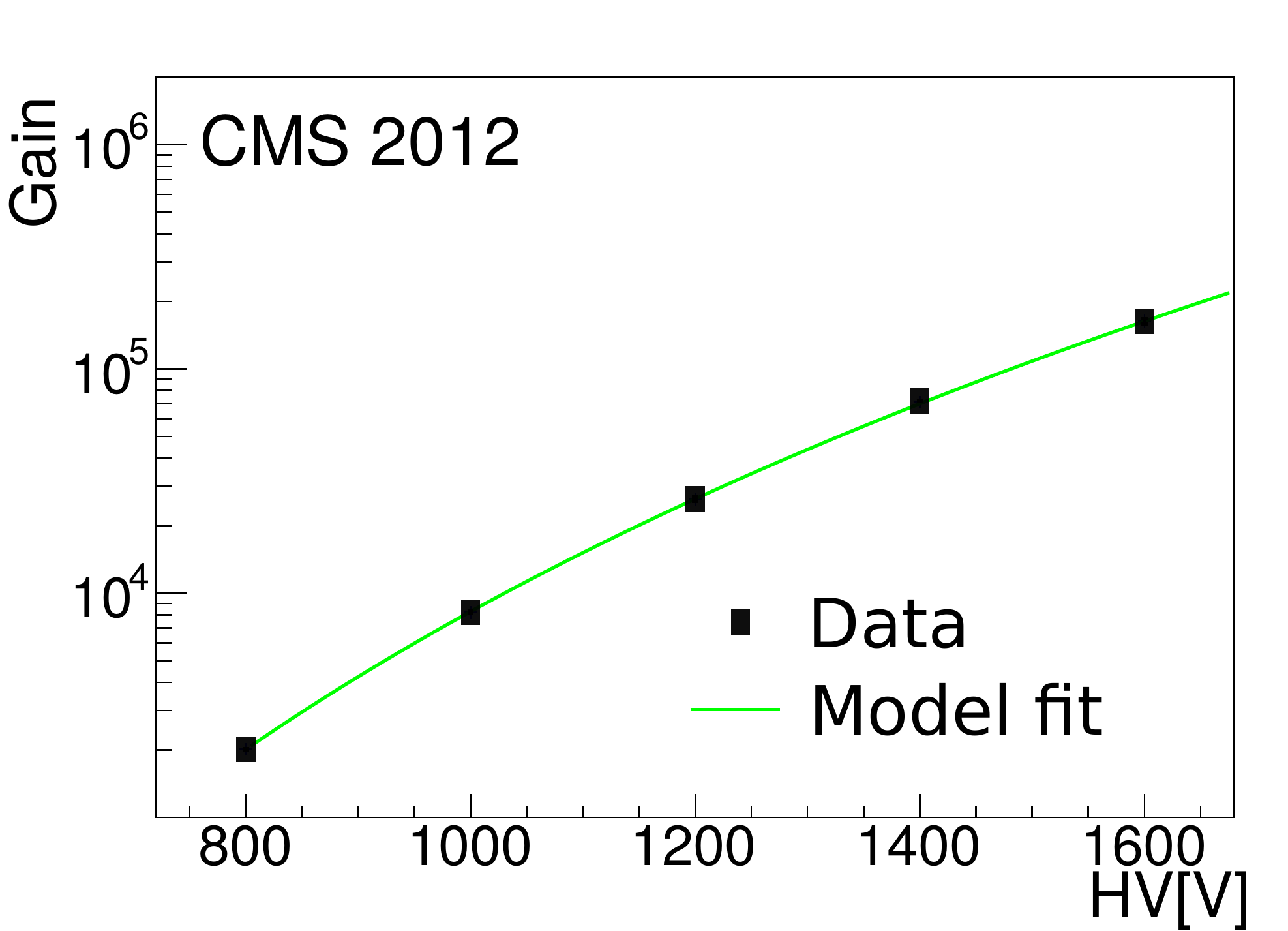}
\includegraphics[width=.49\textwidth]{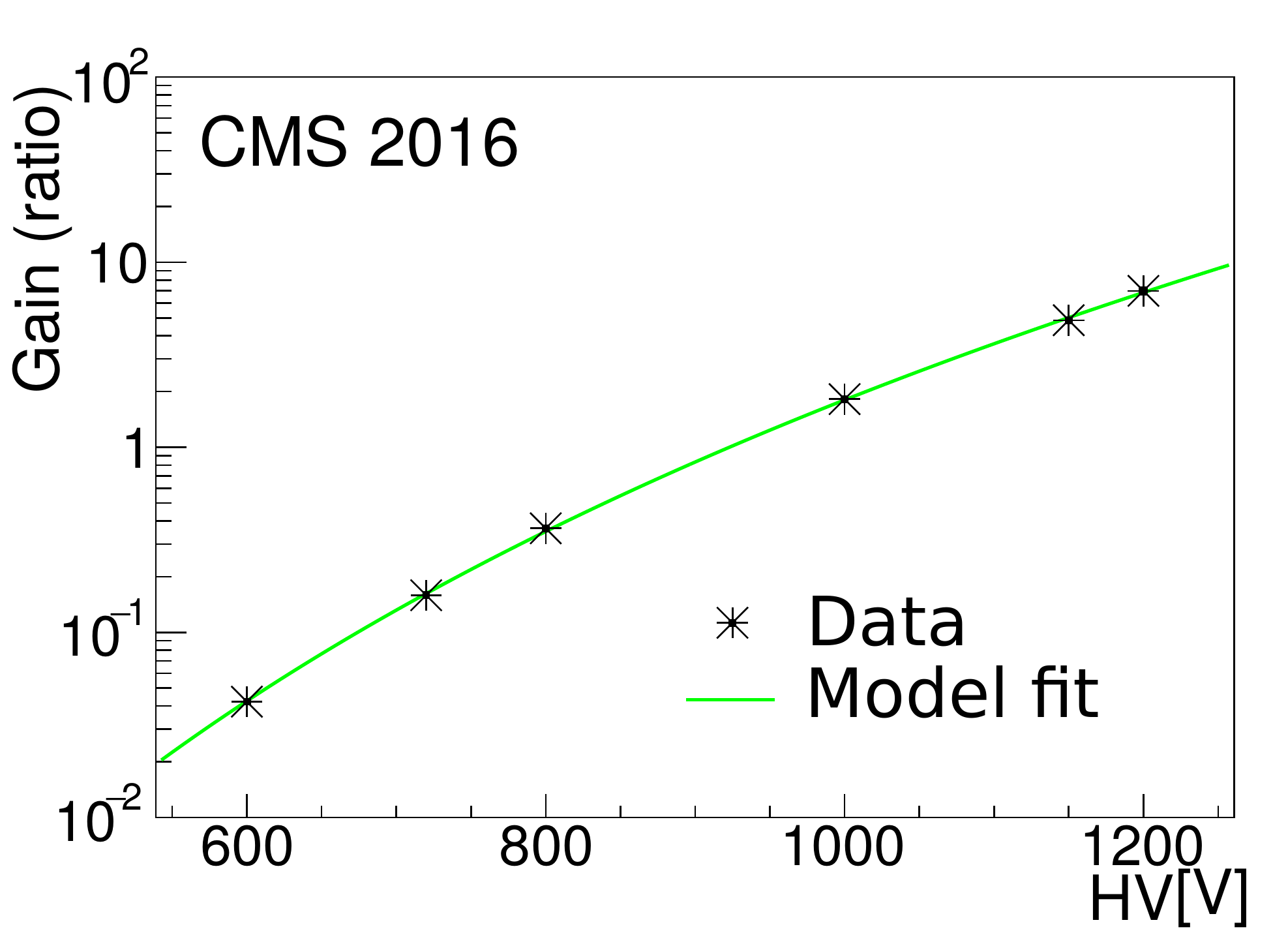}
\caption{Response of the CASTOR PMT in tower 1, module 3 versus the
  high-voltage setting prior to data taking exposed to LED pulses (\cmsLeft)
  and in physics events (\cmsRight). The (statistical) uncertainties
  are too small to be seen. 
  For the latter, no absolute gain measurement was
  performed, and the ratio with respect to the gain at the voltage of
  950\unit{V} is shown. The model fit is of the form $g(V)={p_0}V^{p_1}$
  with two free parameters $p_0$ and $p_1$. 
\label{fig:PMTcharacterization}}
\end{figure}

One option is to rely on the PMT characterization measurements
performed in the laboratory prior to their installation in the
experiment. In Fig.~\ref{fig:PMTcharacterization}~(\cmsLeft), the measured
dependence of the gain on the high-voltage value is depicted for one
channel. Very precise data exist for almost all PMTs.  However,
potential aging effects are not included. The use of the results is
also limited since they are not performed under exact data-taking
conditions, and this restricts its usage to data sets recorded without
magnetic field.

A second option is to use LHC collision data to perform a direct
in-situ measurement of the dependence of the PMT gain on the high voltage
(Fig.~\ref{fig:PMTcharacterization}, right).  This is
restricted to special situations where stable collisions are recorded
that are not useful for most physics analyses.  The ideal
scenario is to use higher-pileup data that generate very strong signals
and, at the same time, are not used for physics analyses with CASTOR.
In 2016, a dedicated voltage scan was performed in such optimal
conditions.  The average pileup in this data sample was $\approx$5, and a
CMS minimum-bias trigger that used coincidences of towers in
the hadron-forward (HF) calorimeters within $3.15<\abs{\eta}<4.90$
was employed to select the events. To account for the
slowly changing LHC beam intensity, the CASTOR data were also
normalized by the measured average energy in the HF detector.

\begin{figure}[b!]
\centering
\includegraphics[width=.49\textwidth]{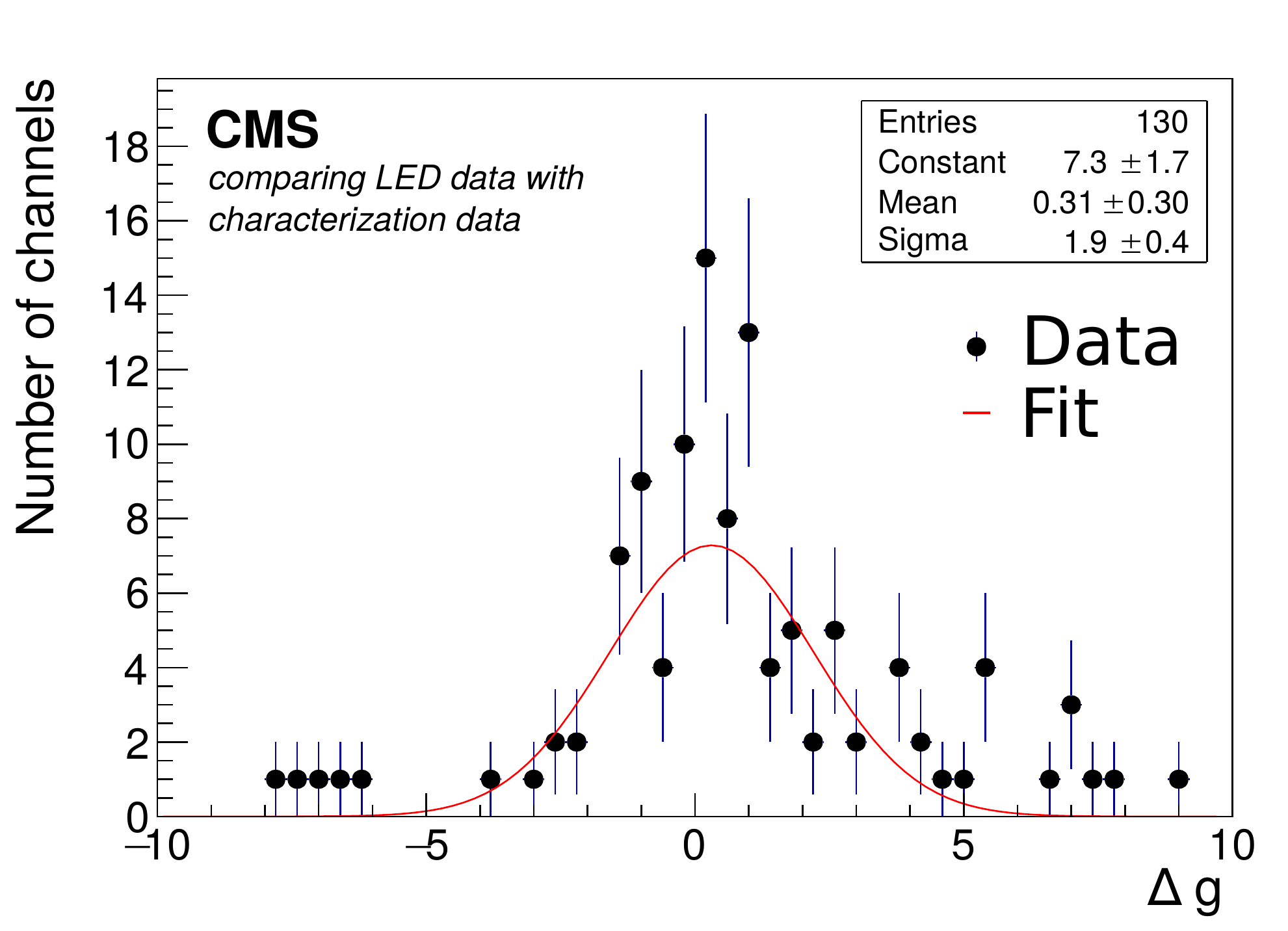}~
\includegraphics[width=.49\textwidth]{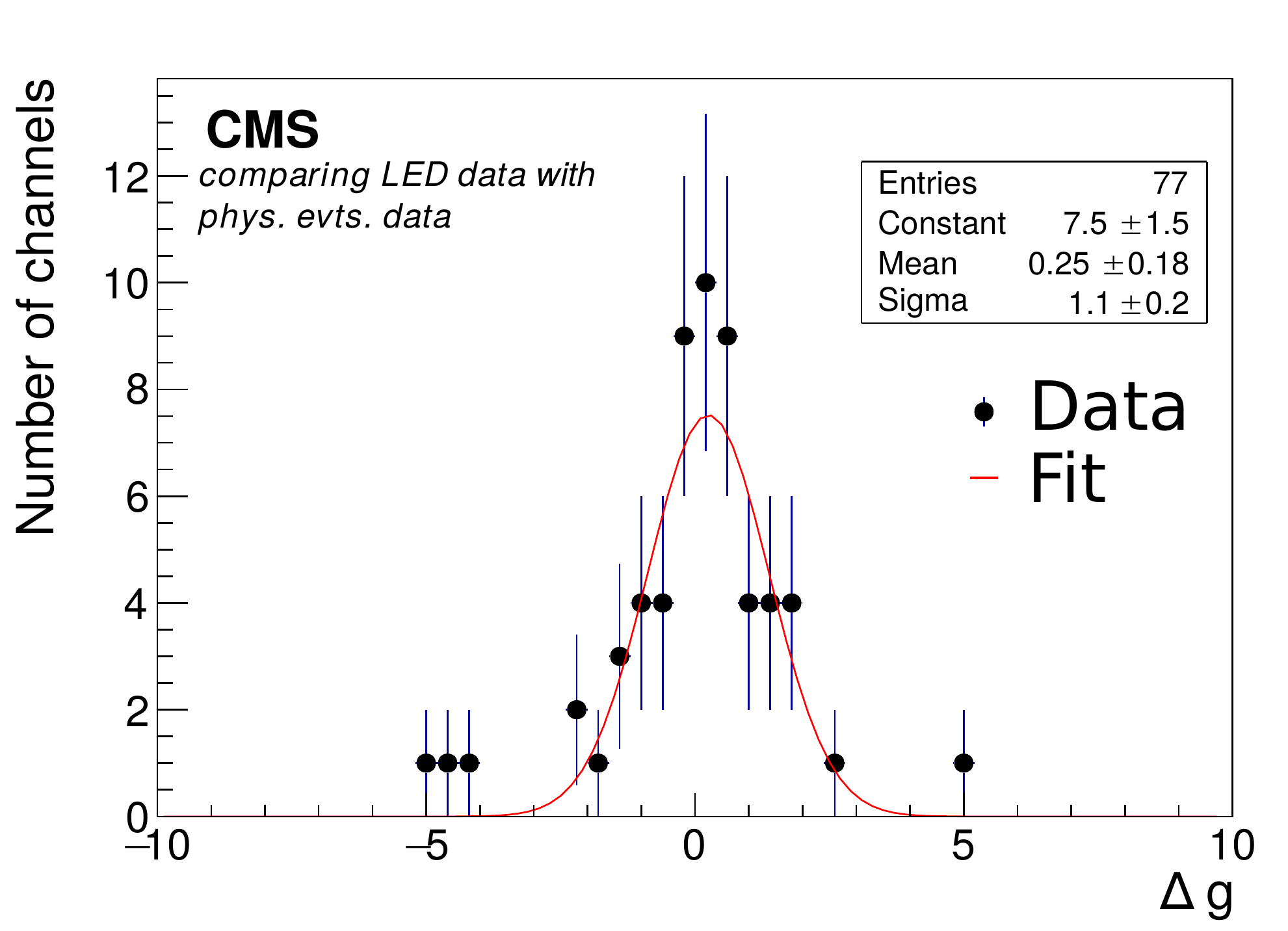}     
\caption{Distributions of the weighed difference of gains, $\Delta
  g=(g_1-g_2)/\sqrt{\sigma_{g_1}^2+\sigma_{g_2}^2}$, for two sets of
  correction factors obtained with two different methods, fitted to a
  Gaussian function. The uncertainties are statistical only.
  Left: Comparison of LED and PMT characterization
  methods at $B=0\unit{T}$ for muon halo and proton physics
  HV settings in 2016.  Right: Comparison of LED and physics events
  methods for muon halo and heavy ion physics HV settings at
  $B=3.8\unit{T}$.  Note that in both distributions the
  number of entries differs from 224 since only a subset of all
  channels has valid data available in both
  measurements. The boxes show the parameters of the red Gaussian fit. 
  \label{FIGPULLDISTRIBUTION}} 
\end{figure}

Finally, a third method relies on a statistical gain estimate
performed using LED pulses after the PMTs are installed in CMS. Such a
statistical method evaluates the fluctuations observed in each channel
as a response to the illumination of the cathode with LED pulses. The
signal seen by a PMT for a LED pulse, $\SLED$, is related to the
PMT gain, $g$, and the number of photoelectrons,
$\Npe$, according to $\SLED=g\Npe$. Thus, the relative variance of
$\SLED$ consists of a contribution from the gain fluctuations
($\sigma_{g}$) and from the Poisson fluctuations of the conversion of
photons to photoelectrons ($\sigma_{\Npe}$), and is
given by
\begin{linenomath} 
\begin{equation}
\left(\frac{\sigma_{\SLED}}{\SLED}\right)^2=\left(\frac{\sigma_{\Npe}}{\Npe}\right)^2 + 
\left(\frac{\sigma_{g}}{g}\right)^2\approx\left(\frac{\sigma_{\Npe}}{\Npe}\right)^2=\frac{1}{\Npe} = \frac{g}{\SLED} \,,
\end{equation}
\end{linenomath} 
where it is assumed that the relative fluctuations of $g$ are much
smaller than the relative conversion fluctuations, and all effects of
bandwidth limitations on the fluctuations are neglected. The gain is
then estimated using $g=\sigma_{\SLED}^{2}/\SLED$.

For a final set of high-precision gain correction factors, usually the
results of at least two of the methods described above are combined.
In Fig.~\ref{FIGPULLDISTRIBUTION}~(\cmsLeft), the distribution of weighted
differences between the gain correction factors obtained with the PMT
characterization and LED methods during {\Pp}{\Pp} operation at $B=0\unit{T}$
is displayed.  The width of this pull distribution is
consistent with unity only at the level of 2 standard deviations
($2\,\sigma$).  The corresponding distribution for heavy ion
collisions at $B=3.8\unit{T}$ is shown in
Fig.~\ref{FIGPULLDISTRIBUTION}~(\cmsRight) comparing correction factors
obtained with physics events and LED measurements. This latter pull
distribution reveals a better statistical consistency between the two
methods. From those results, it seems that the uncertainties obtained
during the PMT characterization setup are slightly underestimated,
although not in a significant way.  Furthermore, since all these
measurements have in general very high statistical precision, of the
order of 1\%, the absolute impact on calibration and data analyses are
negligible. We conclude that the statistical method using LED data,
which has the broadest range of applicability, is well suited for gain
correction factors whether or not a magnetic field is present.

\subsection{Channel-by-channel intercalibration}
\label{IntercalibrationParagraph}

The relative channel-by-channel intercalibration is carried out with
recorded beam halo muons originating from the LHC beam. These muons
traverse the CASTOR towers longitudinally, and lose only about 10\GeV
of energy traversing the calorimeter.  For the high-energy muons of
the beam halo, with energies in the hundreds of \GeV or above, such
energy loss is to a very good approximation insignificant.  Radiative
muon energy losses above the critical energy are an important
contribution at such high energies, but on average they contribute
equally to all channels of the calorimeter. Almost all halo muons used
for the calibration were taken during the period when the LHC is filled
with new protons at a constant beam energy of 450\GeV. Halo muons are
an excellent and stable probe for the relative calibration over the
entire lifetime of CASTOR.

\begin{figure}[b!]
 \centering
 \includegraphics[width=.4\textwidth]{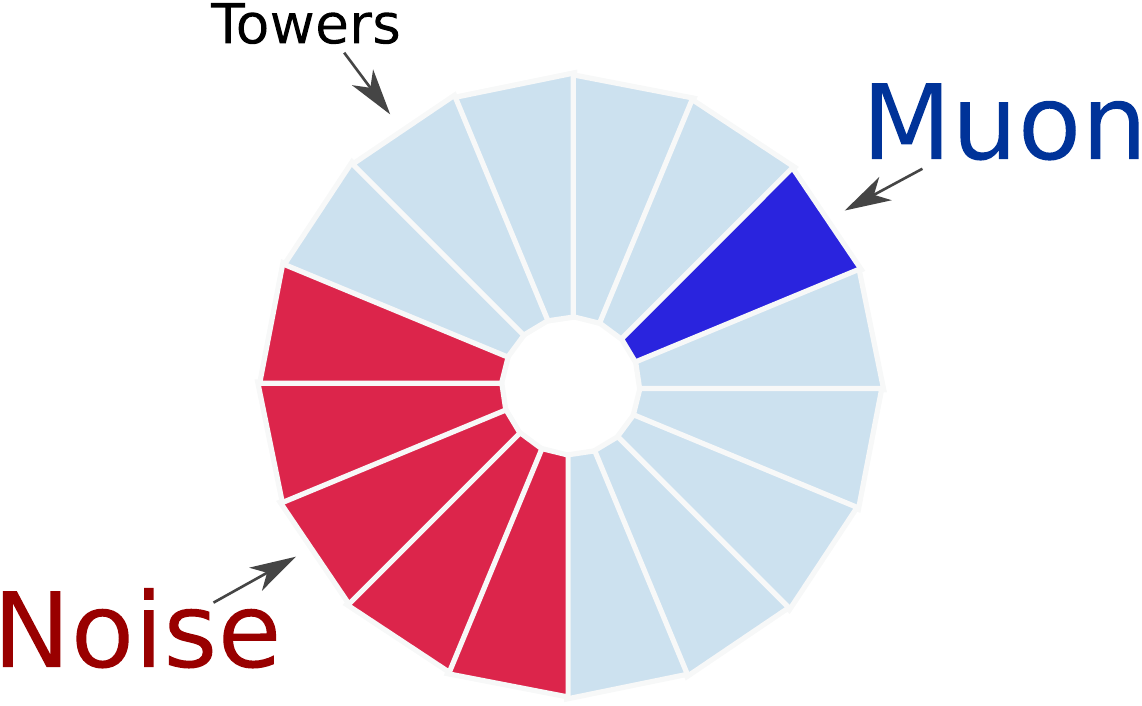}
 \caption{Schematic front view of CASTOR indicating its 16 towers. In this example
          event a muon candidate is identified in the blue tower. The calorimeter
          noise is then updated from the same event in the red towers. }
 \label{fig:noise_update}
\end{figure}

Beam halo muons are recorded during regular CMS data taking in
the periods when the LHC is in
interfill and circulating beam modes. A dedicated CASTOR
hardware trigger is activated for these runs to record events that are
sent to the L1 CMS trigger.  For this trigger, each tower is split
into four groups of three consecutive modules. Because of the hardware
design of the trigger, only the front 12 channels in each tower are
included in the trigger logic, excluding the two channels in the rear.
This may lead to a small bias for these last rear channels that must
be addressed by the subsequent event reconstruction and data analysis.
The trigger requires one channel (two channels in 2015/2016) above
noise level in at least three of these groups, and that there be no
other channel with energy above the noise level anywhere else in the
calorimeter. The typical trigger rate of this configuration is around
10 to 100\unit{Hz}, depending on the number of protons in the LHC.  The
separation of signal from noise during triggering is challenging since
the signal corresponds to just about one observed photoelectron per
channel, requiring very precise channel-by-channel estimates of the
noise level and baseline.  To reliably detect muons, the
high voltage is typically increased to 1800\unit{V} during the dedicated
halo muon runs.  However, during LHC Run 2, muon data were partly
taken at the specific physics HV setting for cross-checks, and to avoid
any additional corrections due to changing PMT gains and detector noise.

\begin{figure}[tpb!]
 \centering
 \includegraphics[width=\textwidth]{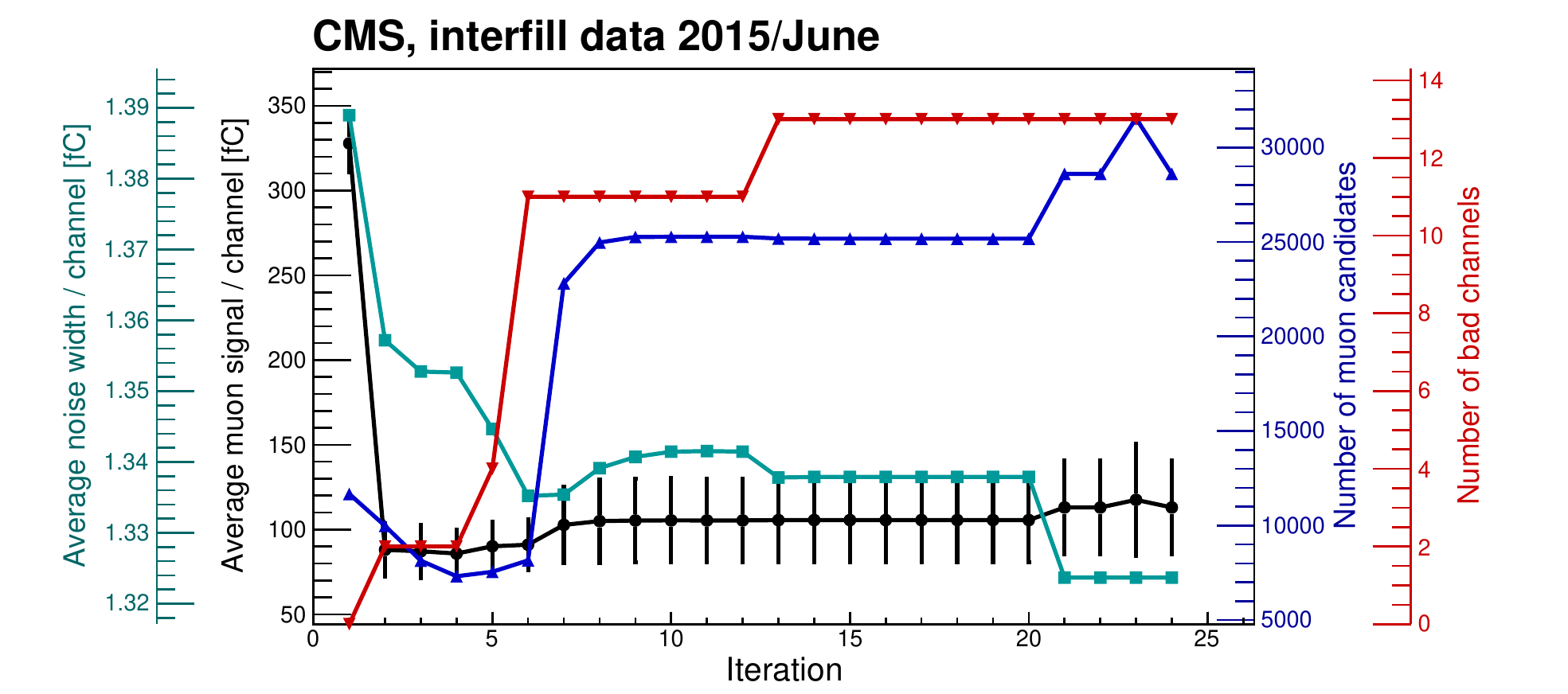}
 \caption{Visualization of the main parameters of the iterative
  offline muon event selection
  process: Average muon signal (black circles), average noise
  (turquoise squares), number of muon candidates (blue triangles), and
  number of bad channels (red inverted triangles) versus number of
  iterations.  The progress is mainly steered by first adjusting the
  number of bad (noisy) channels. The muon and noise responses quickly
  converge to stable levels. In the last four iterations, global
  parameters of the procedure are fine tuned. The shown uncertainties are statistical only, however, they are highly correlated in each graph. \label{fig:muonIter}}
\end{figure}

A detailed offline event analysis is required to prepare the collected
data for intercalibration. For this purpose, for each of the 224
channels, the no-beam noise thresholds must be accurately determined. 
Using this information, an offline
zero suppression of the data is applied,
using a threshold of $2\,\sigma$ above the noise.
A high-quality exclusive muon candidate is defined as a
single tower with at least 4 non-zero-suppressed
channels ($\nmumin$) in an event with at most 6 additional
non-zero-suppressed channels ($\nextramax$) in the rest of CASTOR.
Furthermore, the
tower containing the muon candidate must
have at least two non-zero-suppressed channels in three of these four
longitudinal regions: modules 1 to 3, modules 4 to 7, modules 8 to 11, and
modules 12 to 14. This requirement is sensitive to the penetrating
nature of muons and rejects low-energy pions that do not reach 
the back of the calorimeter.

Since the muon event selection uses energies very close to the
pedestals, a precise determination of the noise levels is of paramount
importance. We found that only an iterative procedure, where the
noise is measured from the same events as those from the actual muon
candidates, provides the required accuracy. For this purpose, when a
muon candidate is found, the calorimeter noise levels for the channels
in the five towers most distant from the muon candidate (opposite in
the transverse direction) are updated.  Thus, the data are used
simultaneously to measure the muon response and the calorimeter noise
levels. This procedure is illustrated in Fig.~\ref{fig:noise_update}.
The threshold levels for
zero suppression are determined from Gaussian fits to these data. If
no statistically significant improvement of the noise levels can be
determined in any of the channels of the calorimeter, the procedure
has converged and is stopped. Convergence typically occurs after 10 to
20 iterations, as demonstrated in Fig.~\ref{fig:muonIter} for data
recorded in June 2015. For this run, the noise levels quickly converge
after about five iterations. It is crucial to closely supervise this
process and to identify noisy channels that have a big impact on the
selection process. The list of bad channels is revised by excluding
the most noisy channels in several steps until no further impact on
the muon response is found.  In the last four points shown in
Fig.~\ref{fig:muonIter}, the average noise and the number of muon
candidates converges to slightly different values, because of a fine
tuning of the parameters $\nmumin$ and
$\nextramax$. The fine tuning is performed considering all
data from 2011 to 2016 simultaneously. This yields a slightly better
global noise level, increased muon data samples, and no significant
change in the measured muon response.

\begin{figure}[btp!]
  \centering
  \includegraphics[width=.6\linewidth]{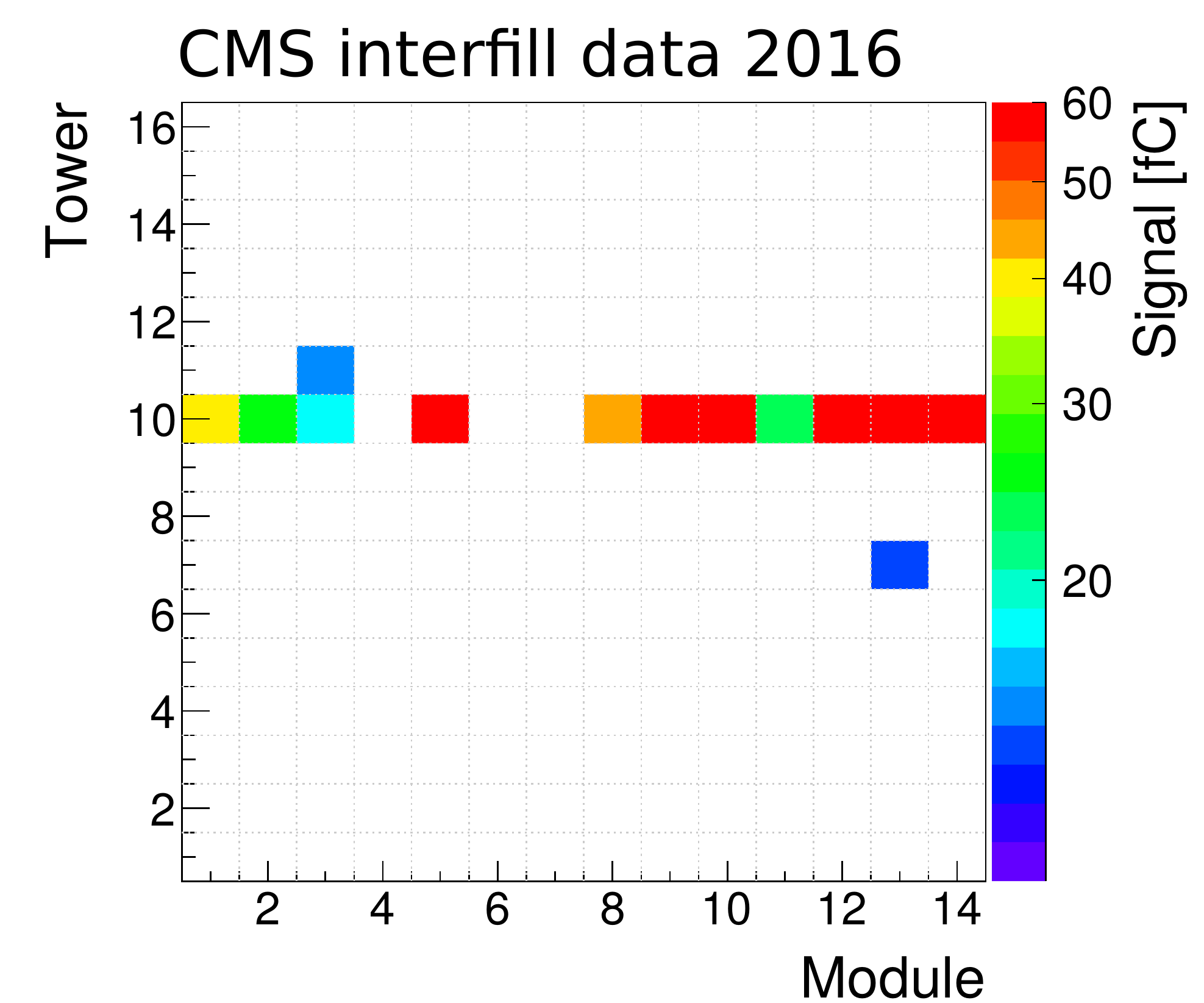}
  \caption{Example
   of a reconstructed halo muon event in the tower ($\phi$) vs.\
   module ($z$) plane recorded during the proton-lead run period in
   2016; channels are zero suppressed at the $2\,\sigma$ noise
   level. } \label{fig:muonSel}
\end{figure}

\begin{figure}[tpb!]
\centering
\includegraphics[width=0.5\textwidth]{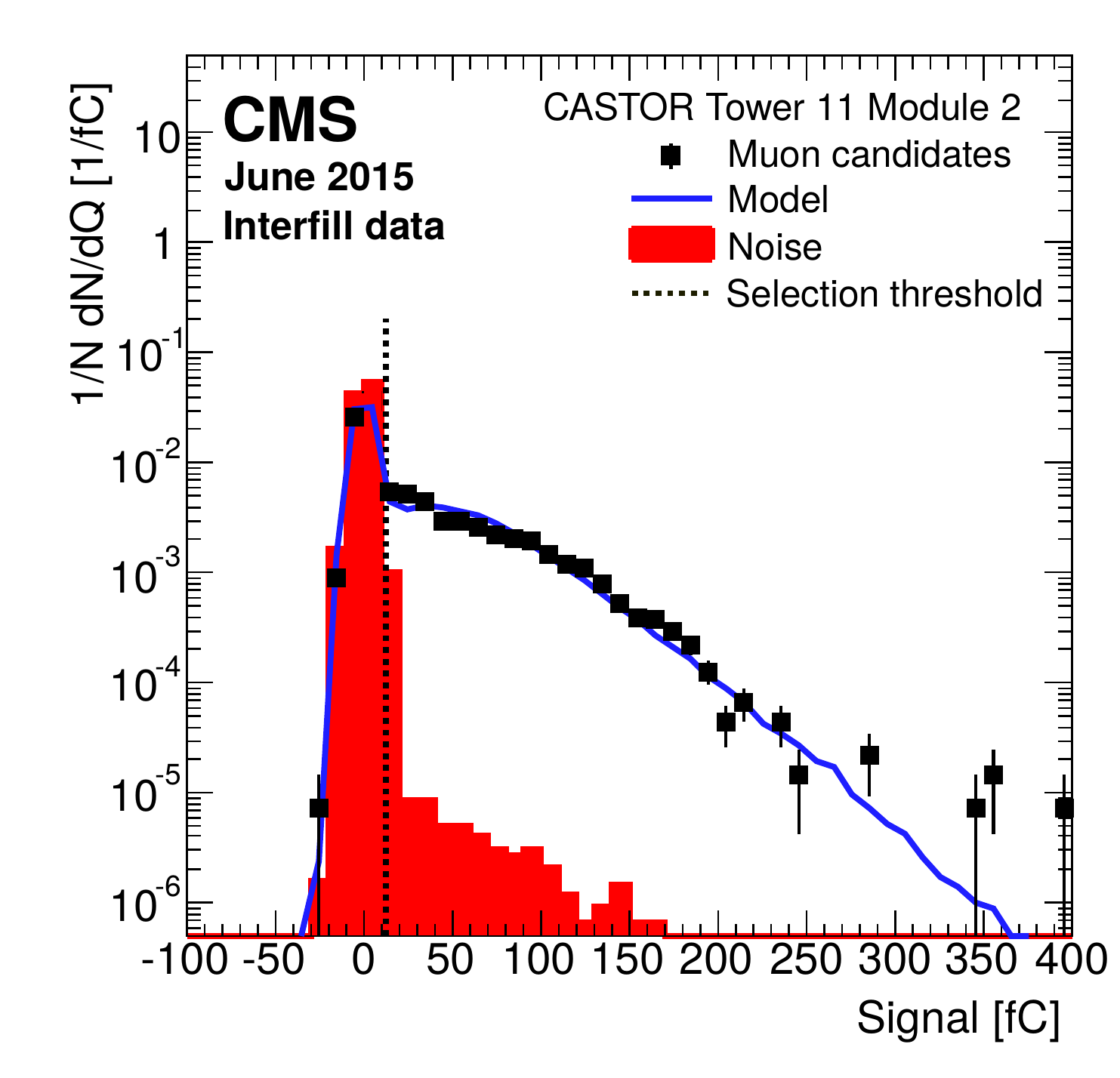}
\caption{Signal spectrum for a typical CASTOR channel after an
  offline isolated muon event selection. The data were recorded in June
  2015 with proton beams and CMS magnet at 0\unit{T}. The dedicated
  muon high-voltage setting is used. The uncertainties on
  the measured data are statistical only. The overlaid noise distribution
  is measured from noncolliding bunch data. The model line
  corresponds to a simplified mesh-type PMT with 15 dynodes,
  amplification/dynode of 2.65, and dynode-miss probability of 0.21,
  for an average number of photoelectrons
  $\langle\Npe\rangle=0.58$.
\label{FIGMUONSIGNALS}}
\end{figure}

In Fig.~\ref{fig:muonSel}, an example of the result of the online and
offline muon selections can be seen. These data were taken with the CMS
magnet at 3.8\unit{T}, which is why many of the modules at depth 7 to
9 yield less signal. In Fig.~\ref{FIGMUONSIGNALS}, the signal and noise
spectrum after offline selection are displayed for a typical channel together
with a prediction based on a simplified model of a fine-mesh PMT,
tuned to the data. The model assumes constant amplification per dynode
including Poissonian fluctuations. For the used fine-mesh PMT, it is
important to consider the probability of electrons missing a
particular dynode, $p_\textrm{miss}$. This leads to a nonideal
low-energy resolution that is of concern to understand the recorded
muon data: while the muons are clearly seen above noise level, there
is no obvious muon peak produced. 

\begin{figure}[t!]
\centering
\includegraphics[width=\textwidth]{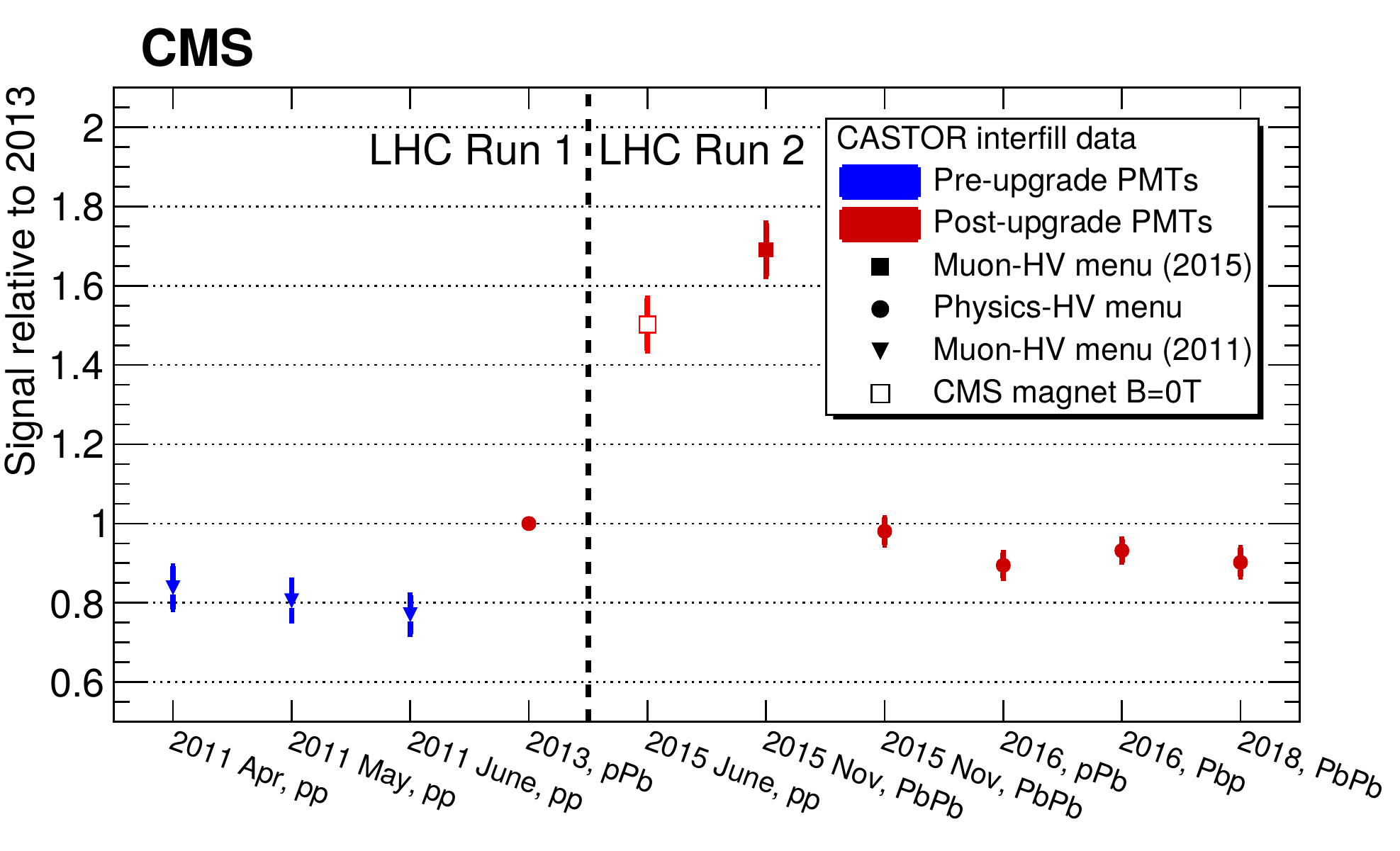}
\caption{Relative muon response averaged over all channels versus time.
  Shown is the
  average of the channel-by-channel ratios relative to the 2013 data.
  The vertical bars indicate the RMS of the distributions.
  All of the distributions are found to be consistent
  with a Gaussian shape.\label{fig:muon_physHV}}
\end{figure}

Since there is a large overlap of muon signal with noise levels, the
average recorded signal is not an ideal estimator of the muon
response. We found that the RMS is a much better estimator
instead. Since there are very rare spurious high-signal noise hits
also present in the recorded muon data, we exclude the 2\% highest
energy deposits for each channel. In most channels this has no visible
effect, but it helps to remove abnormal fluctuations in a few
channels. Thus, the muon response is defined to be the truncated RMS
per channel after rejecting the 2\% highest energy events.

These muon data provide a very powerful probe of the stability of the
calorimeter over time.  The time dependence of the gain for each
channel is studied using the average from muons collected during 2013
(Run 1) as a reference.  This is also of paramount importance for
maintaining a stable energy scale over the time span of the various
runs, from 2011 to 2018.  In Fig.~\ref{fig:muon_physHV}, the relative
change of the muon response during all data-taking periods is shown.
This is done by normalizing the average channel-by-channel ratio of
the muon response to that from the 2013 data.  In the 2011 data, the
loss of sensitivity due to radiation damage is visible. The step
between 2011 and 2013 coincides with the upgrade of the calorimeter
with new PMTs, which obviously had a positive effect on the
sensitivity.  June 2015 is the only period where the
halo muon data were recorded with 0\unit{T} magnetic field in
CMS. It is particularly interesting to compare the data collected in 2013 with
those in 2016, since they were all recorded with nominal magnetic
field and almost the same (physics) HV settings. The
stability is better than 10\% over three years. The change in
response is thought to be related to residual differences in the
magnetic field structure, detector aging, geometry, and HV
settings.  Contrary to other components of the CMS
experiment, the CASTOR calorimeter did not experience any significant
aging effects in Run 2.  The spread of the channel gains can be well
described by Gaussian distributions. The observed differences are
corrected with the described calibration procedure.  The response to
muons is used as an absolute reference scale.  Using a bootstrapping
method (case resampling) to estimate the statistical uncertainties of the
intercalibration constants indicates an uncertainty of $\approx$15\%
in each channel.

\subsection{Absolute energy scale}

\begin{figure}[tb!]
  \centering
  \includegraphics[width=0.6\linewidth]{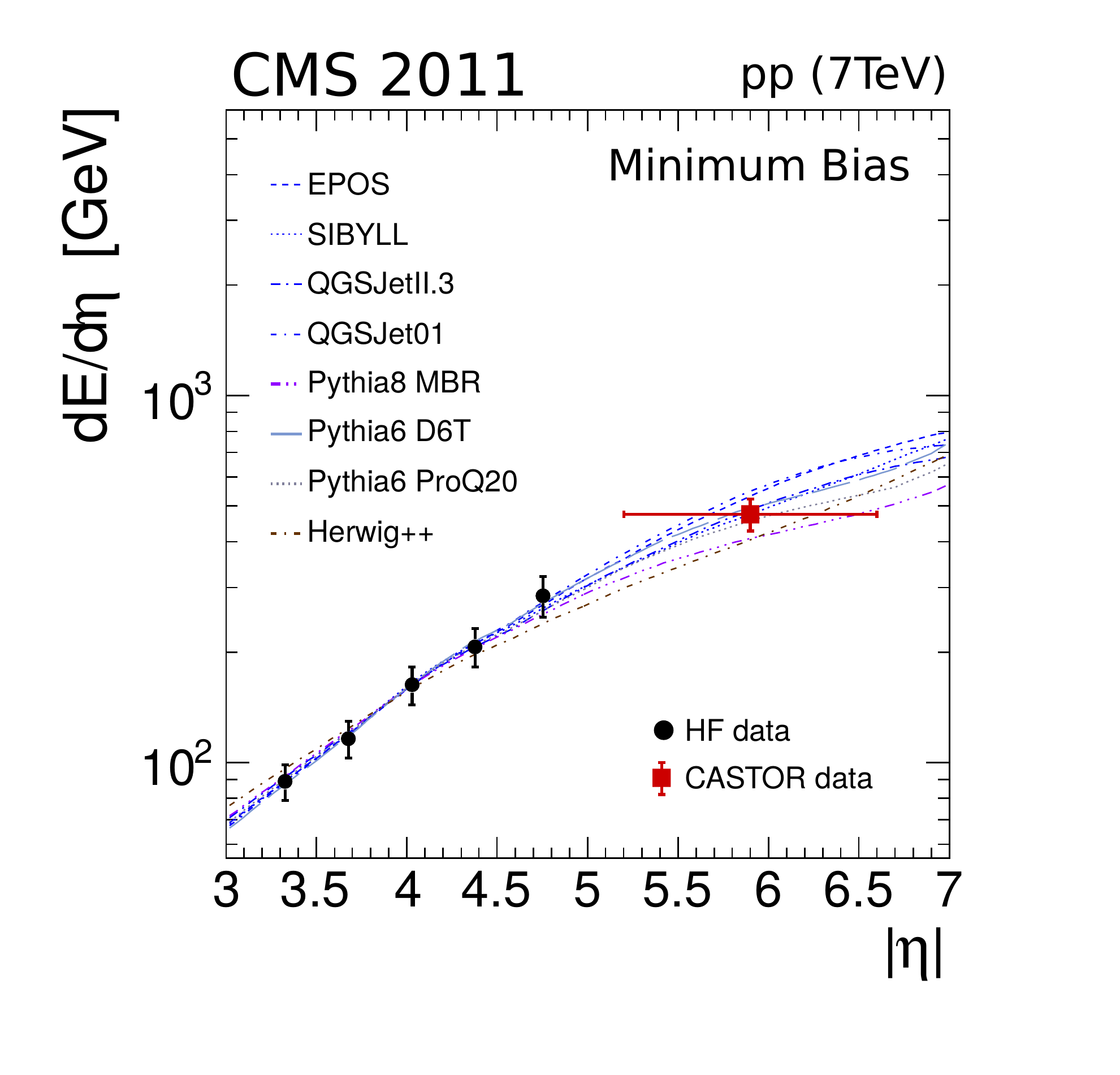}
  \caption{Pseudorapidity
  dependence of the forward energy flow in {\Pp}{\Pp} collisions at
  $\sqrts=7\TeV$ measured in HF~\cite{Chatrchyan:2011wm} and CASTOR,
  compared to MC simulations extrapolated to CASTOR.
  The uncertainties on the data are dominated by systematic
  effects. For the CASTOR datum the horizontal error bar indicates
  the $\eta$ range of the calorimeter.  \label{fig:eta_dependence}}
\end{figure}

The initial fundamental calibration of CASTOR was done with test-beam
measurements at the CERN SPS using electrons, charged
pions, and muons of well defined energies ranging from 10 to
300\GeV. The corresponding results are published in
Ref.~\cite{Andreev:2010zzb}. Good energy linearity as well as resolution were
found for electron beams, and the ratio between charged pion and electron
responses at the same particle energy was measured. The latter is a
fundamental property of any calorimeter and is linked to
noncompensation of energy losses in hadronic showers due to nuclear
breakup, and production of neutrons, muons, etc.

However, it is not sufficient to determine the energy scale in the test-beam
environment.  The conditions when installed in CMS are quite
different from those of the test-beam measurement. For example, the
presence of magnetic fields changes the PMT gains in a nontrivial
way. Correction factors have been determined for the difference in
gain, yielding an estimated energy scale in Run 1 of
0.016\unit{\GeV/fC} for the proton physics high-voltage settings used
in 2010.

An independent and more realistic calibration is performed from
collision data recorded when CASTOR was installed in CMS by
cross-calibrating its response to that of the HF calorimeter
($3.15<\abs{\eta}<4.90$).  The pseudorapidity dependence of the energy
deposited in the HF calorimeter in minimum-bias events (defined as {\Pp}{\Pp}
collisions at generator level with at least one stable particle produced
in both HF sides over $3.9<\abs{\eta}<4.4$) studied in
Ref.~\cite{Chatrchyan:2011wm}, has been used for this task.  In
Fig.~\ref{fig:eta_dependence}, we show the measured HF data extrapolated
to the CASTOR region together with different MC predictions from \PYTHIA~6.424~\cite{Sjostrand:2006za}
(D6T~\cite{Field:2008zz} and ProQ20~\cite{Buckley:2009bj}
tunes), \PYTHIA~8.145 MBR, {\HERWIG}++~\cite{Corcella:2000bw},
and commonly used models in cosmic ray physics: \textsc{QGSJet01}~\cite{Kalmykov:1994ys}, 
\textsc{QGSJetII}~\cite{Ostapchenko:2005nj}, \textsc{Sibyll2.1}~\cite{Ahn:2009wx}, and \textsc{Epos-LHC}.
The \PYTHIA~6 D6T tune is derived from charged particle multiplicities measured
by the UA5 experiment at the CERN S$\Pp\PAp$S-collider.  The
ProQ20 tune describes the CDF data at $\sqrts=0.63$ and
1.8\TeV, using LEP results to tune the parton-to-hadron
fragmentation.  The \PYTHIA~8 generator includes a new MPI model
interleaved with parton showering. In the \PYTHIA~8 generator, the
treatment of diffraction is improved compared
to \PYTHIA~6. The {\HERWIG}++ generator uses a different parton
fragmentation model than \PYTHIA. All cosmic ray MC models share the
underlying Gribov's Reggeon field theory
framework~\cite{Gribov:1968fc}, but different implementations of
perturbative parton scatterings and
diffraction~\cite{dEnterria:2011twh}. The full set of models chosen
provides thereby a very comprehensive range of predictions for
minimum-bias {\Pp}{\Pp} collisions at the LHC.

\begin{figure}[tpb!]
\centering
\includegraphics[width=.49\textwidth]{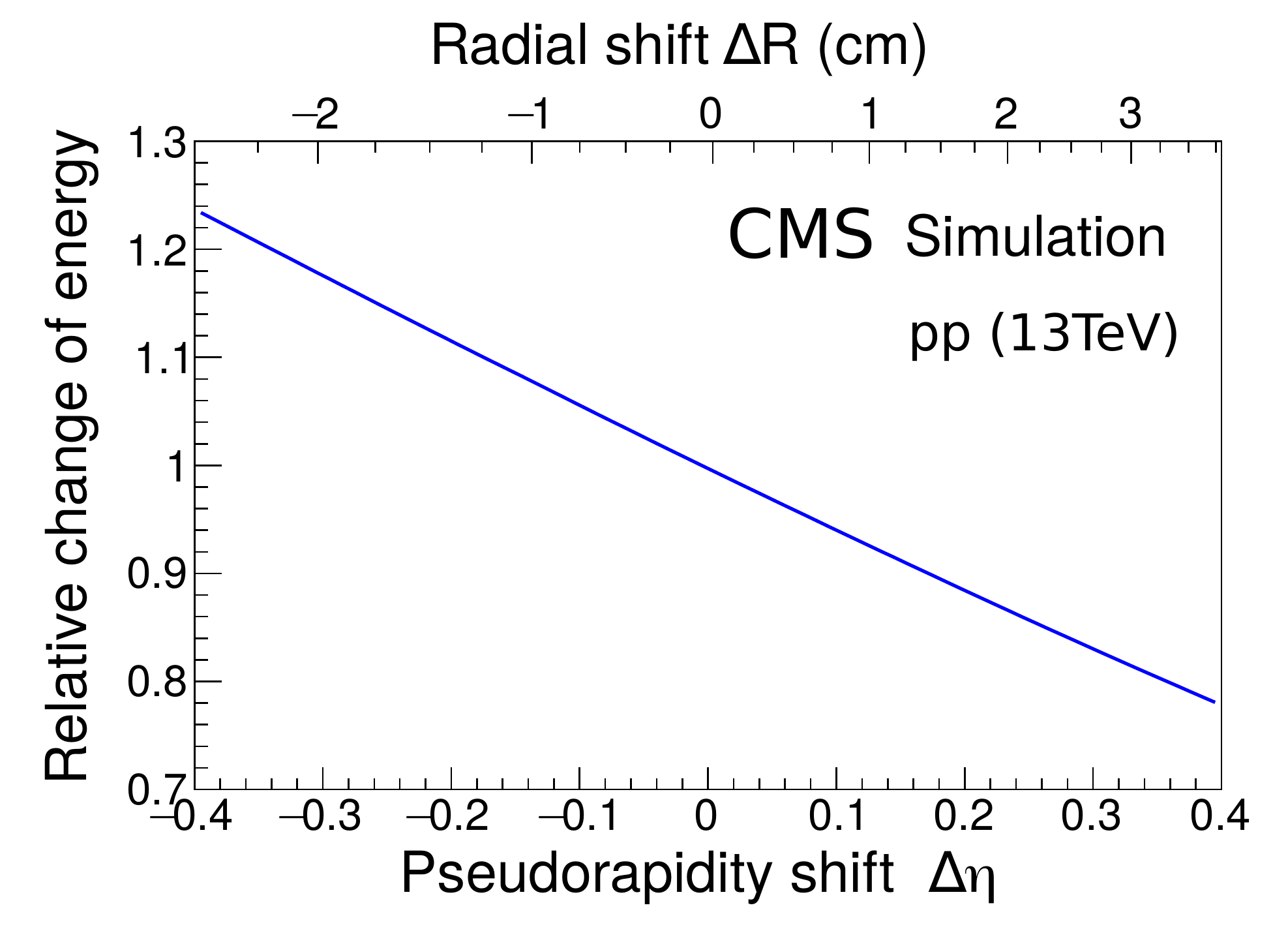}~
\includegraphics[width=.49\linewidth]{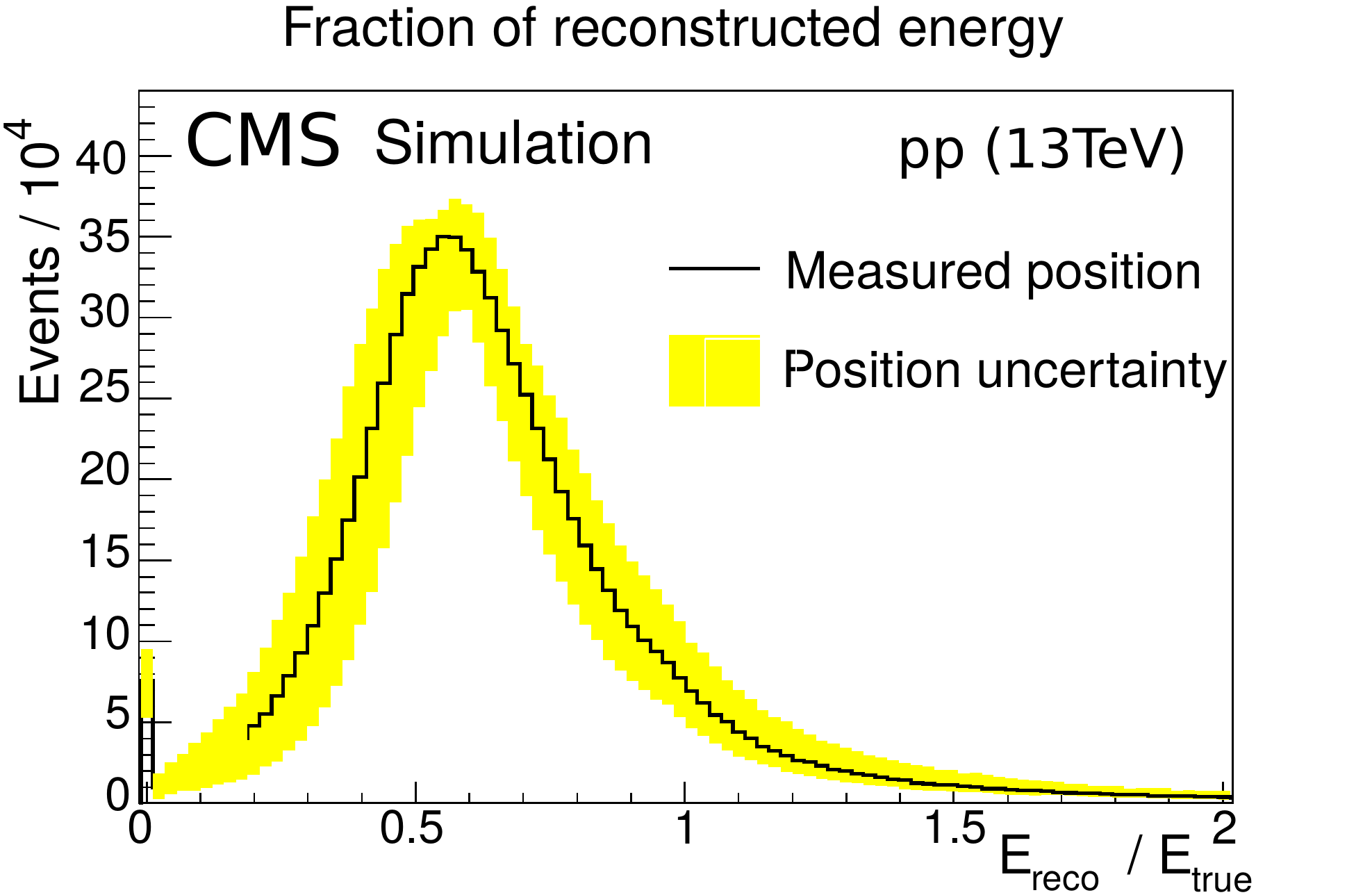}
\caption{Left: Impact of a pseudorapidity shift $\Delta\eta$ (due
  to a transverse CASTOR
  detector displacement $\Delta R$) on the energy response for
  simulated {\Pp}{\Pp} collisions at $\sqrts=13\TeV$.  Right: Distribution
  of the ratio between reconstructed and true energy measured
  in {\Pp}{\Pp} minimum-bias simulations generated with \PYTHIA~8 (CUETP8M1 tune).
  The average
  of about 0.6 is due to the noncompensating properties of
  CASTOR. The yellow error band indicates the uncertainty due to the
  alignment.  \label{fig:pos_impact}}
\end{figure}

The measured $\eta$ dependence of energy deposits in HF is
extrapolated to the CASTOR $\eta$ range. The model predictions are
scaled to best reproduce the measured $\eta$ dependence in HF for this
purpose.  All model predictions are calculated with the
generator-level definition using the online and offline event
selections defined by the HF measurement: at least one charged particle
on both sides of CMS over $3.9<\abs{\eta}<4.4$, and a minimal lifetime of
particles of c$\tau>10\mm$, excluding muons and neutrinos.  The
spread of the model predictions in the CASTOR acceptance provides an
estimate of the model dependence of this extrapolation.  The
extrapolated average energy deposit in the CASTOR acceptance found by
this procedure is
$\Epp(7\TeV)\equiv\rd E/\rd\eta_\mathrm{CASTOR}=665\GeV$
with an RMS of 60\GeV (10\%) for the 7\TeV minimum-bias proton-proton
data. The uncertainties in the absolute calibration of the HF as well as
in the HF generator-level correction are 10\%, and the
total uncertainty in the energy estimate in CASTOR is thus
$\Delta(\Epp(7\TeV))=\sqrt{(10\%)^2+(10\%)^2}\approx14\%$.
The average charge signal recorded in the five frontmost modules for
this minimum-bias data is $\Spp(7\TeV)=21\,000\unit{fC}$.
Since only these front modules of the calorimeter are used, about 75\%
of the energy of hadronic showers are contained.  Furthermore, two
channels of these front modules (module 1 in towers 5 and 6) are not
included in the measurement due to stringent quality criteria on
noise.  The calorimeter response factor
$\Cpp(7\TeV)=\Egen/\Edet=1.978$ is obtained
from a \GEANTfour-based~\cite{Agostinelli:2002hh} simulation of CASTOR
using minimum-bias events generated using \PYTHIA~6 (tune Z2), where the
simulated detector configuration and conditions were matched to those
used during data taking.  Here, $\Egen$ is the total
energy from \PYTHIA in the acceptance range and
$\Edet$ is the corresponding energy from the
simulation.  The calibrated energy $\Ecal$ measured
by CASTOR for any type of collisions, with either protons and/or lead ions, at
any center-of-mass energy, is then given by
\begin{linenomath}
 \begin{equation}
  \Ecal = \Sdet\; \frac{\Epp(7\TeV)}{\Spp(7\TeV)} \; \frac{C}{\Cpp(7\TeV)} \; \gHV\;,
  \label{eq:cal}
 \end{equation}
\end{linenomath}
where $\Sdet$ is the measured signal of CASTOR, $C$ is the
corresponding calorimeter response factor, and $\gHV$ are the
gain changes due to any difference in high-voltage settings.
Regarding the uncertainties in this assignment: the measurements of
$\Sdet$, $\Spp(7\TeV)$, and $\gHV$ have
negligible statistical uncertainties, the estimation of
$\Epp(7\TeV)$ is based on HF and has a 14\%
uncertainty.  It is important to note that the numerator and
denominator of the $C$/$\Cpp(7\TeV)$ ratio have uncertainties
that are highly correlated.  The individual absolute model-dependent
uncertainties of $C$ and $\Cpp(7\TeV)$ are not relevant, but
rather the uncertainty in the ratio of the two. Since these
uncertainties are not dominant, and their concrete evaluation requires
detailed studies for any new data-taking scenario, we assume that the
fractional uncertainty in $\Ecal$ is equal to that
of the fully uncorrelated uncertainties in $C$.  This neglects the
existing correlations and, thus, the resulting uncertainty is slightly
conservative.  From extensive detector simulation studies, the CASTOR
response $C$ has an uncertainty of $\approx$5\% arising from the
noncompensation in the energy response of the detector and the
predicted relative composition of hadrons and electromagnetic
particles in minimum-bias events for all colliding systems at the LHC
(combination of protons and lead nuclei) and different nucleon-nucleon
center-of-mass energies (2.76, 5.02, 7, and 13\TeV).  The effect of the geometry and
alignment of CASTOR with respect to the LHC luminous region can have
an important impact on $C$, as discussed also in the next
section. This is the largest source of uncertainty for the
determination of $C$. For example, the uncertainty due to imperfections
in the detector alignment during the 13\TeV proton-proton collisions in
June 2015 is 6.5\% (Fig.~\ref{fig:pos_impact}, right).
Also the number and location of dead channels in the calorimeter can,
in principle, have a large effect on $C$.  However, once they are
incorporated into the simulation of the detector, their effect is
negligible.

Thus, the absolute energy scale in CASTOR is determined by the energy
measured in the HF detector, combined with a model-dependent extrapolation using the
shape of event generator predictions and assumptions on the stability
of the determination of $C$, with a total precision of about
$\sqrt{(14\%)^2+(5\%)^2+(6.5\%)^2}\approx17\%$.  The approach leading
to Eq.~(\ref{eq:cal}) can also be used for the absolute calibration
factor corresponding to the scenario of a pure electron beam
\begin{linenomath}
\begin{equation}
  k_\mathrm{cal,e}=\frac{\Epp(7\TeV)}{\Spp(7\TeV) \; \Cpp(7\TeV)}
  = \frac{665\GeV}{21\,000\unit{fC} \;\; 1.978}
  = (0.0160\pm0.0027)\GeV/\unit{fC} ,
\end{equation}
\end{linenomath}
which  is  consistent with the corresponding test-beam measurement.

\section{Geometry and alignment}
\label{sec:align}

Because of the very forward location of CASTOR, even small movements of
the calorimeter can significantly affect its $\eta$ acceptance.  A
$\Delta R=1\mm$ (10\mm) radial shift affects the inner
acceptance of $\eta=-6.6$ by $\Delta\eta= 0.025$ (0.3) and the outer
acceptance of $\eta=-5.2$ by $\Delta\eta=0.01$ (0.07).  Since CASTOR
is located in a region where the energy deposition per collision
depends strongly on $\eta$, even small changes in the acceptance can
have a large effect on the expected energy
deposit~(Fig.~\ref{fig:pos_impact}, left). The position of the
calorimeter must be known with a precision of a few mm for a precise
interpretation of the measurements.  Fig.~\ref{fig:pos_impact}~(\cmsRight)
illustrates the impact on the energy response of the measured
position uncertainty during the data taking in June 2015.
This leads to uncertainties for reconstructed energies on hadron
level of up to 6.5\%.

The precise determination of the location of CASTOR is difficult since
the whole structure of the forward region of CMS moves by several
centimeters when the CMS magnet is ramped up.  This also leads to
nonnegligible movements of the otherwise nonmagnetic calorimeter,
which are probed with different infrared (Excelitas VTR24F1H) and
potentiometer (Active Sensors CLS1313) positioning sensors. Such
probes are particularly critical to keep a safe distance between the
CASTOR and the fragile beam pipe.

\begin{figure}[tpb!]
\centering
\includegraphics[width=.6\textwidth]{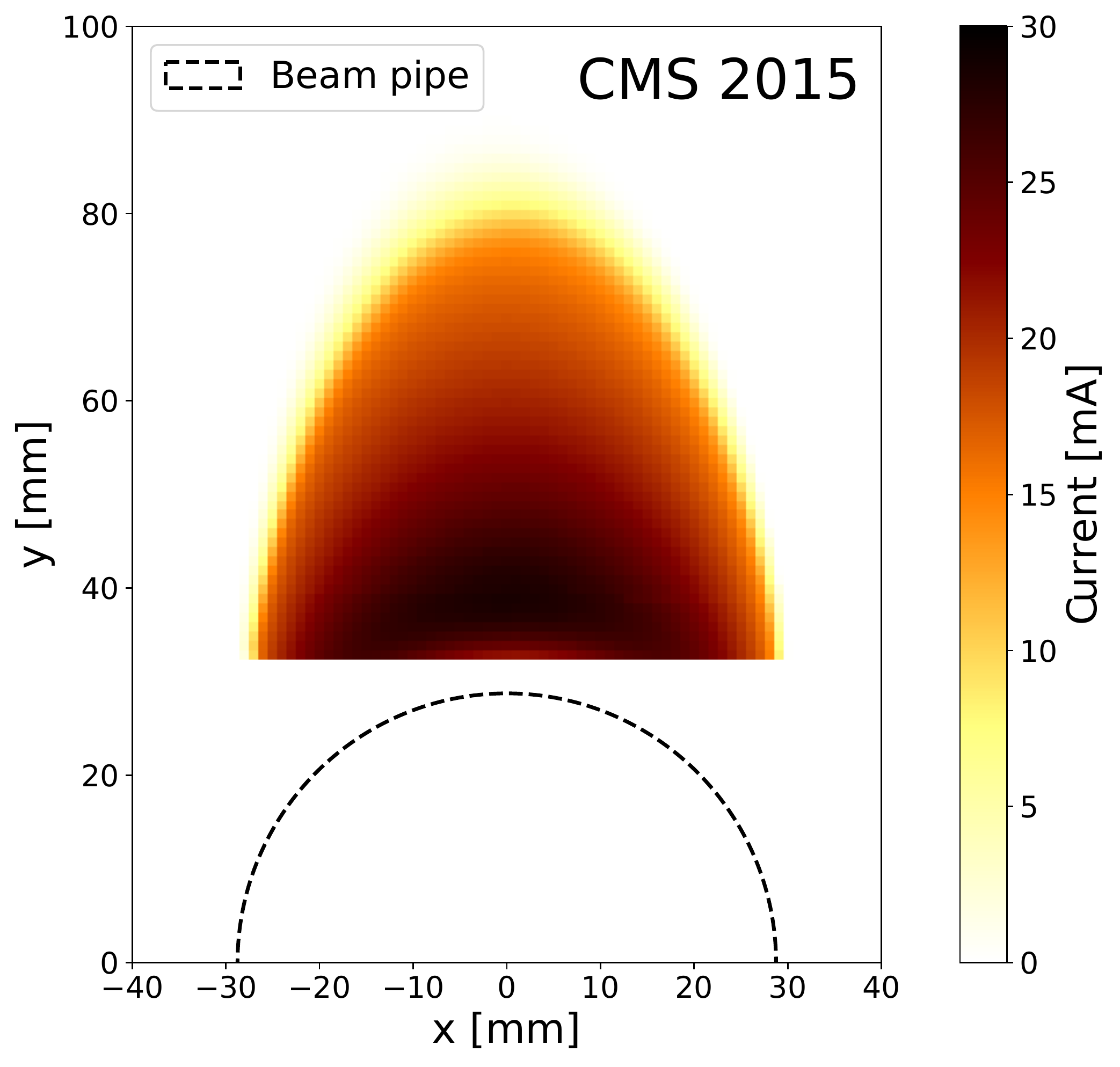}
\caption{Infrared distance sensor calibration data in the $(x,y)$ plane. Shown
  is the measured response during a bidimensional position scan in
  front of a beam pipe mockup. The used sensors do not work for very
  close proximities to the beam pipe due to self-shadowing effects, however, this is well
  outside of the range where the sensors are used in CMS. This effect
  can be seen in the calibration data at $x=0$ and close to $y=30\mm$.
\label{fig:sensor_data}}
\end{figure}

The position of CASTOR is measured after final installation, when there
is negligible magnetic field, by a survey team with laser targets to a
precision of $<$1\mm. Movements of the calorimeter during the magnetic
field ramp-up are observed with the positioning sensors. An absolute
position at full magnetic field is therefore determined with
information from the positioning sensors and the laser measurements.

\begin{figure}[tpb!]
  \centering
  \includegraphics[width=0.6\textwidth]{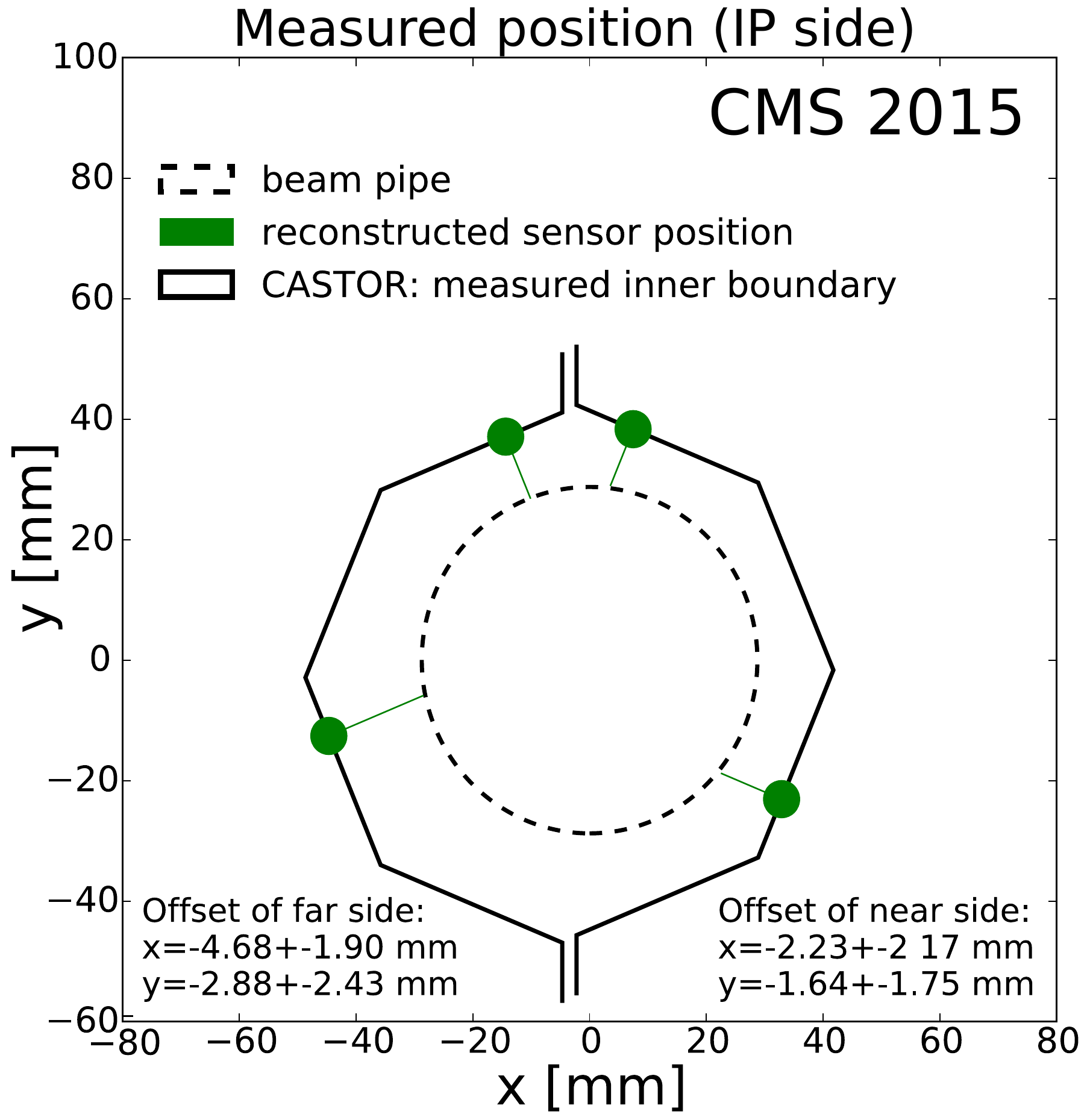}
  \caption{Illustration
  of the global fit result of the position of both CASTOR halves in
  the $(x,y)$ plane.  The beam pipe is indicated as a dashed
  circle. The measured distances of the sensors are shown as lines
  attached to the round markers that indicate the reconstructed sensor
  positions.  The potentiometer sensors are not shown here but are
  also used as constraints for the
  alignment.}  \label{fig:sensors_fit}
\end{figure}

On the side facing the interaction point, two infrared sensors are
mounted on each half that measure the distance to the beam pipe.  Two
potentiometer sensors measure the opening between the two halves of
CASTOR at $y=\pm19\mm$. On the other side, there are three
infrared sensors and three potentiometer sensors that measure relative
$x$, $y$, and $z$ movements. All sensor measurements are combined in a
global goodness-of-fit $\chi^2$ minimization for which the $x$- and
$y$-positions of the two halves are free parameters. The output
voltage of the sensors is calibrated to absolute distances with a
custom setup, which yields an uncertainty of about 1\mm for both
sensor types. This is achieved with an automated procedure on a
two-dimensional (2D) positioning table for the infrared sensors, and
by measuring the distance with  the fully contracted potentiometer
sensor. In Fig.~\ref{fig:sensor_data}, the measured response of
one infrared sensor in the 2D scan is shown.  The surface of the beam pipe
mockup is shown as dashed line. The uncertainties propagated to the
position found with the minimization procedure are lower when the infrared
sensors are pointing perpendicular to the beam
pipe. Figure~\ref{fig:sensors_fit} shows the result of this procedure
for the position of CASTOR after a magnet cycle in April 2015. A
position with an uncertainty in $x$ and $y$ of about 2\mm is
found.

\begin{figure}[tpb!]
  \centering
  \includegraphics[width=.52\textwidth]{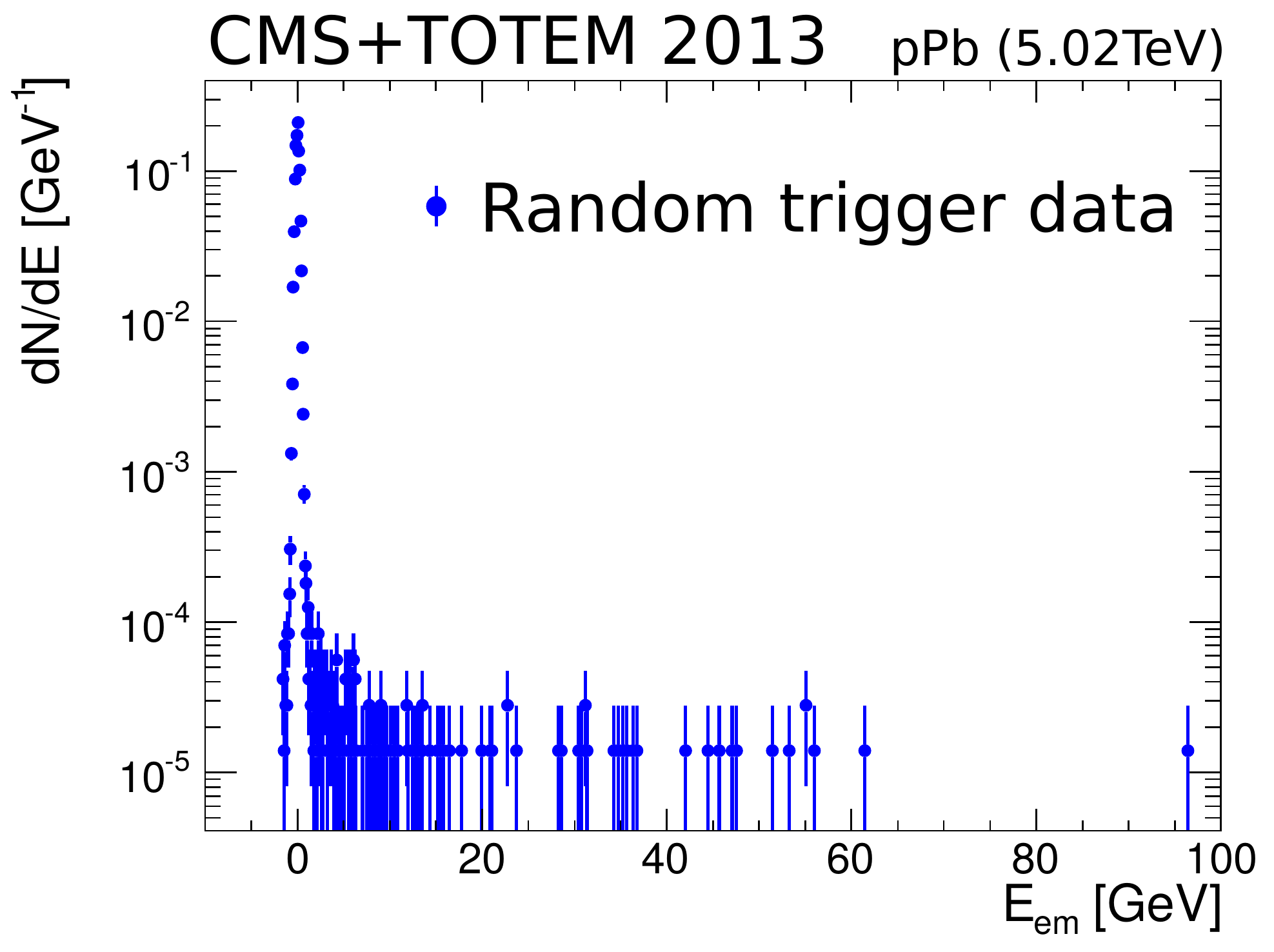}~
  \includegraphics[width=.48\textwidth]{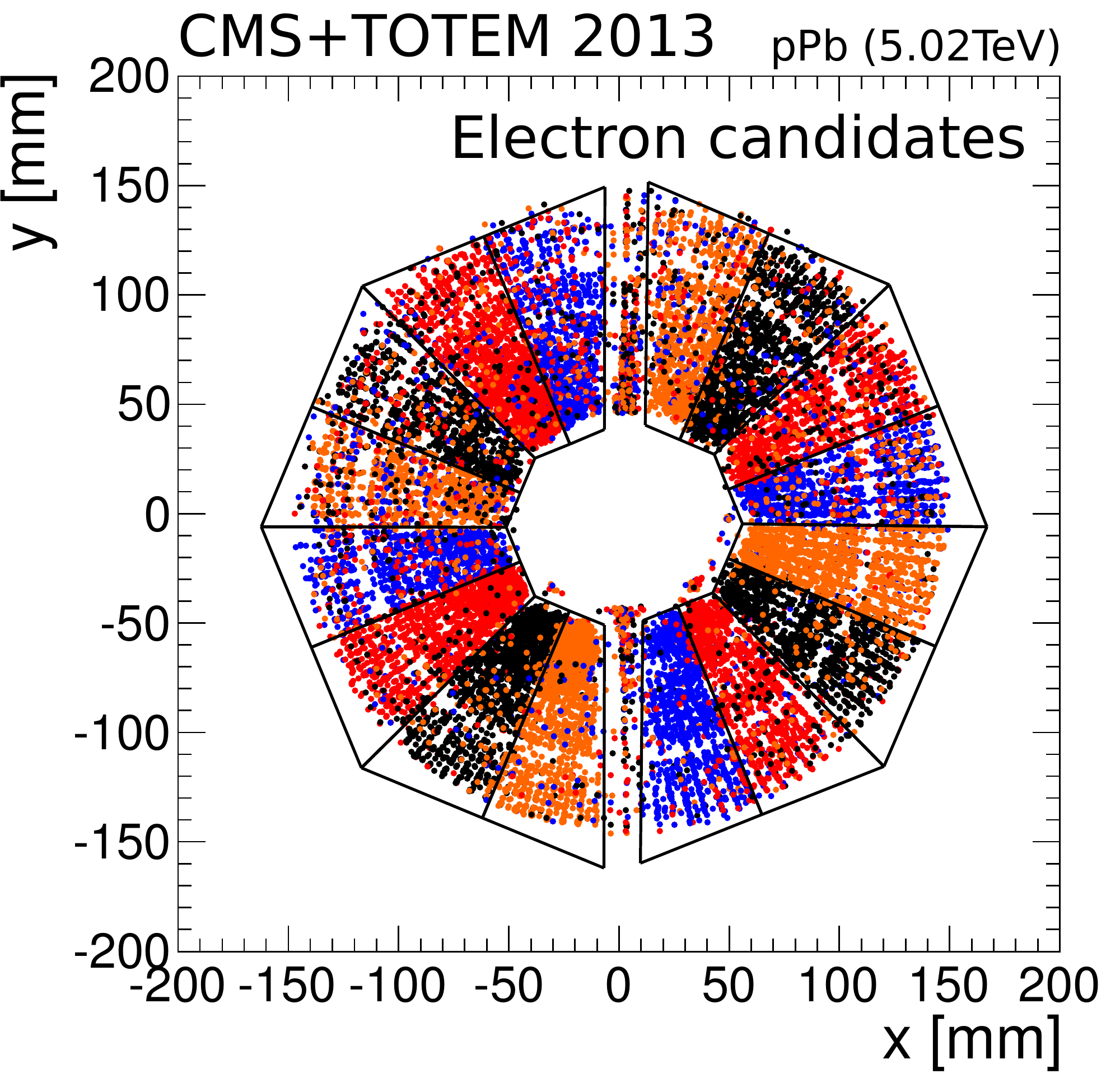}
  \caption{Left: Noise distribution of e.m.\ clusters in data used for the
  alignment of CASTOR with TOTEM T2 tracks, measured with a random trigger
  in {\Pp}Pb events with no collisions~\cite{CMS-DP-2014-014}.
  The RMS of the distribution is $<$1\GeV;
  the uncertainties are statistical only. 
  Right:
  Projected (x,y) coordinates of single isolated T2 tracks onto the CASTOR front surface~\cite{CMS-DP-2014-014}.
  Only tracks (electron candidates) that coincide with a single isolated e.m.\ cluster in CASTOR are plotted. 
  The points are colored alternatively depending on which calorimeter tower has registered the 
  deposited energy. The black contours indicate the optimal positions of the two halves of CASTOR as 
  derived from the analysis of these data. The empty ring at a radius of about 110\mm is caused by the 
  shadow of a conical beam pipe section.}
  \label{fig:t2tracks}
\end{figure}

At the same time, other independent methods to determine the position
of the calorimeter with LHC collision data have been implemented and
cross-checked with the above procedure. Firstly, during 2013 a special
trigger was developed to record single electrons in CASTOR during the
{\Pp}Pb data taking. The trigger made use of the TOTEM T2 tracking
station~\cite{Hilden:2009zz} in front of CASTOR (see Fig.~\ref{fig:photo2}).
The trigger decision
requires a very small track multiplicity in T2 and an isolated
electromagnetic cluster in the calorimeter.
The offline event selection requires a single e.m.\ cluster above 10\GeV
with all other towers being below 5\GeV in total energy, representing a very strong
isolation criterion as well as a very efficient hadron veto.
The analysis of these data
yielded a very accurate picture of the alignment relative to the T2
detector. The measured geometric correlation of T2 tracks with electromagnetic energy
clusters in CASTOR is shown in  Fig.~\ref{fig:t2tracks}~(\cmsRight).
The ring with reduced track density corresponds to the shadowing induced by the conical
beam pipe in front of T2 and CASTOR. The number of background events in this
plot becomes visually enhanced since the background is uniformly
distributed, whereas the signal is highly clustered on top of each
of the CASTOR towers.
The main source of this background is calorimeter noise, since the
energy threshold to select these electron events is 10\GeV per tower, whereas the calorimeter
noise tails contribute at the level of $<$10$^{-4}$ per event (Fig.~\ref{fig:t2tracks}, left).
It can also be seen that the acceptance of T2 is slightly smaller than that of CASTOR.
The geometric correlation between track and energy cluster is then used for an alignment fit of CASTOR.
From this alignment fit, we find that both detector halves are
separated by 22\mm, a situation that was not exactly known prior to this measurement.

Furthermore, minimum-bias collision data at $\sqrts=0.9$, 2.76, and 7\TeV have been compared to
corresponding detailed detector simulations. The energy distributions
in depth and azimuth depend on the position and tilts of the two
calorimeter halves. The position of CASTOR in simulations was
therefore optimized to best describe the distribution found in data.
There is agreement within the experimental uncertainties among the
different approaches.

\section{Detector simulation and validation} 
\label{sec:sim}

Typical particle physics measurements need detector simulations to
calculate the acceptance, correction factors, and bin migration
effects.  Having at hand a precise simulation and validation of the
CASTOR detector response is thereby of fundamental importance for
physics analyses. Both topics are discussed next.

\subsection{Simulation}

\begin{figure}[tpb!]
  \centering
  \includegraphics[width=0.49\textwidth]{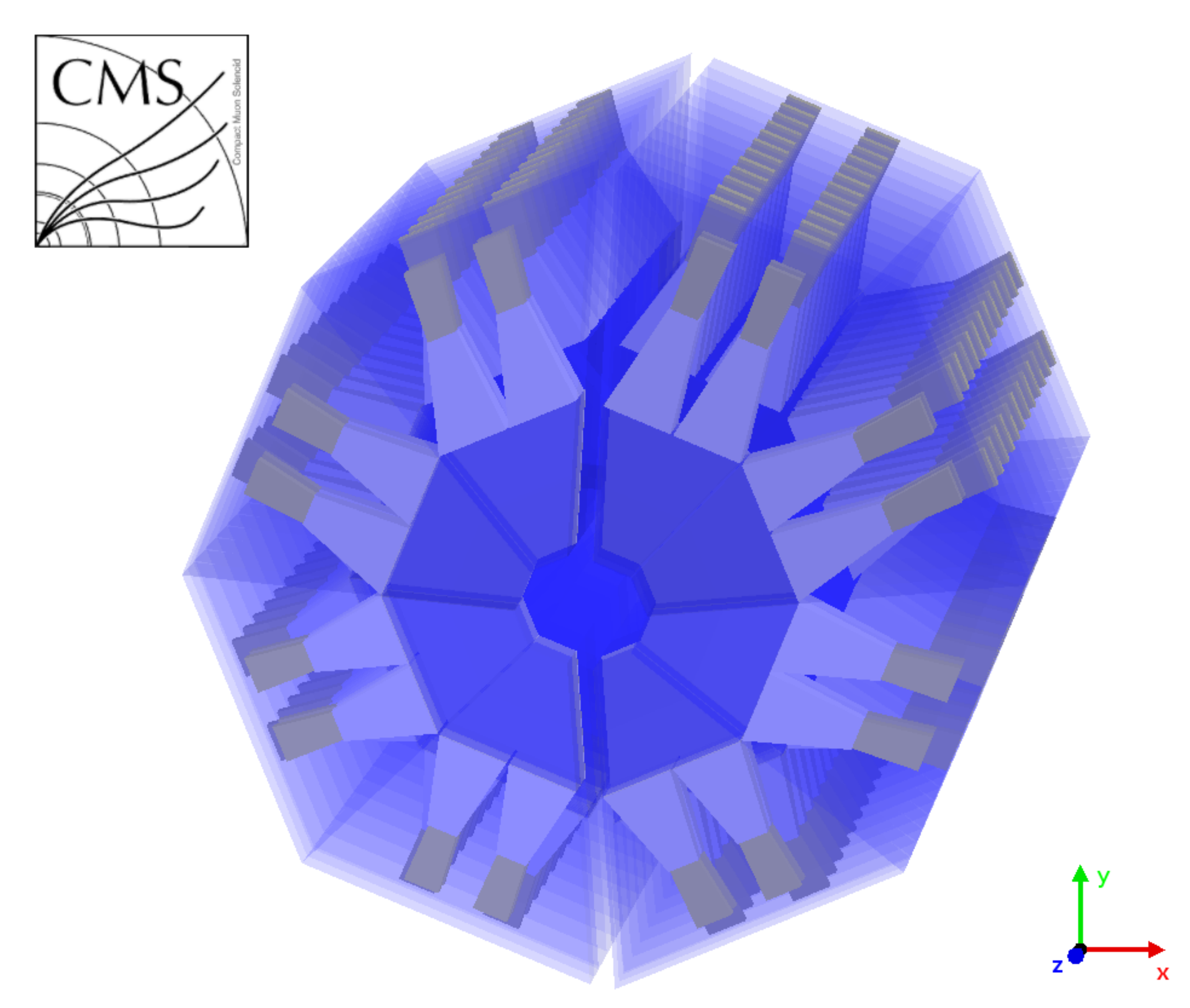}
  \caption{Visualization
  of the CASTOR detector geometry in simulations. The inner trapezoids
  in darker blue are the active volumes, while the outer volumes
  describe the light guides and the PMTs. The near and far halves are
  placed at the measured position corresponding to the data taking in
  2013. An opening of 2.2\unit{cm} along the $x$-axis is visible
  between the two halves.  \label{fig:cas_fitted_pos}}
\end{figure}

The detector simulation is fully integrated into the offline software
package of the CMS experiment CMSSW~\cite{Innocente:2001nx}.  The
particles produced by the event generator are transported to the
CASTOR detector using the \GEANTfour framework~\cite{Agostinelli:2002hh}.
Hereby, interactions with detector materials and the beam pipe in
front of the calorimeter, as well as the magnetic field produced by
the CMS solenoid that extends into the very forward region, are
simulated.  The geometry implementation of CASTOR is shown in
Fig.~\ref{fig:cas_fitted_pos}. The two halves of the calorimeter are
shown in the geometry corresponding to the 2013 {\Pp}Pb
data-taking period.  Electromagnetic and/or hadronic particle showers
are detected in the sensitive volumes of the detector.  The quartz and
tungsten plates that define the volume of the CASTOR detector are
accurately and separately modeled for each detector-half to
allow the description of tilts and movements of the individual halves.
Various physics settings have been compared to SPS beam
data~\cite{Andreev:2010zzb}.  For the test-beam data, the best
agreement with full \GEANTfour simulations of the detector is obtained
with the ``QGSP\_FTFP\_BERT\_EML'' physics setting~\cite{Apostolakis:2018ieg}, 
which gives a good description of the longitudinal shower profiles. However, it
underestimates the effect of noncompensation for hadronic
particles. Thus, a further residual scaling factor of about 0.8,
applied solely to the response generated by incoming hadrons, is
implemented in the simulation to match the measured test-beam data.

{\tolerance=800
The number of Cherenkov photons, $\NgamCh$, produced by
charged particles traversing the quartz plates is estimated by
considering Poissonian fluctuations and a conversion of
$\rd E/\rd x=1.24\unit{eV}/\unit{cm}$ of emitted energy
per unit distance,
in the relevant wavelength interval from 400 to 700\unit{nm}. The
photons are converted into charge by a simplified analytical model of
the photon transport in the quartz, light guides, and PMTs. The first
energy emission is described by an incident-angle-dependent
probability to reach the light guides, $\pdqz$. Combined
with the quantum efficiency of the PMT,
$\QE=19\%$, and the PMT response factor,
$\felec=0.307$, necessary to obtain a simulated signal
for electrons in agreement with the SPS beam measurement, one obtains
the number of photoelectrons as
\begin{linenomath}
 \begin{equation}
  \Npe=\NgamCh\,\QE\,\pdqz\,\felec~.
 \end{equation}
\end{linenomath}
The number of photoelectrons is then converted to charge according to 
\begin{linenomath}
 \begin{equation}
   Q = \Npe \, \gpmt \, ,
 \end{equation}
\end{linenomath}
with $\gpmt=4.01\unit{fC}/\mathrm{p.e.}$ assuming an
instantaneous signal processing (no pulse shape) and, thus, only
detected over one 25\unit{ns} readout bin. Since data taking with the
CASTOR detector is performed with low-intensity beams and large
spacing between filled bunches, this assumption is sufficient. With
these values for the factors $\gpmt$ and
$\felec$ the simulated response of CASTOR is
consistent with the measured energy of incoming individual electrons
in beam tests.\par}

\begin{figure}[tpb!]
\centering
\includegraphics[width=.49\textwidth]{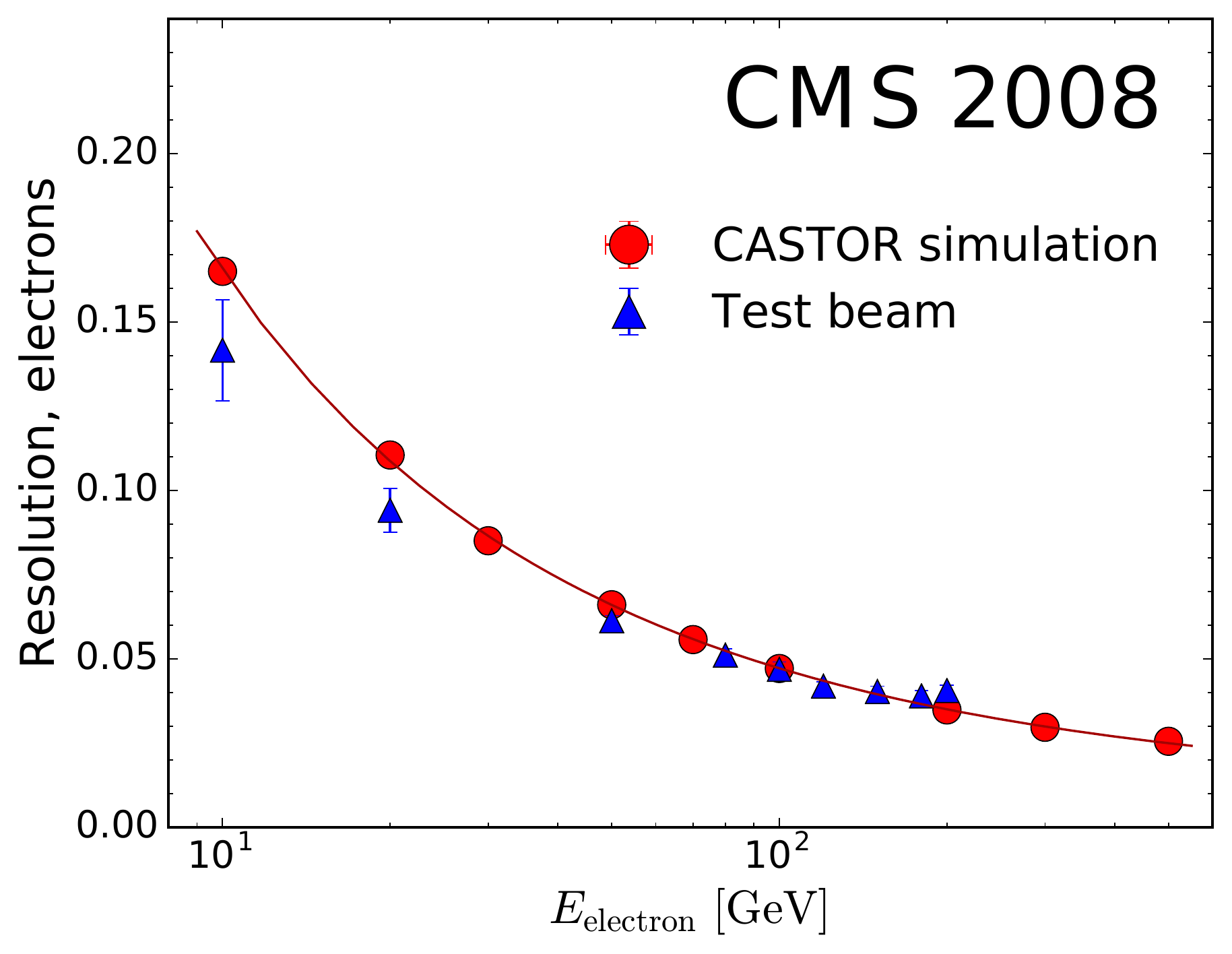}~
\includegraphics[width=.49\textwidth]{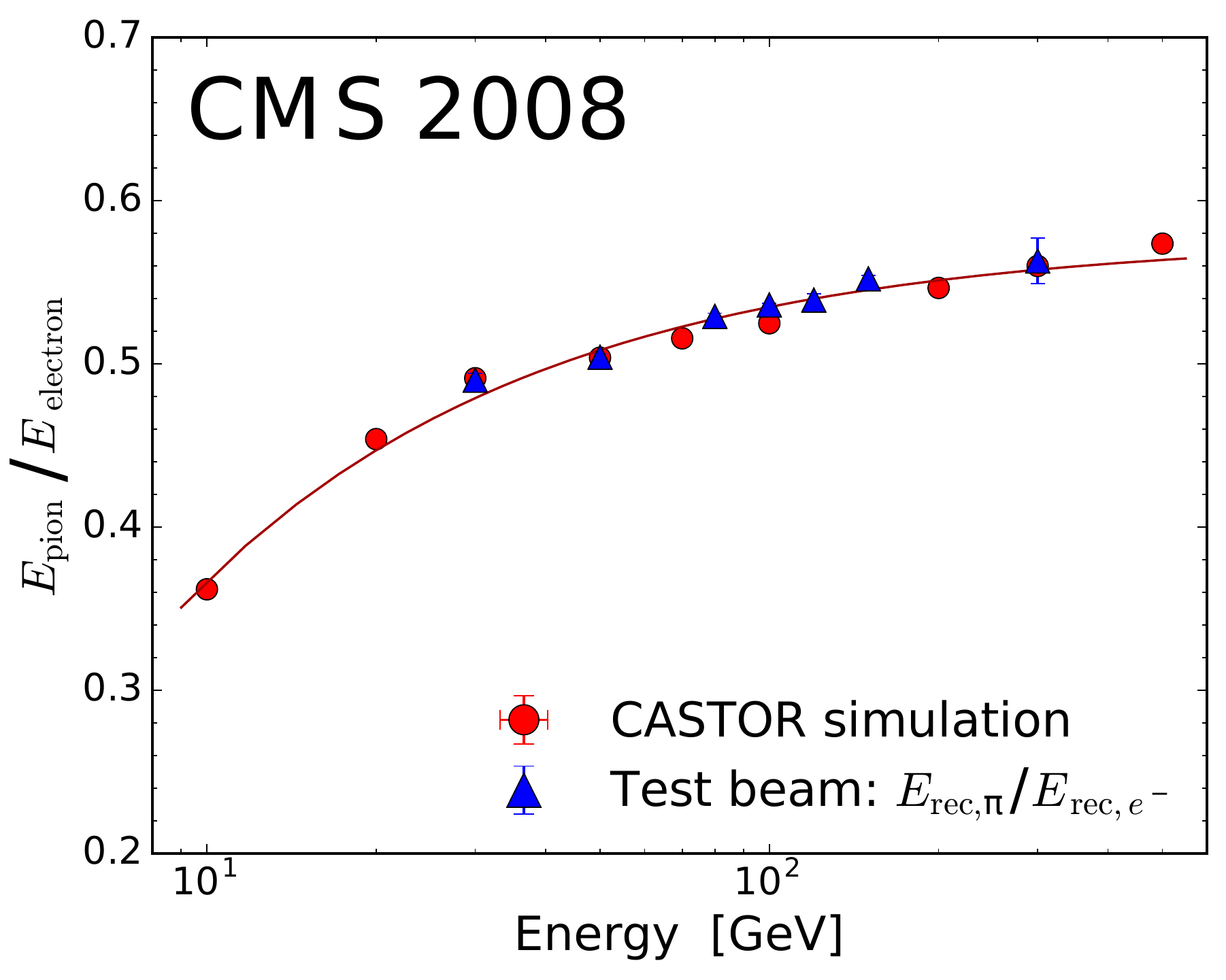}
\caption{Comparison of detector simulations (circles) to test-beam
  measurements (triangles). Resolution of electron energy reconstruction
  (\cmsLeft), and observed charged pion/electron response ratio (\cmsRight) as a
  function of test-beam energy. The uncertainties plotted are statistical only,
  and the fitted curves are shown to guide the eye. 
\label{fig:testbeam}}
\end{figure}

\begin{figure}[tbp]
\centering
\includegraphics[width=0.47\textwidth]{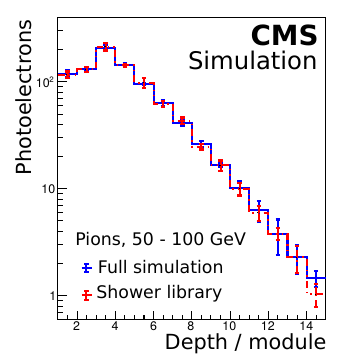}~
\includegraphics[width=0.47\textwidth]{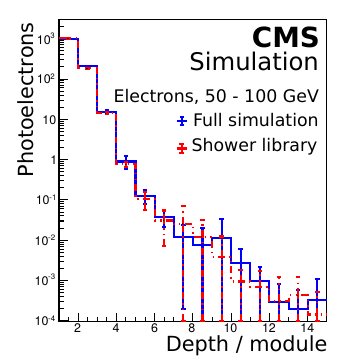}
\caption{Comparison of the full (blue histogram) and fast (red histogram)
         simulation responses (number of photoelectrons as a function
         of detector depth) for charged pion (\cmsLeft) and electron (\cmsRight)
         showers penetrating CASTOR with energies from 50 to 100\GeV.
         Error bars around data points indicate statistical
         uncertainties.}
         \label{fig:sl_validation}
\end{figure}

In Fig.~\ref{fig:testbeam}, the simulation of the CASTOR standalone
detector is compared to test-beam data for electrons (\cmsLeft) and charged pions
(\cmsRight), illustrating the good agreement between data and simulations.
In CMS detector simulations, the required time per generated event is
an important factor, since the available computing resources are
limited. For CASTOR, such computing constraints are important since
particles and jets with very high energies, close to that of the beam,
need to be simulated. The subsequent large and complex hadronic
showers developing in the calorimeter require significant computing
time to be simulated in full detail.  To increase the performance, a
fast version of the simulations has been implemented using a tabulated
shower library.  Instead of tracking each individual particle, the
energy deposit of particles in each readout channel is tabulated in 37
inequally spaced bins in energy ($1\le E(\GeV)\le1500$), five equally
spaced bins in azimuth angle ($0\le \phi\le\pi/4$), and seven equally
spaced bins in pseudorapidity. About 50 events are tabulated per phase
space bin. In Fig.~\ref{fig:sl_validation} the validation of the
shower library with respect to full simulations for the energy bins
from 50 to 100\GeV is shown.
When a particle that is tracked in \GEANTfour is found
to enter the CASTOR volume and fulfills basic requirements, such as a
minimum energy of 1\GeV or a limited incident angle, it is replaced by
a randomly chosen tabulated shower of its corresponding phase space
bin. This results in a faster simulation time by factors of 3 to 25 (2
to 16) for charged pions (electrons) at energies from 1.0 to 1.5\TeV.  A
disadvantage of the shower library is that the geometry of CASTOR
cannot be changed or shifted for systematic studies, since this would
require new shower libraries. Thus, for dedicated systematic studies
the full simulations are required.

\subsection{Validation}

\begin{figure}[tpb!]
\centering
\includegraphics[width=.6\textwidth]{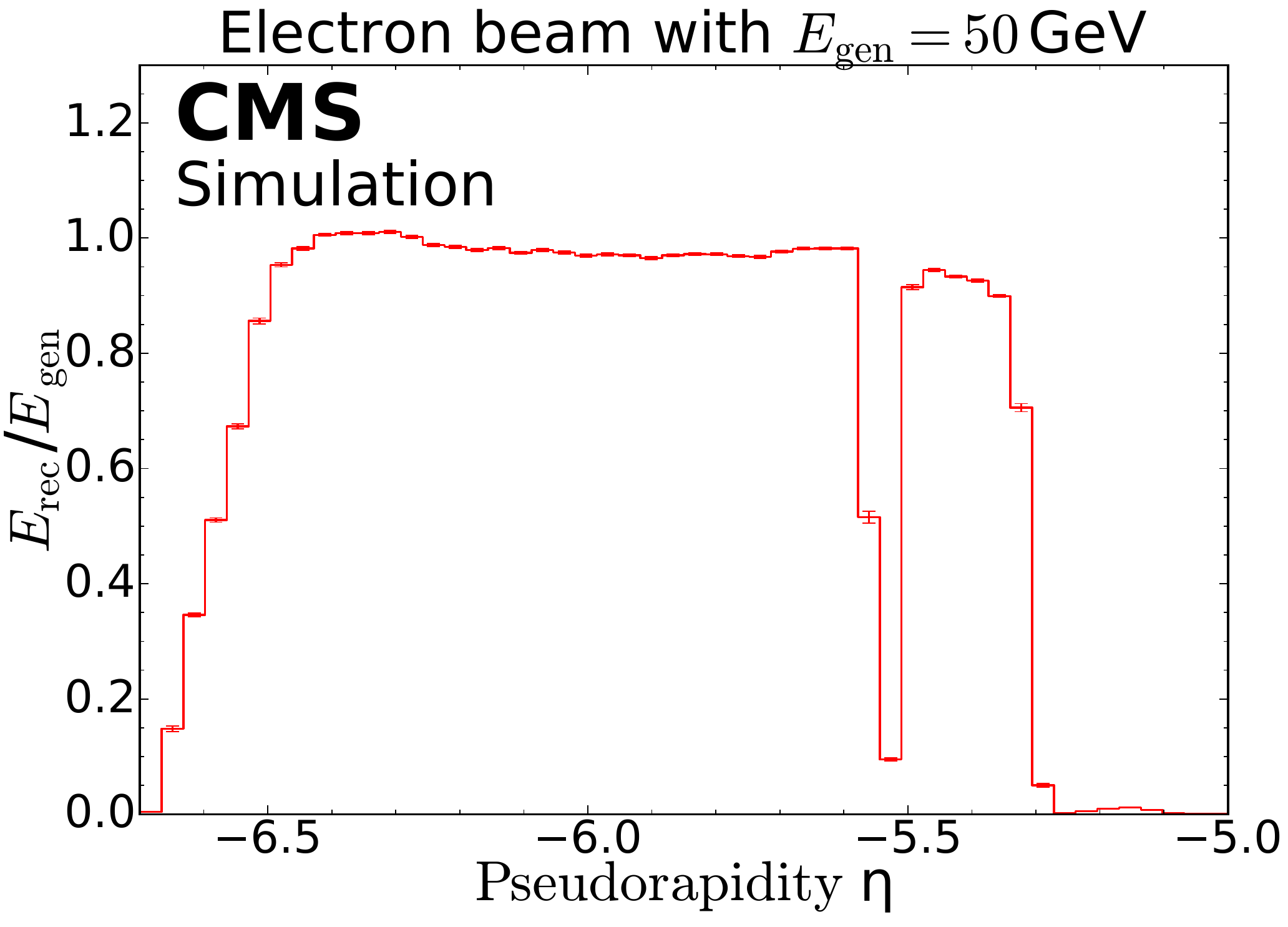}
\caption{Pseudorapidity dependence of reconstructed energies
  in CASTOR for 50\GeV electrons
  emitted from the nominal CMS interaction point.  The CASTOR
  simulation includes all CMS passive and active materials and the
  beam pipe.  The dip around $\eta=-5.5$ corresponds to the
  projection of the conical beam pipe.
\label{fig:castor_eta}}
\end{figure}

The quality of the simulations of proton and nuclear collisions in the
very forward direction at the LHC, dominated by semihard and
nonperturbative dynamics (MPI, beam remnants, \etc), cannot be
cross-checked with first-principles calculations.  The event
generator predictions in this phase space region are purely
phenomenological in nature, and the description of the measurements
depends on MC parameter tuning, which in the very forward direction
essentially corresponds to data extrapolations.  Different simulated
event characteristics need to be consistently compared to the measured
data to cross-check their validity. The simulations can be obtained
from a variety of MC generator predictions to determine which ones
provide a better description of the data in the relevant phase space.
The observed discrepancies in detailed comparisons between simulated
and real detector responses can be used to estimate the uncertainty
related to the understanding of the detector response. Also, the level
of model dependence of a specific process can be used to estimate the
related systematic uncertainties in physics measurements.  The
integration of CASTOR inside CMS has a further impact on the response
of the calorimeter, due to the additional scattering of particles in the
materials of the CMS detector around it.
The simulated response to single electrons as a
function of pseudorapidity, is shown in Fig.~\ref{fig:castor_eta}. The
dip around $\eta\approx-5.5$ is due to the projection of the conical
beam pipe towards the interaction point.

\begin{figure}[tpb!]
\centering  
\subfloat[Proton-lead collisions at $\sqrtsNN=5.02\TeV$
          recorded in 2013 and 2016 at $B=3.8\unit{T}$. ]{
  \includegraphics[width=.49\textwidth]{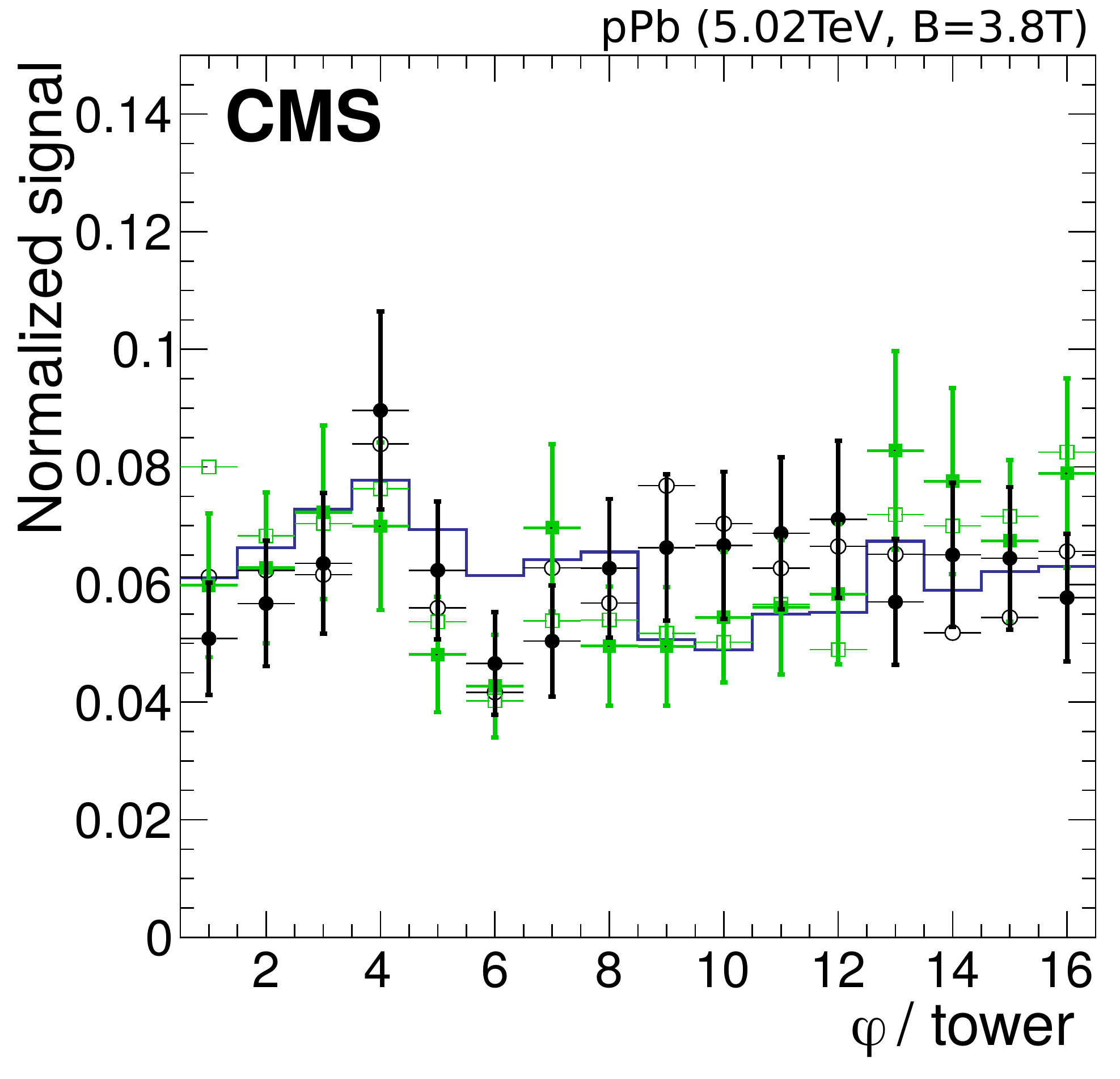}
  \includegraphics[width=.49\textwidth]{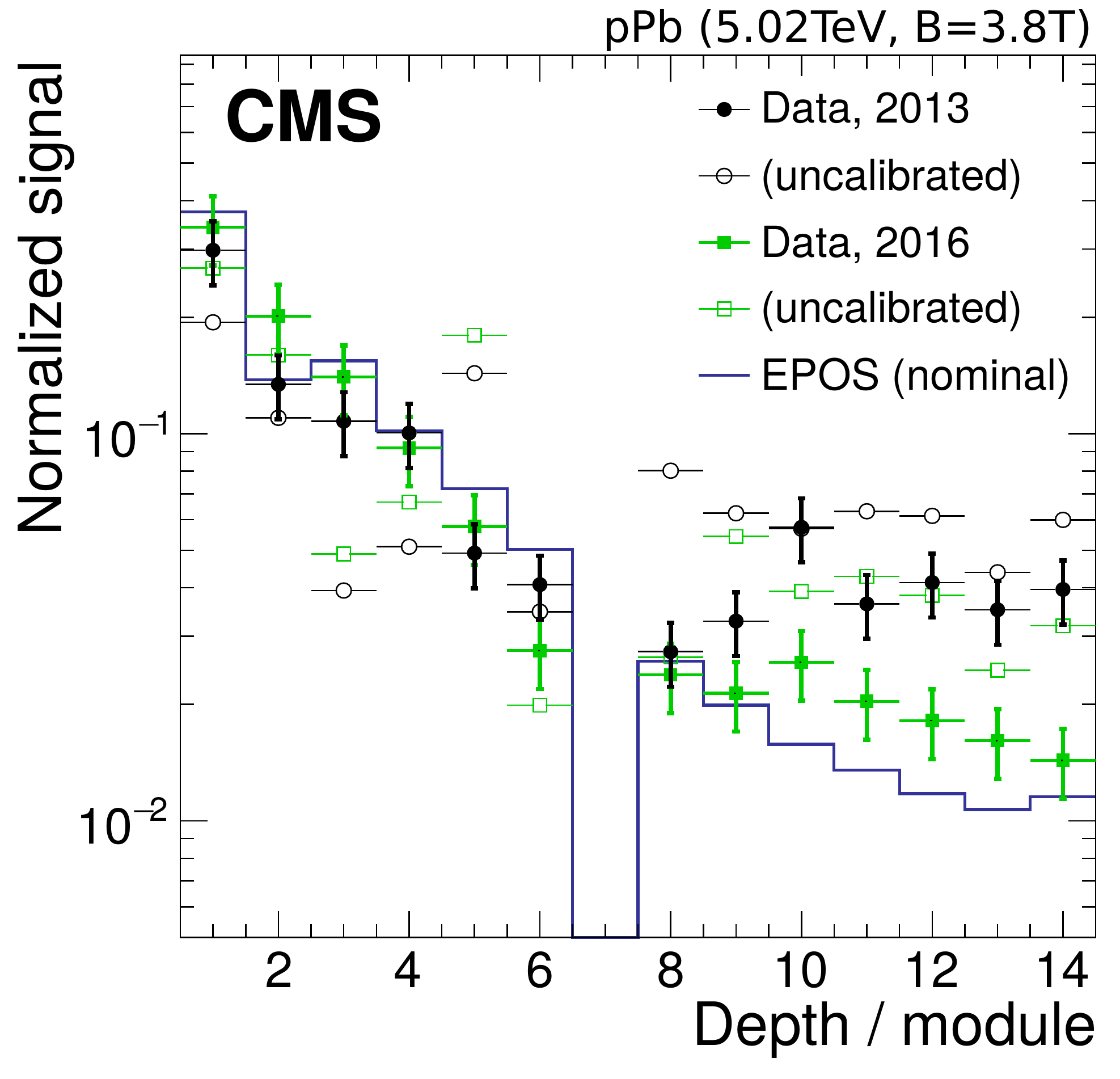}
  }\\
\subfloat[Proton-proton collisions at $\sqrts=13\TeV$ from 2015
          at $B=0\unit{T}$.]{
  \includegraphics[width=.49\textwidth]{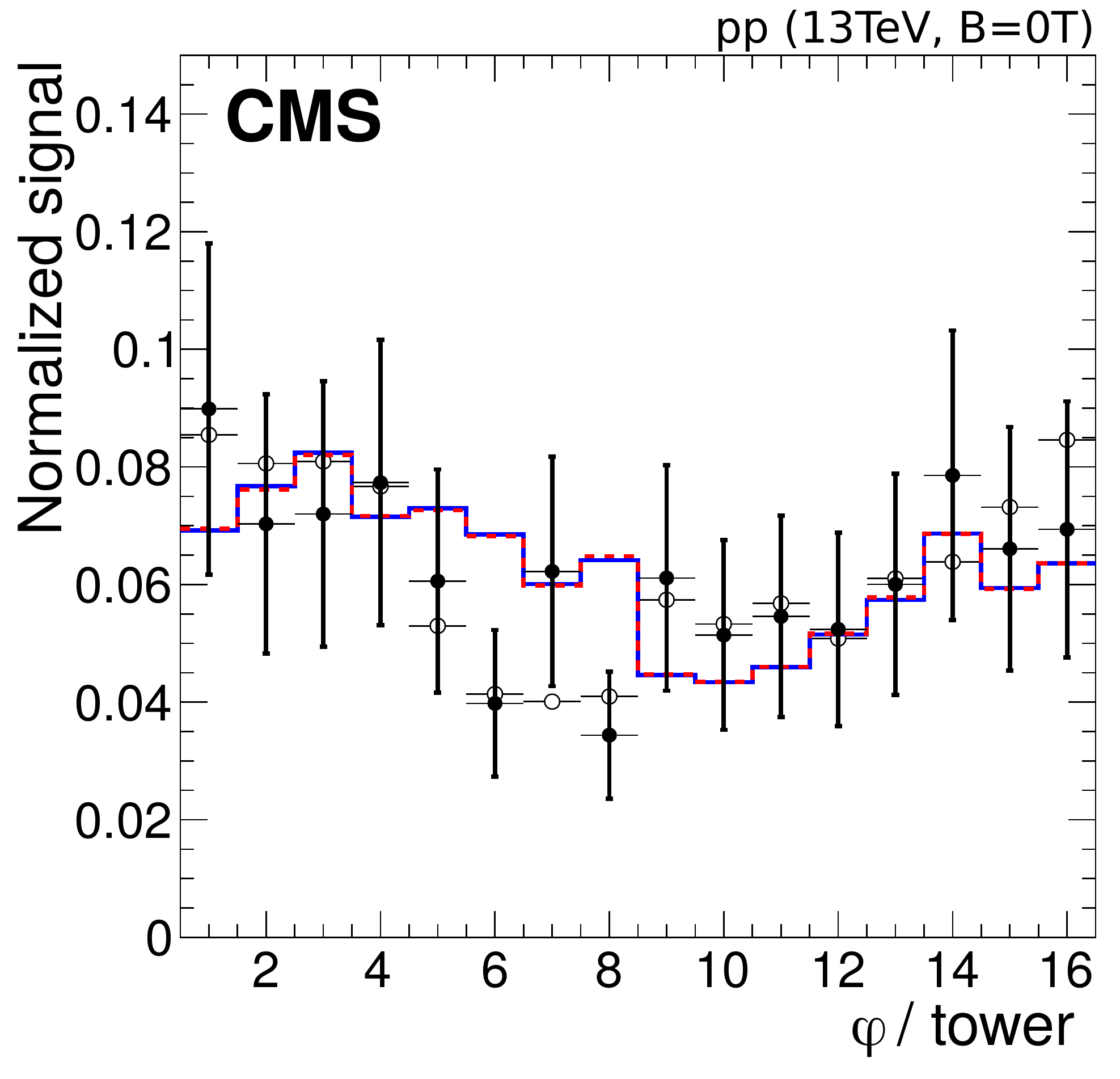}
  \includegraphics[width=.49\textwidth]{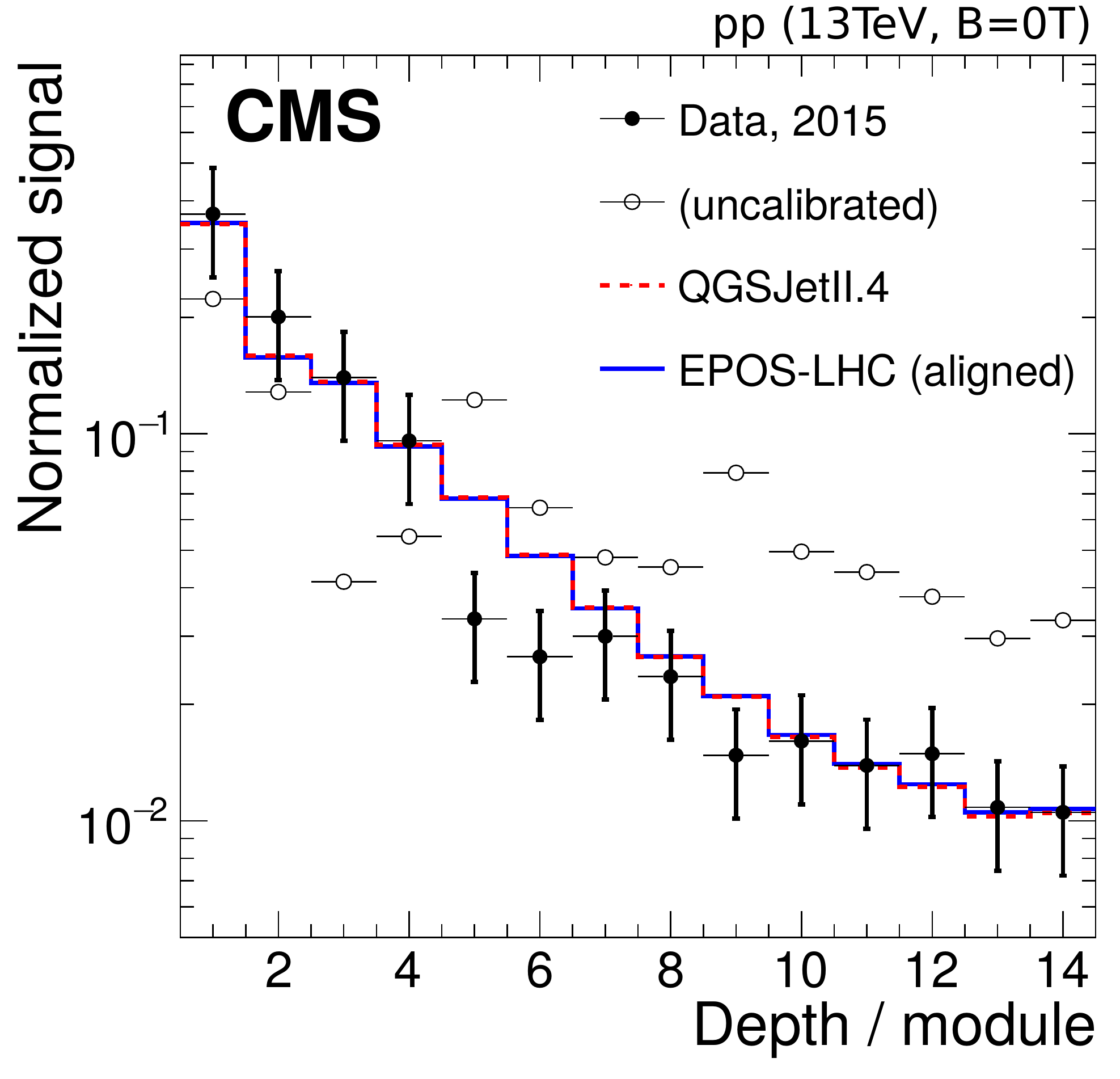}
  }
\caption{CASTOR signal response in calibrated (filled symbols) 
  and uncalibrated (open symbols) data compared to the detector
  simulations (histograms), projected into the azimuthal (\cmsLeft) and
  longitudinal (\cmsRight) direction. Upper: Comparison of minimum-bias {\Pp}Pb
  collision data recorded in 2013 and 2016 with the \textsc{Epos-LHC}
  simulations (with the nominal MC geometry). The impact of the
  magnetic field around module
  7 is clearly visible. 
  Lower: Comparison of the minimum-bias {\Pp}{\Pp} collision data at
  $\sqrts=13\TeV$ to the \textsc{QGSJetII} and \textsc{Epos-LHC} (with
  geometry tuned to the 2015 data alignment) simulations. 
  The CMS magnet was off while taking those data.
  The uncertainties of the calibrated data in all plots
  are statistical only, and originate almost entirely from the limited
  beam halo statistics for intercalibration. There is an additional
  scaling factor of 1.25 on the uncertainties of the calibrated data,
  as discussed in the text.
\label{fig:validation}}
\end{figure}

Figure~\ref{fig:validation} presents several comparisons of
minimum-bias data, collected in {\Pp}Pb (upper panels) and {\Pp}{\Pp} (lower panels)
collisions at the LHC, to detector simulations. Minimum-bias events
are selected that have at least one HF tower with more than 10\GeV of
energy on both sides of CMS, and exactly one reconstructed collision
vertex. The open symbols show the data before calibration, while the
filled symbols show them after normalization to the beam halo muon
response.  Bad channels are removed in data as well as in simulations.  The
data and MC distributions are absolutely normalized in order to
perform a shape-sensitive comparison with the aim to study the
performance of the intercalibration procedure.  This is done since the
average energy per collision is a highly model-dependent quantity and
is not necessary for the understanding of the
intercalibration. Furthermore, we exploit that the underlying physics
at the LHC is, on average, symmetric in the azimuth angle. However, a
small sinus-shaped modulation in $\phi$ is seen in the simulations in
Fig.~\ref{fig:validation}~(\cmsLeft) that is due to the known relative
alignment of the LHC luminous region (``beam spot'') to the CASTOR
calorimeter. Some parts of the detector are a bit closer in the
transverse plane to the luminous region than others. In the very
forward region, this represents a sizable effect on the detector
acceptance. Such a modulation is also observed in the data. After the
full calibration of collision data with beam halo muons, the
statistical agreement of the azimuthal distributions between data 
and realistic simulations of the detector response is only fair.
This suggests that the point-to-point uncertainties are underestimated, possibly
because of additional unspecified sources of statistical and systematic uncertainties.
This is corrected by 
scaling the error bars to yield $\chi^2/N_\textrm{d.f.}=1$ 
(where $N_\textrm{d.f.}$ are the number of degrees of freedom)
in the azimuthal shape distributions with respect to the realistic full detector
simulations (with events generated with the \textsc{Epos-LHC} model). 
We determine a residual additional calibration
uncertainty at the tower level of about 25\%, corresponding to the
shown vertical bars.  This uncertainty on the tower level translates into
a possible impact on the energy scale normalization of the full CASTOR of
the order of 25\%/$\sqrt{16}\approx6.5\%$, consistent with
alignment-related uncertainties in the absolute energy scale.  Reasons
for such residual mismatches can be numerous and cannot be
disentangled from the available data. Possible reasons include
differences in the signal generation between beam halo muons (as used
for the calibration) and minimum-bias collisions, statistical
fluctuations, geometrical as well as mechanical effects, etc. For any specific
physics analysis of the data, these effects must be studied
and quantified explicitly.

The projection in the longitudinal direction indicates a very good
calibration from the front towards the rear of the calorimeter over
almost two orders of magnitude in signal strength, showing that the
calibration with the beam halo muon response yields a good measurement
of the shower shape. The uncertainties shown in the shape of the
longitudinal projected data are scaled with the same factor as
described in the previous paragraph. 

In the proton-lead data, there is some data-MC discrepancy in the rear
part of the calorimeter, which is due to known alignment issues in
2013 relative to the LHC beam pipe, affecting the rear part of
CASTOR. Thus, the data in modules 8 to 14 recorded in 2013 must be
treated with special care. Dedicated detector simulation studies are
needed to yield the required accuracy for physics analysis.

It is also important to realize that the detailed shower development
in the CASTOR calorimeter is very sensitive to the underlying physics
processes under study and cannot be assumed to be well known at such
high energies. Thus, the comparisons in depth are only indicative and
a $\chi^2$ test would not be a good measure of the detector
performance. For physics measurements that use CASTOR data, the
uncertainties in the calibration procedure must be specifically
studied and quantified.

\subsection{Analysis backgrounds and noise levels}

There are two typical applications where the CASTOR detector is
used for event classification measurements. First, tag events with
forward particle production to provide the best possible
acceptance for inelastic collisions~\cite{Sirunyan:2018nqx}. 
Second, tag events with no activity that are related to processes
with large rapidity gaps~\cite{Khachatryan:2015gka}.  The ability to
identify these conditions are both related to the noise levels in the
calorimeter. Noise levels were already discussed at the level of
single readout capacitors per channel as a tool of data quality
control in Section~\ref{section:NoiseandBaseline}. For physics
analysis, this study must be extended (in units of energy) to account
for noise effects in the reconstruction of calibrated physics objects.

\begin{figure}[htpb!]
  \centering
  \includegraphics[width=0.49\textwidth]{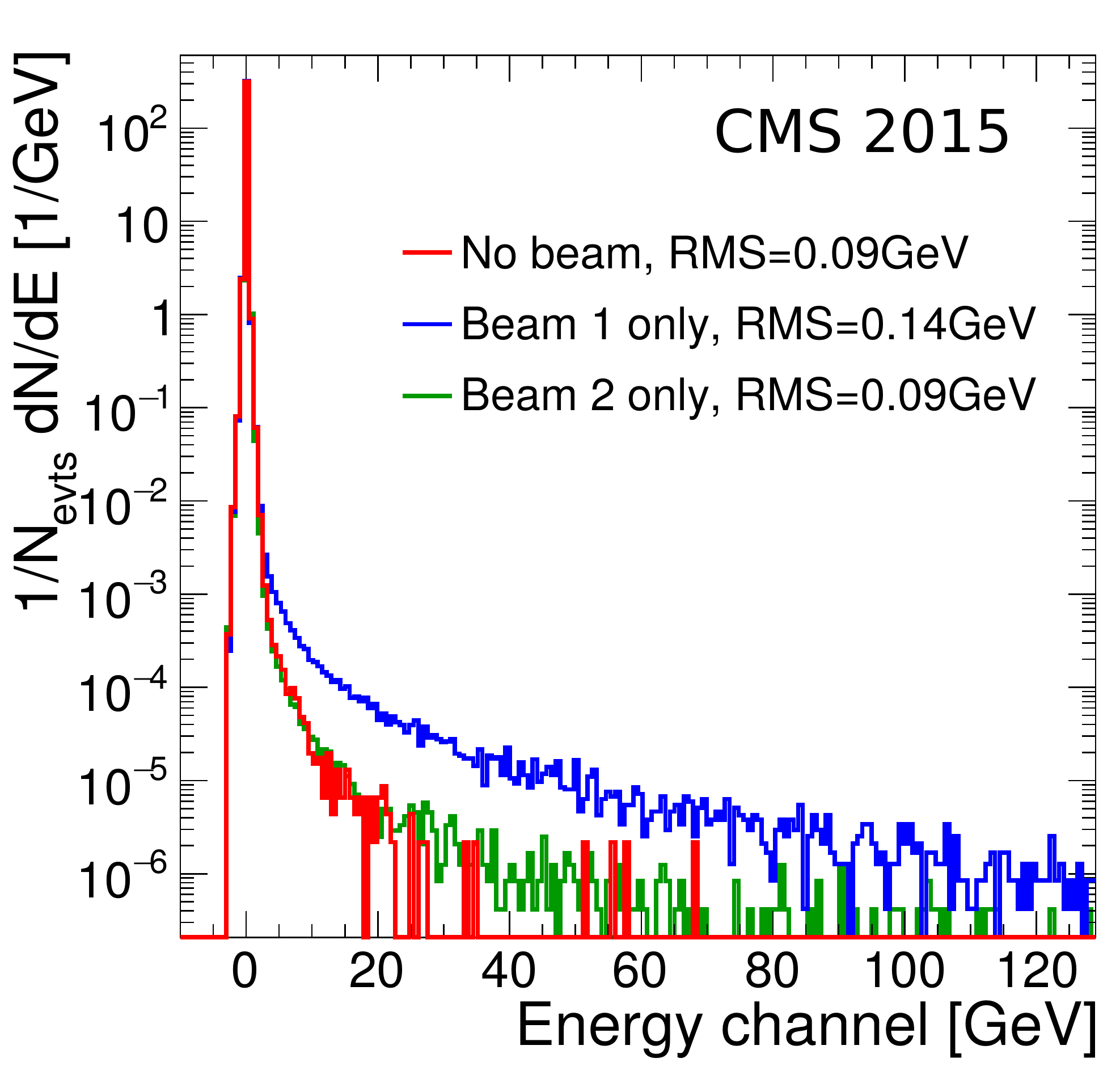}~
  \includegraphics[width=0.49\textwidth]{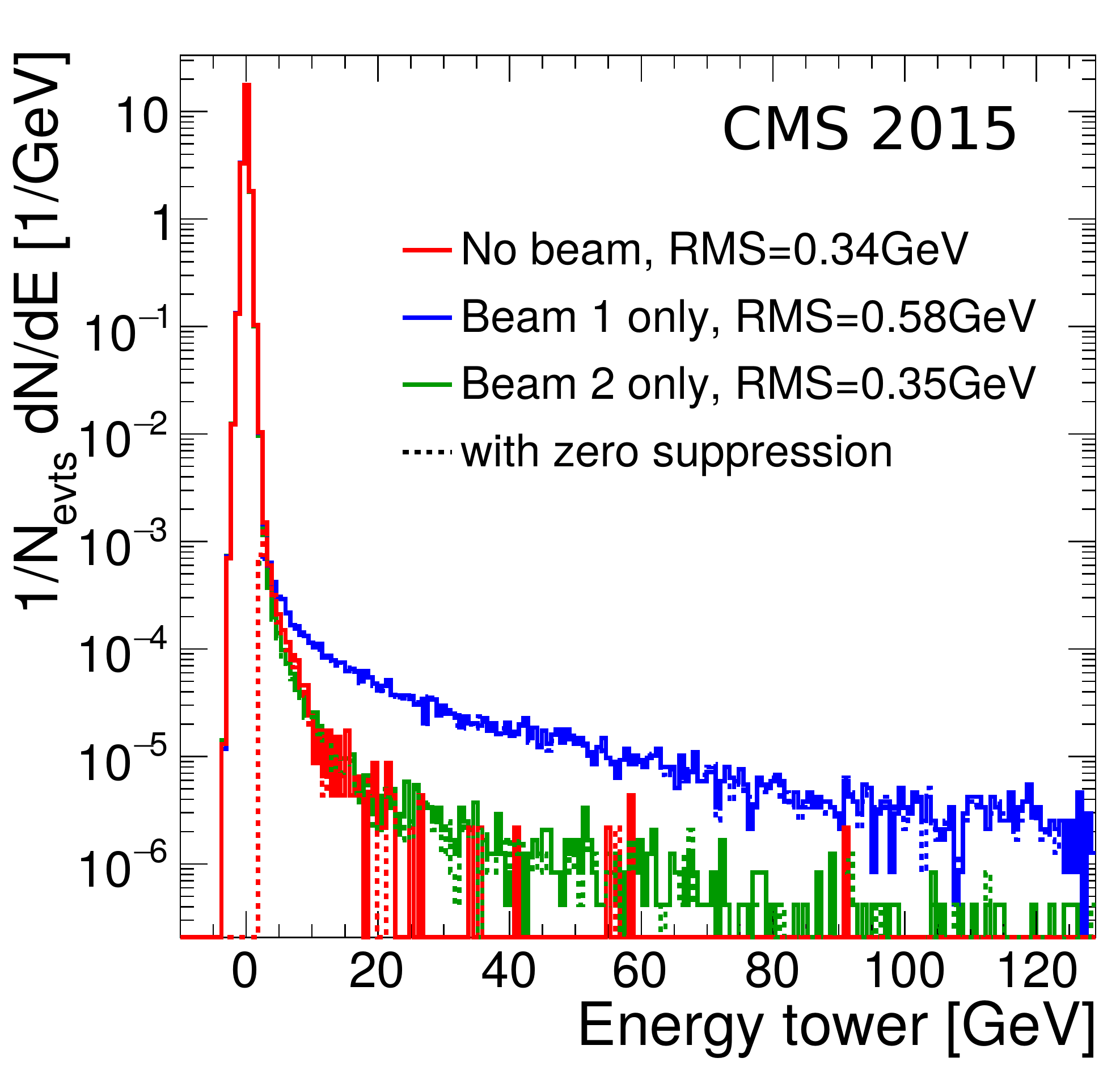}\\
  \includegraphics[width=0.49\textwidth]{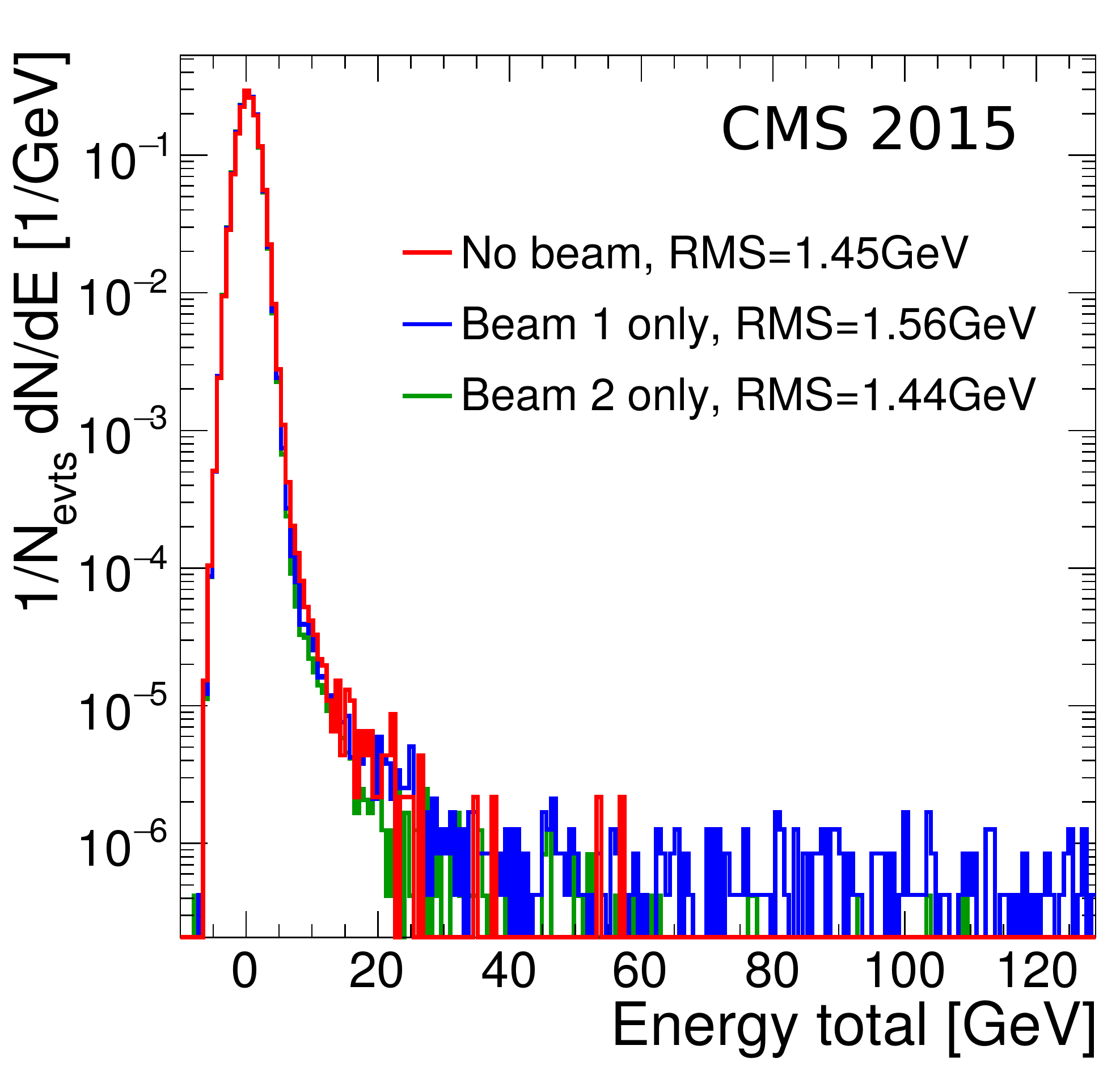}~
  \includegraphics[width=0.49\textwidth]{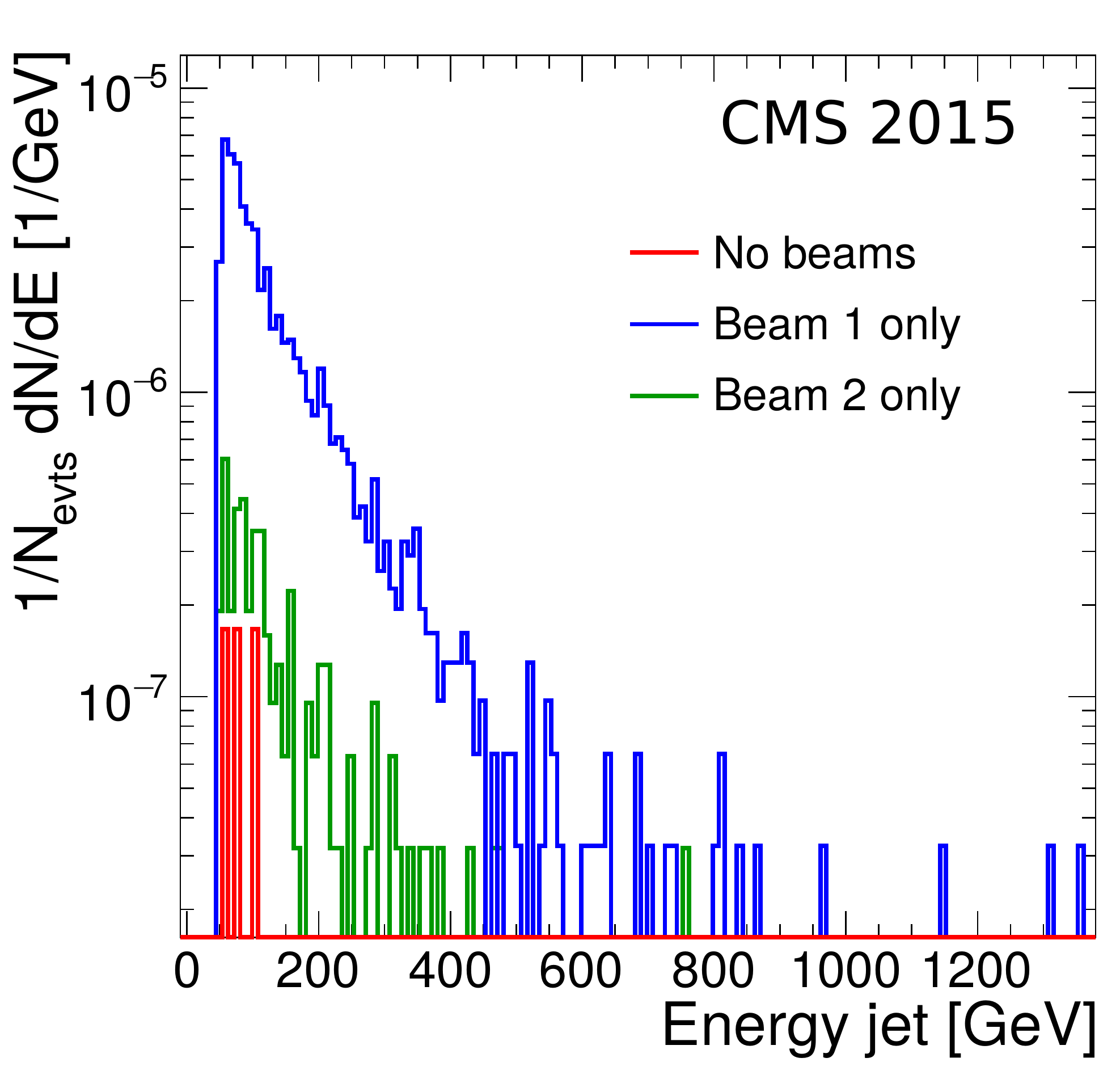}
 \caption{Measured
  CASTOR noise distributions for reconstructed physics objects at the
  channel (upper left), tower (upper right), full-CASTOR (lower left),
  and jet (lower right) levels.  The data are from June 2015 during
  proton-proton collisions at $\sqrts=13\TeV$. Towers are normally
  zero suppressed although, here, towers both with (solid) and without
  (dashed) zero suppression are shown. Since jets are zero suppressed,
  the distribution is non-Gaussian, and no RMS values are reported in this case.
\label{fig:noise}}
\end{figure}

Noise studies are best performed with data taken in periods without
LHC collisions, or alternatively with dedicated triggers for event
selection. In Fig.~\ref{fig:noise}, we show the noise levels per
channel~(upper left), per tower~(upper right), and total in CASTOR~(lower
left), as well as the rate of energy deposits misidentified as jets (lower
right).  These are examples obtained from data recorded in 2015 in
proton-proton collisions at $\sqrts=13\TeV$ with a CMS trigger requiring the
absence of at least one of the two LHC beams per recorded bunch
crossing.
Because of the length of the pulse shape, particles from immediately preceding collisions
can produce signals in the bunch crossing of interest. Hence, to avoid noise
from out-of-time pileup, the data from bunch crossings that are closer in time
than 250\unit{ns} to any bunch of the nominal LHC filling scheme are excluded.
The leakage of signal just outside
the nominally colliding bunches is important to consider for LHC
operations with small bunch separations. Some CASTOR data were
recorded with bunch separation as small as 50\unit{ns}.  Extra care
must be taken in such scenarios, although the overall performance does
not change in a significant way.  Readout noise is seen in the no-beam
data, while beam-related backgrounds can be studied with one of the
beams present.  Since CASTOR is installed in the negative-$\eta$ side of the
CMS experiment, the noise and timing of backgrounds from each LHC beam
is different. Only the LHC beam 1 has a direction that can contribute
to backgrounds for physics analyses, since the beam background
will arrive timed-in with the secondaries from actual collisions at the IP.
Recording and studying such data
is very important for many physics analyses that are sensitive to
noise and background levels.  The noise level in units of energy
mainly depends on the PMT gain settings, the channel intercalibration,
and the absolute energy scale calibration. However, the main
contribution is electronics noise from the cables and the digitizers,
and is very stable.

For {\Pp}{\Pp} collision HV settings, the typical analysis energy
thresholds for physics measurements are about 0.4\GeV per channel,
1.5\GeV per tower, around 5\GeV for the full CASTOR detector, and a few
hundred \GeV for jets. These values correspond roughly to the measured
$3\,\sigma$ noise levels.  The noise probability per event drops
significantly faster towards the higher energy thresholds than
observed in the jet rates from particle collision data
(as seen by comparing Fig.~\ref{fig:noise}, lower right, with
Fig.~\ref{fig:jettrigger}, right).  Even at the lowest energy
thresholds, the fraction of noise in the reconstructed jet sample
never exceeds $10^{-3}$.  The typical \pt of jets at the
lower threshold of about 500\GeV is about
$500\GeV/\cosh(-6.0)\approx2.5\GeV$. All these results confirm that
the small noise levels in CASTOR make it a suitable detector for
full cross section measurements, for event type classification, as
well as for jet reconstruction studies.

\section{Summary} 
\label{sec:summ}

The physics motivation, detector design, triggers, calibration,
alignment, simulation, and overall performance of the CASTOR
calorimeter project have been described.  The detector, located in an
extremely challenging environment at very forward rapidities in the
CMS experiment, has proven to be a valuable apparatus for multiple
physics measurements in proton and nuclear collisions at the CERN LHC.
The physics motivations for the detector construction, as well as
measurements carried out during the Run 1 and 2 operations at the LHC,
have been outlined.  The overall detector design has been reviewed,
emphasizing the features that allow its operation under a significant
radiation exposure, required fast optical response, high occupancy,
magnetic fields, and severe space constraints.

The calibration of CASTOR is a major challenge. 
Three different methods (offbeam, onbeam, and LED pulsers) used to set
the gain correction factors of the CASTOR photosensors have been reviewed.
The
absolute energy scale is determined with a total precision of about
17\%, which represents the
dominant source of systematic uncertainty for many measurements. 
Despite the difficulties to precisely determine the absolute
energy scale of the calorimeter, beam halo muons have proven a
valuable tool to intercalibrate all channels, to monitor the
stability of the detector, and to determine the noise levels.  The
alignment of CASTOR, which has an important impact on the calorimeter
energy scale, data-simulation agreement, and on the actual detector
acceptance, has been surveyed with two types of sensors, and
cross-checked with measurements taken concurrently with the TOTEM T2
tracker in front of CASTOR. An alignment precision of the order of
1\mm was achieved.

Different comparisons of the data to Monte
Carlo simulations confirm that the reconstructed objects can be used
for various physics analyses, from event-type classification to
more advanced measurements (jets, single-particle spectra, etc.).
Since CASTOR covers a unique phase space at the LHC, and since there
are no future plans for the installation of a detector with comparable
performance, the  exploitation of the recorded data can
provide many opportunities for interesting measurements in proton and
nuclear collisions at the LHC.

\begin{acknowledgments}

\hyphenation{
Bundes-ministerium Forschungs-gemeinschaft Forschungs-zentren
Rachada-pisek}

We congratulate our colleagues in the CERN accelerator departments for the excellent performance of the LHC and thank the technical and administrative staffs at CERN and at other CMS institutes for their contributions to the success of the CMS effort. In addition, we gratefully acknowledge the computing centers and personnel of the Worldwide LHC Computing Grid for delivering so effectively the computing infrastructure essential to our analyses. 
Moreover, the CASTOR detector project received important support
and explicit funding from the Helmholtz Association (Germany) via
grants HN-NG-733 and the joint Helmholtz-Russian grant HRJRG-002,
as well from the ``A. G. Leventis''
Foundation (Greece).
We are also grateful to our Colleagues from the TOTEM Collaboration for
the combined data taking and the fantastic work on the T2+CASTOR
combined data analysis. 
Finally, we acknowledge the enduring support for the construction and operation of the LHC and the CMS detector provided by the following funding agencies: the Austrian Federal Ministry of Education, Science and Research and the Austrian Science Fund; the Belgian Fonds de la Recherche Scientifique, and Fonds voor Wetenschappelijk Onderzoek; the Brazilian Funding Agencies (CNPq, CAPES, FAPERJ, FAPERGS, and FAPESP); the Bulgarian Ministry of Education and Science; CERN; the Chinese Academy of Sciences, Ministry of Science and Technology, and National Natural Science Foundation of China; the Colombian Funding Agency (COLCIENCIAS); the Croatian Ministry of Science, Education and Sport, and the Croatian Science Foundation; the Research and Innovation Foundation, Cyprus; the Secretariat for Higher Education, Science, Technology and Innovation, Ecuador; the Ministry of Education and Research, Estonian Research Council via PRG780, PRG803 and PRG445 and European Regional Development Fund, Estonia; the Academy of Finland, Finnish Ministry of Education and Culture, and Helsinki Institute of Physics; the Institut National de Physique Nucl\'eaire et de Physique des Particules~/~CNRS, and Commissariat \`a l'\'Energie Atomique et aux \'Energies Alternatives~/~CEA, France; the Bundesministerium f\"ur Bildung und Forschung, the Deutsche Forschungsgemeinschaft (DFG) under Germany's Excellence Strategy -- EXC 2121 ``Quantum Universe" -- 390833306, and Helmholtz-Gemeinschaft Deutscher Forschungszentren, Germany; the General Secretariat for Research and Technology, Greece; the National Research, Development and Innovation Fund, Hungary; the Department of Atomic Energy and the Department of Science and Technology, India; the Institute for Studies in Theoretical Physics and Mathematics, Iran; the Science Foundation, Ireland; the Istituto Nazionale di Fisica Nucleare, Italy; the Ministry of Science, ICT and Future Planning, and National Research Foundation (NRF), Republic of Korea; the Ministry of Education and Science of the Republic of Latvia; the Lithuanian Academy of Sciences; the Ministry of Education, and University of Malaya (Malaysia); the Ministry of Science of Montenegro; the Mexican Funding Agencies (BUAP, CINVESTAV, CONACYT, LNS, SEP, and UASLP-FAI); the Ministry of Business, Innovation and Employment, New Zealand; the Pakistan Atomic Energy Commission; the Ministry of Science and Higher Education and the National Science Centre, Poland; the Funda\c{c}\~ao para a Ci\^encia e a Tecnologia, Portugal; JINR, Dubna; the Ministry of Education and Science of the Russian Federation, the Federal Agency of Atomic Energy of the Russian Federation, Russian Academy of Sciences, the Russian Foundation for Basic Research, and the National Research Center ``Kurchatov Institute"; the Ministry of Education, Science and Technological Development of Serbia; the Secretar\'{\i}a de Estado de Investigaci\'on, Desarrollo e Innovaci\'on, Programa Consolider-Ingenio 2010, Plan Estatal de Investigaci\'on Cient\'{\i}fica y T\'ecnica y de Innovaci\'on 2017--2020, research project IDI-2018-000174 del Principado de Asturias, and Fondo Europeo de Desarrollo Regional, Spain; the Ministry of Science, Technology and Research, Sri Lanka; the Swiss Funding Agencies (ETH Board, ETH Zurich, PSI, SNF, UniZH, Canton Zurich, and SER); the Ministry of Science and Technology, Taipei; the Thailand Center of Excellence in Physics, the Institute for the Promotion of Teaching Science and Technology of Thailand, Special Task Force for Activating Research and the National Science and Technology Development Agency of Thailand; the Scientific and Technical Research Council of Turkey, and Turkish Atomic Energy Authority; the National Academy of Sciences of Ukraine; the Science and Technology Facilities Council, UK; the US Department of Energy, and the US National Science Foundation.

Individuals have received support from the Marie-Curie program and the European Research Council and Horizon 2020 Grant, contract Nos.\ 675440, 724704, 752730, and 765710 (European Union); the Leventis Foundation; the A.P.\ Sloan Foundation; the Alexander von Humboldt Foundation; the Belgian Federal Science Policy Office; the Fonds pour la Formation \`a la Recherche dans l'Industrie et dans l'Agriculture (FRIA-Belgium); the Agentschap voor Innovatie door Wetenschap en Technologie (IWT-Belgium); the F.R.S.-FNRS and FWO (Belgium) under the ``Excellence of Science -- EOS" -- be.h project n.\ 30820817; the Beijing Municipal Science \& Technology Commission, No. Z191100007219010; the Ministry of Education, Youth and Sports (MEYS) of the Czech Republic; the Lend\"ulet (``Momentum") Program and the J\'anos Bolyai Research Scholarship of the Hungarian Academy of Sciences, the New National Excellence Program \'UNKP, the NKFIA research grants 123842, 123959, 124845, 124850, 125105, 128713, 128786, and 129058 (Hungary); the Council of Scientific and Industrial Research, India; the HOMING PLUS program of the Foundation for Polish Science, cofinanced from European Union, Regional Development Fund, the Mobility Plus program of the Ministry of Science and Higher Education, the National Science Center (Poland), contracts Harmonia 2014/14/M/ST2/00428, Opus 2014/13/B/ST2/02543, 2014/15/B/ST2/03998, and 2015/19/B/ST2/02861, Sonata-bis 2012/07/E/ST2/01406; the National Priorities Research Program by Qatar National Research Fund; the Ministry of Science and Higher Education, project no. 02.a03.21.0005 (Russia); the Tomsk Polytechnic University Competitiveness Enhancement Program; the Programa de Excelencia Mar\'{i}a de Maeztu, and the Programa Severo Ochoa del Principado de Asturias; the Thalis and Aristeia programs cofinanced by EU-ESF, and the Greek NSRF; the Rachadapisek Sompot Fund for Postdoctoral Fellowship, Chulalongkorn University, and the Chulalongkorn Academic into Its 2nd Century Project Advancement Project (Thailand); the Kavli Foundation; the Nvidia Corporation; the SuperMicro Corporation; the Welch Foundation, contract C-1845; and the Weston Havens Foundation (USA).
\end{acknowledgments}

\bibliography{auto_generated}
\cleardoublepage \appendix\section{The CMS Collaboration \label{app:collab}}\begin{sloppypar}\hyphenpenalty=5000\widowpenalty=500\clubpenalty=5000\vskip\cmsinstskip
\textbf{Yerevan Physics Institute, Yerevan, Armenia}\\*[0pt]
V.~Khachatryan, A.M.~Sirunyan$^{\textrm{\dag}}$, A.~Tumasyan
\vskip\cmsinstskip
\textbf{Institut f\"{u}r Hochenergiephysik, Wien, Austria}\\*[0pt]
W.~Adam, F.~Ambrogi, T.~Bergauer, M.~Dragicevic, J.~Er\"{o}, A.~Escalante~Del~Valle, R.~Fr\"{u}hwirth\cmsAuthorMark{1}, M.~Jeitler\cmsAuthorMark{1}, N.~Krammer, L.~Lechner, D.~Liko, T.~Madlener, I.~Mikulec, F.M.~Pitters, N.~Rad, J.~Schieck\cmsAuthorMark{1}, R.~Sch\"{o}fbeck, M.~Spanring, S.~Templ, W.~Waltenberger, C.-E.~Wulz\cmsAuthorMark{1}, M.~Zarucki
% ...
\vskip\cmsinstskip
\dag: Deceased\\
1:  Also at Vienna University of Technology, Vienna, Austria\\
% ...
\end{sloppypar}
%%% END EDITABLE REGION %%%
% skeleton_end
\end{document}